\documentclass[11pt,fleqn]{article}
\usepackage{amsmath}
\usepackage{amssymb}
\usepackage{mathtools}
\usepackage{empheq}
\usepackage[margin=1in,bottom=1.1in]{geometry}
\usepackage{newtxtext}
\usepackage{newtxmath}
\usepackage{dsfont}
\usepackage{wrapfig}
\usepackage{caption} 
\usepackage{subcaption}
\usepackage{mathrsfs}	
\usepackage{cite} 
\usepackage{enumitem}
\usepackage{array}
\usepackage{booktabs}
\usepackage{xurl}
\usepackage{nicefrac}
\usepackage{oubraces}
\usepackage[linkcolor=blue,urlcolor=blue,citecolor=magenta,colorlinks=true]{hyperref}
\usepackage[noabbrev,capitalize]{cleveref}
\AddToHook{cmd/appendix/before}{\crefalias{section}{appendix}}
\usepackage{scalerel}
\usepackage{xcolor}
\usepackage[nolist]{acronym}
\usepackage{braket}
\usepackage{mleftright}\mleftright
\usepackage{booktabs}
\usepackage{accents}
\usepackage{xfrac}
\usepackage{enumitem}
\usepackage[normalem]{ulem}
\usepackage{soul}
\usepackage{microtype}
\usepackage[bottom]{footmisc}
\usepackage{eqparbox}
\makeatletter
\NewDocumentCommand{\eqmathbox}{o O{c} m}{%
  \IfValueTF{#1}
    {\def\eqmathbox@##1##2{\eqmakebox[#1][#2]{$##1##2$}}}
    {\def\eqmathbox@##1##2{\eqmakebox{$##1##2$}}}
  \mathpalette\eqmathbox@{#3}
}
\makeatother
\makeatletter
\RequirePackage{pgfopts}%
\newif\ifpgf@tp@ISO
\pgf@tp@ISOtrue
\newif\ifpgf@tp@boldmath
\pgf@tp@boldmathtrue
\newif\ifpgf@tp@thickmuskip
\pgf@tp@thickmuskiptrue
\newif\ifpgf@tp@defeq
\pgf@tp@defeqtrue
\newif\ifpgf@tp@sbraket
\pgf@tp@sbrakettrue
\newif\ifpgf@tp@sbraket
\pgf@tp@sbrakettrue
\newif\ifpgf@tp@transpose
\pgf@tp@transposetrue
\newif\ifpgf@tp@misc
\pgf@tp@misctrue
\ifpgf@tp@ISO
    \newcommand{\ii}{\mathop{}\!\mathrm{i}\!\mathop{}}%
    \newcommand{\ee}{\mathrm{e}}%
\fi
\ifpgf@tp@boldmath
    \g@addto@macro\bfseries{\boldmath}
\fi
\ifpgf@tp@thickmuskip
\fi
\ifpgf@tp@defeq
    \newcommand*{\defeq}{\mathchoice{\mathrel{\rlap{%
                     \raisebox{0.24ex}{$\m@th\cdot$}}%
                     \raisebox{-0.24ex}{$\m@th\cdot$}}%
                     =}{\mathrel{\rlap{%
                     \raisebox{0.24ex}{$\m@th\cdot$}}%
                     \raisebox{-0.24ex}{$\m@th\cdot$}}%
                     =}{\mathrel{\rlap{%
                     \raisebox{0.08ex}{\small$\m@th\cdot$}}%
                     \raisebox{-0.28ex}{\small$\m@th\cdot$}}%
                     =}{\mathrel{\rlap{%
                     \raisebox{0.08ex}{\tiny$\m@th\cdot$}}%
                     \raisebox{-0.28ex}{\tiny$\m@th\cdot$}}%
                     =}}
    \newcommand*{\eqdef}{\mathchoice{=\mathrel{\rlap{%
                     \raisebox{0.24ex}{$\m@th\cdot$}}%
                     \raisebox{-0.24ex}{$\m@th\cdot$}}}{%
					 =\mathrel{\rlap{%
                     \raisebox{0.24ex}{$\m@th\cdot$}}%
                     \raisebox{-0.24ex}{$\m@th\cdot$}}}{%
					 =\mathrel{\rlap{%
                     \raisebox{0.08ex}{\small$\m@th\cdot$}}%
                     \raisebox{-0.28ex}{\small$\m@th\cdot$}}}{%
					 =\mathrel{\rlap{%
                     \raisebox{0.08ex}{\tiny$\m@th\cdot$}}%
                     \raisebox{-0.28ex}{\tiny$\m@th\cdot$}}}%
                     }
\fi          
\ifpgf@tp@sbraket
    {\catcode`\|=\active
    \xdef\sBraket{\protect\expandafter\noexpand\csname sBraket \endcsname}
    \expandafter\gdef\csname sBraket \endcsname#1{\begingroup
        \ifx\SavedDoubleVert\relax
        \let\SavedDoubleVert\|\let\|\BraDoubleVert
        \fi
        \mathcode`\|32768\let|\BraVert
        \left\lbrack{#1}\right\rbrack\endgroup}
    }
    \def\BraVert{\@ifnextchar|{\|\@gobble}
        {\egroup\,\mid@vertical\,\bgroup}}
    \def\BraDoubleVert{\egroup\,\mid@dblvertical\,\bgroup}
    \let\SavedDoubleVert\relax
\fi
\ifpgf@tp@transpose
    \newcommand*{\transpose}{%
    {\mathpalette\@transpose{}}%
    }
    \newcommand*{\@transpose}[2]{%
    \raisebox{\depth}{$\m@th#1\intercal$}%
    }
\fi
\ifpgf@tp@misc 
    \newcommand{\Group}[2]{\ensuremath{\mathrm{#1}(#2)}}
    \DeclareMathOperator{\re}{Re}
    \DeclareMathOperator{\im}{Im}
    
    \DeclareMathOperator{\diag}{diag}

\fi
\makeatother
\definecolor{c1}{RGB}{0,119,187}
\definecolor{c2}{RGB}{51,187,238}
\definecolor{c3}{RGB}{0,153,136}
\definecolor{c4}{RGB}{238,119,51}
\definecolor{c5}{RGB}{204,51,17}
\definecolor{c6}{RGB}{238,51,119}
\definecolor{c0}{RGB}{187,187,187}
\colorlet{2_1}{c5} 
\colorlet{2_2}{c3}
\colorlet{2_3}{c4}
\colorlet{2_4}{c1}

\newcommand{\rep}[1]{\ensuremath{{\boldsymbol{#1}}}}
\newcommand{\crep}[1]{\ensuremath{\overline{\boldsymbol{#1}}}}
\newcommand{\CenterObject}[1]{\ensuremath{\vcenter{\hbox{#1}}}}

\newcommand{\TowerW}{\ensuremath{W}}
\newcommand{\TowerN}{\ensuremath{N}}
\newcommand{\TowerR}{\ensuremath{R}}
\newcommand{\TowerI}{\ensuremath{I}}
\newcommand{\OrbW}{\ensuremath{s_W}}
\newcommand{\OrbN}{\ensuremath{s_N}}
\newcommand{\ModularWeight}{\ensuremath{n}}

\newcommand{\Xb}[0]{\overline{X}}
\newcommand{\Yb}[0]{\overline{Y}}
\newcommand{\Zb}[0]{\overline{Z}}
\numberwithin{equation}{section}

\newcommand{\mytitle}{Flavor Symmetries and Winding Modes}
\begin{document}
\begin{titlepage}
\thispagestyle{empty}\bigskip
\begin{flushright}
UCI-HEP-TR-2025-06%
\end{flushright}
\bigskip
\begin{center}
{\Huge\sffamily\bfseries\mytitle}%
\par\bigskip\renewcommand*{\thefootnote}{\fnsymbol{footnote}}

  Xueqi Li\rlap{,}$^{a}$\footnote{E-mail: \texttt{xueqi.li@uci.edu}}~~
  Xiang-Gan Liu\rlap{,}$^{a}$\footnote{E-mail: \texttt{xianggal@uci.edu}}~~
  Hans Peter Nilles\rlap{,}$^{b}$\footnote{E-mail: \texttt{nilles@th.physik.uni-bonn.de}}~~
  Michael Ratz$^{a}$\footnote{E-mail: \texttt{mratz@uci.edu}}~~and~~
  Alex Stewart$^{a}$\stepcounter{footnote}\stepcounter{footnote}\footnote{Email: \texttt{ajstewa1@uci.edu}}
  \\*[20pt]
  \begin{minipage}{0.75\linewidth}
  \begin{center}
  {\itshape\small$^a$ Department of Physics and Astronomy, University of California, Irvine,\\ CA~92697-4575, USA}\\[5mm]
  {\itshape\small$^b$ Bethe Center for Theoretical Physics and Physikalisches Institut
der Universit\"at Bonn,\\
  Nussallee~12, 53115~Bonn, Germany}
  \end{center}
  \end{minipage}
\end{center}  
\bigskip

\begin{abstract}
 Modular flavor symmetries have been proposed as a new way to address the flavor problem. 
 It is known that they can emerge from string compactifications. 
 We discuss this connection in detail, and show how the congruence subgroups of $\Group{SL}{2,\mathds{Z}}$, which underlie many modular flavor symmetries, emerge from stringy duality symmetries by orbifolding. 
 This requires an analysis of massive states, which reveals a picture that is more intricate than the well-known situation on the torus. 
 It involves towers of states of different quantum numbers, related by modular transformations. 
 Members of different towers become massless at different points in moduli space. 
 We also show that, at least in the $\mathds{Z}_3$ orbifold, the string selection rules can be understood as discrete remnants of continuous gauge symmetries. 
 Non-Abelian discrete flavor symmetries arise as relics of various, relatively misaligned, continuous Abelian gauge symmetries. 
 The generators of these \Group{U}{1} symmetries give rise to $\mathcal{CP}$-violating  Clebsch--Gordan coefficients. 
 If the modulus settles close to a critical point, the corresponding gauge bosons may be light enough to be searched for at future colliders. 
\end{abstract}
\end{titlepage}
\clearpage\renewcommand*{\thefootnote}{\arabic{footnote}}

\section{Introduction}
\label{sec:Introduction}

Modular symmetries play a key role in string theory \cite{Giveon:1994fu,DHoker:2022dxx}. 
If string theory is to describe our world, it should allow us to address the open questions of the \ac{SM}. 
One of these questions is the flavor problem, which concerns the understanding of the fermion masses, mixing angles and $\mathcal{CP}$ phases. 
In fact, the vast majority of continuous parameters of the \ac{SM} reside in the flavor sector, and so there is no fully convincing framework that can explain these parameters, let alone with a precision that is comparable to the experimental error bars.

Modular symmetries have been applied to the flavor problem \cite{Feruglio:2017spp}, which has led to the scheme of modular flavor symmetries (see e.g.\ \cite{Feruglio:2019ybq,Almumin:2022rml,Kobayashi:2023zzc,Ding:2023htn,Nilles:2023shk,Ding:2024ozt,Nilles:2024iqp} for reviews and more references). 
This framework employs the so-called finite modular groups along with the corresponding \ac{VVMF}~\cite{Liu:2021gwa}. 
In most of the examples, the finite modular groups emerge as the quotients of $\Gamma\defeq\Group{SL}{2,\mathds{Z}}/\mathds{Z}_2$ and some appropriate congruence subgroup, $\Gamma(N)$. 

What is the origin and interpretation of such modular flavor symmetries?
In the \ac{BU} framework the action typically has an $\Group{SL}{2,\mathds{Z}}$ symmetry. 
This raises the question what modular flavor symmetries are if they are not symmetries of the action. 
One may also wonder whether are there additional, firm predictions allowing us to test this scheme. 

Since modular flavor symmetries have an explicit realization in string theory \cite{Chun:1989se,Lauer:1989ax,Lerche:1989cs,Chun:1991js,Quevedo:1996sv}, it is instructive to study these questions in this \ac{TD} framework. 
Many important aspects can be understood in the Narain formalism \cite{Nilles:2020kgo,Nilles:2020gvu,Baur:2021bly,Baur:2024qzo}. 
However, in the analyses thus far, the focus has been on states that remain massless for all values of the moduli. 
It is the purpose of this work to point out that massive states play a key role in the detailed understanding of modular flavor symmetries. 
As we shall see in detail, modular transformations relate towers of states. 
More specifically, we will see that transformations within the congruence subgroup $\Gamma(N)$ are maps between physically equivalent towers whereas the other $\Group{SL}{2,\mathds{Z}}$ transformations relate tower states carrying different quantum numbers. 

The fact that different towers are related by modular transformations has important implications for the understanding and implications of so-called ``traditional'' flavor symmetries. 
Furthermore, a more detailed analysis of modular and traditional flavor symmetries provides us with important insights on the $\mathcal{CP}$ transformation. 
Whether or not $\mathcal{CP}$ is broken also depends on whether there are relations between the masses of massive states. 
As we shall see, the massive states are also instrumental when establishing that the traditional flavor symmetries are gauged.  

Finally, towers of states are believed to be essential for the consistency of a theory of quantum gravity (see e.g.\ \cite{Palti:2019pca}). 
Understanding the properties of these towers on orbifolds rather than tori therefore has implications far beyond the flavor problem.  

The outline of this paper is as follows. 
In \Cref{sec:Modular_transformations} we will collect some basic facts on modular transformations. 
\Cref{sec:Orbifolding} is devoted to a detailed analysis of the moduli space of the K\"ahler modulus $\varrho$, and fields that become massless at special values of $\varrho$.
In \Cref{sec:Towers} we study the towers of massive states and how they are related by modular transformations. 
\Cref{sec:CP} concerns the $\mathcal{CP}$ transformation. 
In \Cref{sec:Applications} contains a discussion as well as applications of our analysis for flavor model building, and \Cref{sec:Summary} consists of our summary. 
Various details are deferred to appendices. 

\section{Modular transformations}
\label{sec:Modular_transformations}

We begin by discussing modular transformations.
In string theory, target space modular transformations are symmetries of mass spectra \cite{Giveon:1994fu}. 
In the simplest case, this could be an exchange of \ac{KK} and winding modes on a circle. 

In this work, we are interested in toroidal orbifold compactifications of the heterotic string \cite{Dixon:1985jw,Dixon:1986jc}.
Before discussing orbifolds in \Cref{sec:Orbifolding}, we first collect some basic facts on modular transformations on the torus with many details being deferred to Appendix~\ref{app:Modular_transformations}.
The so-called modular transformations act on the moduli of the torus, $\tau$ and $\varrho$,
\begin{subequations}\label{eq:modulusTrans}
\begin{align}
 \tau&\mapsto\frac{a_\tau\,\tau+b_\tau}{c_\tau\,\tau+d_\tau}\;,\\
 \varrho&\mapsto\frac{a_\varrho\,\varrho+b_\varrho}{c_\varrho\,\varrho+d_\varrho}\;.  
\end{align}   
\end{subequations}
Here, 
\begin{equation}
 a_i\,d_i-b_i\,c_i=1  
\end{equation}
for $i\in\{\tau,\varrho\}$.
As detailed in \Cref{app:Modular_transformations}, a closed string on a 2-dimensional torus carries two \ac{KK} quantum numbers, $n_1$ and $n_2$, as well as two winding numbers, $w^1$ and $w^2$. 
The modular transformations on the torus act on the \ac{KK} and winding numbers as   
\begin{equation}\label{eq:modular_transformation_KK_and_winding_numbers}
 \begin{pmatrix}
  w^1\\ w^2\\  n_1\\ n_2
 \end{pmatrix}
 \mapsto 
 \mathsf{M}\cdot\begin{pmatrix}
  w^1\\ w^2\\  n_1\\ n_2
  \end{pmatrix}
\end{equation}
with some appropriate integer matrix $\mathsf{M}$.
The simultaneous transformations \eqref{eq:modulusTrans} and \eqref{eq:modular_transformation_KK_and_winding_numbers} leave the mass of a string state invariant, cf.\ our discussion in Appendix~\ref{app:Modular_transformations}.
The group of modular transformations on the torus is generated by the matrices (cf.\ \cite{Nilles:2020gvu})
\begin{subequations}
  \label{eq:STmatrices}
\begin{align}
\mathsf{S}_\tau&=\begin{pmatrix*}[r]
  0 & -1 & 0 & \phantom{-}0 \\
 1 & 0 & 0 & 0 \\
 0 & 0 & 0 & -1 \\
 0 & 0 & 1 & 0 
\end{pmatrix*}\;, & 
\mathsf{T}_\tau&=\begin{pmatrix*}[r]
  \phantom{-}1 & -1 & \phantom{-}0 &\phantom{-} 0 \\
 0 & 1 & 0 & 0 \\
 0 & 0 & 1 & 0 \\
 0 & 0 & 1 & 1 \\
\end{pmatrix*}\;,\\ 
\mathsf{S}_\varrho&=\begin{pmatrix*}[r]
  \phantom{-}0 & 0 & \phantom{-}0 & -1 \\
 0 & 0 & 1 & 0 \\
 0 & -1 & 0 & 0 \\
 1 & 0 & 0 & 0 \\
\end{pmatrix*}\;, &
\mathsf{T}_\varrho&=\begin{pmatrix*}[r]
  1 & \phantom{-}0 & \phantom{-}0 & \phantom{-}0 \\
 0 & 1 & 0 & 0 \\
 0 & -1 & 1 & 0 \\
 1 & 0 & 0 & 1 \\
\end{pmatrix*}
 \;.
\end{align}  
\end{subequations}
These matrices satisfy the relations
\begin{subequations}\label{eq:S_T_relations_torus}
\begin{align}
\mathsf{S}^4_\tau&=(\mathsf{S\,T})_\tau^3=\mathsf{S}^2_\tau\,\mathsf{T}_\tau\,\mathsf{S}^{-2}_\tau\,\mathsf{T}^{-1}_\tau=\mathsf{S}^4_\varrho=(\mathsf{S\,T})_\varrho^3=\mathsf{S}^2_\varrho\,\mathsf{T}_\varrho\,\mathsf{S}^{-2}_\varrho\,\mathsf{T}^{-1}_\varrho=\mathds{1}_4\;,\label{eq:S_T_relations_torus_a}\\
\mathsf{S}^2_\tau&=\mathsf{S}^2_\varrho=-\mathds{1}_4\;.\label{eq:S_T_relations_torus_b}
\end{align}
\end{subequations}
From \eqref{eq:S_T_relations_torus_a}, it follows that there are two $\Group{SL}{2,\mathds{Z}}$ groups where, as is evident from \eqref{eq:S_T_relations_torus_b}, the $\mathsf{S}^2$ transformation is identified. 
That is, the modular group of the torus is given by $\bigl(\Group{SL}{2,\mathds{Z}}_\tau\times\Group{SL}{2,\mathds{Z}}_\varrho\bigr)/\mathds{Z}_2^{\mathsf{S}^2}$.
As we shall show in \Cref{sec:Orbifolding}, orbifolding reduces this symmetry to a congruence subgroup of $\Group{SL}{2,\mathds{Z}}$.

Before discussing this in detail, let us collect some basic facts on the upstairs theory on the torus, focusing on $\tau=\omega\defeq\ee^{2\pi\ii/3}$. 
This leads to a $\Group{U}{1}\times\Group{U}{1}$ theory that gets enhanced to \Group{SU}{3} at $\varrho=\omega$. 
The detailed discussion can be found e.g.\ in \cite{Font:2020rsk}.  
The key point for our discussion is that, on the torus, there is \emph{one} special point in $\varrho$ moduli space at which additional states, including gauge fields, become massless. 
These states stem from one set of \ac{KK} and winding mode towers. 
As we shall see next, in the downstairs theory, i.e.\ after orbifolding, the situation is more intricate. 

\section{Orbifolding and residual gauge symmetries}
\label{sec:Orbifolding}

To obtain a chiral gauge theory in four dimensions, we orbifold the torus.
As is well known, orbifolding breaks symmetries, removes states, and leads to the appearance of twisted sectors \cite{Dixon:1985jw,Dixon:1986jc}. 

Here, we focus on the $\mathds{Z}_3$ orbifold, in which we mod out a transformation of order 3.
This transformation acts on the string coordinates $X^1$ and $X^2$ of the torus as
\begin{equation}\label{eq:orbifold_theta_def}
 X^+\defeq \frac{1}{\sqrt{2}}\left(X^1+\ii X^2\right)\xmapsto{~\theta~}\omega^2\,X^+\;,  
\end{equation}
where $\omega\defeq\ee^{2\pi\ii/3}$.
Geometrically, this operation is a rotation by $2\pi/3$.
The modular transformations discussed in \Cref{sec:Modular_transformations} refer to the theory on the torus. 
Orbifolding breaks parts of the symmetries and projects out states. 
In order to be able to mod out a $\mathds{Z}_3$ symmetry, the modulus $\tau$ has to be ``frozen'' at $\omega$.

In slightly more detail, under the $\mathds{Z}_3$ orbifold action $\theta$,
\begin{subequations}\label{eq:orbifold_action_def}
\begin{align}
  \begin{pmatrix}
    w^1\\ w^2
  \end{pmatrix}
  &\xmapsto{~\theta~} \begin{pmatrix*}[r]
    \phantom{-}0 &-1\\ 1 & -1
  \end{pmatrix*}\,\begin{pmatrix}
    w^1\\ w^2
  \end{pmatrix}\eqdef \widehat{\theta}_w\,\begin{pmatrix}
    w^1\\ w^2
  \end{pmatrix}\;,\\ 
  \begin{pmatrix}
    n_1\\ n_2
  \end{pmatrix}
  &\xmapsto{~\theta~} \begin{pmatrix*}
    -1 &-1\\ \phantom{-}1 & 0
  \end{pmatrix*}\,\begin{pmatrix}
    n_1\\ n_2
  \end{pmatrix}\eqdef \widehat{\theta}_n\,\begin{pmatrix}
    n_1\\ n_2
  \end{pmatrix}\;.
\end{align}  
\end{subequations}
As far as the winding and \ac{KK} numbers are concerned, this transformation equals $(\mathsf{S\,T})_\tau$.
Note, however, that the orbifold action is not identical to this transformation as it also involves \eqref{eq:orbifold_theta_def}, see \Cref{sec:ST_transformation_and_orbifold_twist} for a more detailed discussion. 

In what follows, we will establish that, after orbifolding, the modular symmetry of the K\"ahler modulus is the congruence subgroup $\Gamma(3)$, which we will define in \Cref{sec:Gamma(3)}, rather than $\Group{SL}{2,\mathds{Z}}$. 
In order to see this, we will show that there exist physically different continuous gauge symmetries $[\Group{U}{1}\times\Group{U}{1}]^{(\varrho_\mathrm{crit})}$ that become exact at different critical values of $\varrho$.
As we shall see, away from the critical points these continuous symmetries give rise to a $\Delta(27)$ discrete symmetry, which gets enhanced to $\Delta(54)$ by the modular symmetry.

\subsection{Twisted states}

A key feature of orbifolds is the existence of twisted sectors. 
In the context of the heterotic string the corresponding strings are only closed upon the orbifold identification \eqref{eq:orbifold_theta_def}. 
They can be associated to space group elements $(\theta^k,m_\alpha\,e_\alpha)$ consisting of the $k$\textsuperscript{th} power of the orbifold twist $\theta$ and integer linear combinations of the basis vectors of the torus $e_\alpha$ (cf.\ \cite{Dixon:1985jw,Dixon:1986jc,Vaudrevange:2008sm,Ramos-Sanchez:2008nwx}).
For the $\mathds{Z}_3$ orbifold, twisted states can have either $k=1$ or $k=2$. 

\begin{figure}[htb]
  \centering\subcaptionbox{Fundamental domains of torus and orbifold (hatched).\label{fig:fundamental_domains}}
  {\includegraphics{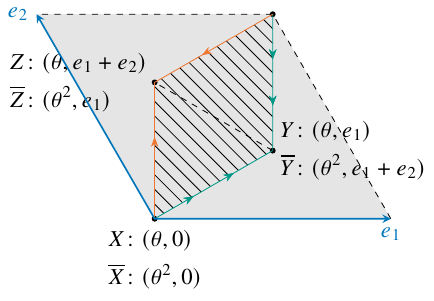}}\qquad
  \subcaptionbox{Orbifold pillow.\label{fig:Z3_orbifold_pillow}}
  {\includegraphics{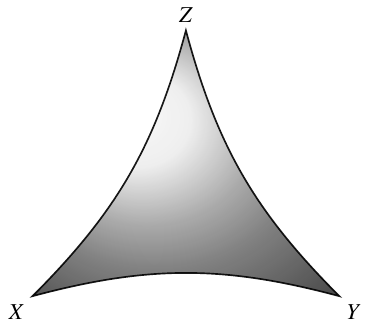}}
  \caption{(a) $\mathds{Z}_3$ plane and (b) orbifold pillow.}
  \label{fig:Z3_orbifold_plane}    
\end{figure}

The $\mathds{T}^2/\mathds{Z}_3$ orbifold has three fixed points with complex coordinates $X=0$, $Y=\ee^{\ii\pi/6}/\sqrt{3}$ and $Z=\ee^{\ii\pi/2}/\sqrt{3}$. Given that 
\begin{subequations}
\begin{align}
  \omega\,Y&=Y-1\;,\\ 
  \omega\,Z&=Z-1-\omega\;,
\end{align}  
\end{subequations} 
one finds that the fixed points $X$, $Y$ and $Z$ are associated with the space group elements $(\theta,0)$, $(\theta,e_1)$ and $(\theta,e_1+e_2)$, respectively, as illustrated in \Cref{fig:fundamental_domains}. 
In the second twisted sector, they correspond to $(\theta^2,0)$, $(\theta^2,e_1+e_2)$ and $(\theta^2,e_1)$, respectively.
There exist bases in which the twisted sector states are localized at the fixed points they correspond to. 
In such a basis, they correspond to the twist fields $\sigma_i$ which have a definite center-of-mass position (cf.\ \cite[discussion around Equation~(8)]{Lauer:1989ax}).
For later convenience, we collect these three twisted states in a triplet, 
\begin{equation}\label{eq:Phi_-2/3}
  \Phi_{-\nicefrac{2}{3}}\defeq\begin{pmatrix}
    X\\  Y\\ Z
  \end{pmatrix}\;.
\end{equation}
Here, the subscript $\ModularWeight^{(\varrho)}_\Phi=-\nicefrac{2}{3}$ indicates the modular weight of these states.
Note that the representation \eqref{eq:Phi_-2/3} is unique up to phase conventions and relabeling of the fixed points. 
The orbifold may then be pictured as a pillow with the edges of the fundamental domain identified as sketched in \Cref{fig:fundamental_domains}, and the corners given by the three fixed points (cf.\ \Cref{fig:Z3_orbifold_pillow}). 
Due to this interpretation we will refer to \eqref{eq:Phi_-2/3} as ``localization eigenstates''.

\subsection{Massless winding modes and gauge bosons}
\label{sec:MasslesWinding}

The so-called untwisted sector consists of orbifold-invariant combinations of states on the torus.
The latter are characterized by left- and right-moving momenta, $p_\mathrm{L}$ and $p_\mathrm{R}$, $\widehat{N} = (w^1, w^2, n_1, n_2)^{\transpose}$, i.e.\ its \ac{KK} and winding numbers, as well as oscillator numbers, $N_\mathrm{L}$ and $N_\mathrm{R}$.
In our discussion we will focus on the \ac{NS} sector.

We are interested in massless gauge fields, for which $p_\mathrm{R}^2=0$, and $N_\mathrm{R}=1/2$.
At $\tau=\omega$ the requirement that $p_\mathrm{R}^2$ vanishes implies that (cf.\ \Cref{eq:p_R_squared})
\begin{equation}\label{eq:p_R_squared=0}
  n_2-\omega\,n_1-\bar{\varrho}\,\bigl(w^1+\omega\,w^2\bigr)=0\;.
\end{equation}
One then distinguishes between winding and oscillator modes. 
On the torus, it is often convenient to associate the oscillator modes with the Cartan generators, and the winding modes with raising and lowering operators. 
However, as is well known, orbifolding leads to \Group{U}{1} symmetries in which the Cartan generators get projected out, such that the gauge fields are in fact winding modes \cite{Dijkgraaf:1989hb}.    
This is, in particular, true in the case at hand \cite{Ibanez:1990ju}.

As is well known, at $\varrho=\omega$, 4D gauge fields become massless \cite{Ibanez:1990ju,Beye:2014nxa}, giving rise to an unbroken $\Group{U}{1}\times\Group{U}{1}$, which we will describe in more detail in \Cref{sec:massless_untwisted_fields_at_omega}.
Before doing so, let us discuss the question whether other points in moduli space give rise to additional gauge symmetries. 
The answer to this question is affirmative.

In order to see this, consider massless winding modes, for which the expression in parentheses, $w^1+\omega\,w^2$, in \eqref{eq:p_R_squared=0} cannot vanish. 
We thus have 
\begin{equation}\label{eq:rho_crit1}
  \varrho=\frac{\omega\,n_2-n_1}{\omega\,w^1+w^2}\;.
\end{equation}
This can be viewed as a $2\times2$ integer matrix $\begin{psmallmatrix} n_2 & -n_1 \\ w^1 & w^2 \end{psmallmatrix}$ acting as a fractional linear transformation on $\omega$. 
The level matching condition \eqref{eq:level_matching} for the massless gauge bosons in consideration, where $N_\mathrm{R}=1/2$ and $p_\mathrm{R}^2=0$, implies that $w^1\, n_1 + w^2\, n_2 = 1$. This means that the determinant of this $2\times2$ integer matrix is 1.
In other words, \eqref{eq:rho_crit1} implies that for some winding modes to become massless, $\varrho$ has to be the image of $\omega$ under an $\Group{SL}{2,\mathds{Z}}_\varrho$ transformation. 

The alert reader may now suspect that these critical values of $\varrho$ are physically equivalent because they are related via modular transformations of $\Group{SL}{2,\mathds{Z}}_\varrho$.
It turns out that this is not the case. 
The fact that not all $\Group{SL}{2,\mathds{Z}}_\varrho$ transformations relate physically equivalent configurations can be established by comparing the properties of the massless winding modes at various special values. 
We start with $\omega+n$ where $n$ is an integer. 
As we shall see next, the physical properties of massless states at $\omega+n$ and $\omega+n'$ are identical only if $n'-n$ is divisible by 3. 
This means that only a subset of the $\Group{SL}{2,\mathds{Z}}_\varrho$ transformations relates physically identical states, which, as we will lay out in detail, leads to the emergence of congruence subgroups of $\Group{SL}{2,\mathds{Z}}_\varrho$.

\subsubsection{Massless untwisted fields at \texorpdfstring{$\varrho=\omega$}{rho=omega}}
\label{sec:massless_untwisted_fields_at_omega}

Let us start by inspecting $\varrho=\omega$.
Regarding $\begin{psmallmatrix} n_2 & -n_1 \\ w^1 & w^2 \end{psmallmatrix}$ as an $\Group{SL}{2,\mathds{Z}}_\varrho$ transformation also allows us infer that there are six combination of \ac{KK} and winding numbers that lead to massless states on the torus. 
They correspond to the stabilizer of $\omega$, $\mathds{Z}_3\times\mathds{Z}_2$, which has 6 elements,
\begin{equation}\label{eq:massless_KK_and_winding_torus}
  \begin{pmatrix}n_2 & -n_1\\ w^1& w^2\end{pmatrix} 
    \in\left\{  
      \pm\begin{pmatrix}1 & 0\\ 0 & 1\end{pmatrix}, 
      \pm\begin{pmatrix}0 & 1\\ -1& -1\end{pmatrix}, 
      \pm \begin{pmatrix}-1 & -1\\ 1 &0 \end{pmatrix}\right\}\;.
\end{equation} 
While these states transform nontrivially under the orbifold action \eqref{eq:orbifold_action_def}, they give rise to 2 orbifold-invariant combinations. 
They correspond to the two gauge fields of a local $\Group{U}{1}\times\Group{U}{1}$ symmetry \cite{Ibanez:1990ju,Beye:2014nxa}, which we will refer to as $[\Group{U}{1}\times\Group{U}{1}]^{(\omega)}$.

In addition, three untwisted ``matter'' fields are massless.
They correspond to the vertex operators~\cite{Ibanez:1990ju,Beye:2014nxa} (cf.~\Cref{app:Ui_omega})
\begin{subequations}\label{eq:untwisted_matter_fields}
\begin{align}
  U_1^{(\omega)} &\defeq \frac{1}{\sqrt{3}}\left( -\ii (\ii \partial X_\mathrm{L}^-) + V_\omega (\widehat{N}_4)^{(\omega)} + V_\omega (-\widehat{N}_4)^{(\omega)} \right) \otimes (\ii \partial X_\mathrm{R}^+)\;,\\
  U_2^{(\omega)} &\defeq \frac{1}{\sqrt{3}}\left( -\ii (\ii \partial X_\mathrm{L}^-) + \omega V_\omega (\widehat{N}_4)^{(\omega)} + \omega^2\, V_\omega (-\widehat{N}_4)^{(\omega)} \right) \otimes (\ii \partial X_\mathrm{R}^+)\;,\\
  U_3^{(\omega)} &\defeq \frac{1}{\sqrt{3}}\left( -\ii (\ii \partial X_\mathrm{L}^-) + \omega^2 V_\omega (\widehat{N}_4)^{(\omega)} + \omega\, V_\omega (-\widehat{N}_4)^{(\omega)} \right) \otimes (\ii \partial X_\mathrm{R}^+)\;.
\end{align}
\end{subequations}
where 
\begin{subequations}
\begin{align}
\ii \partial X^{\pm}_\mathrm{L/R} &\defeq \frac{1}{\sqrt{2}} \left(\ii\partial X^1_\mathrm{L/R} \pm  \ii \partial X^2_\mathrm{L/R}\right) \;,
\end{align}
\end{subequations}
and the $V_{\omega^i}(\widehat{N})^{(\varrho)}$ are various orbifold invariant combinations of untwisted states, as detailed in \Cref{app:orb_inv_vertex_ops}. 
At $\varrho = \omega$, the relevant \ac{KK} and winding numbers are $\widehat{N}_4 = (-1,0,-1,1)^{\transpose}$, where the $4$ anticipates the labeling of these states by different representations of \Group{\Delta}{54}.
As discussed in more detail in \Cref{sec:untwisted_vertex_operators_on_orbifold,sec:untwisted_vertex_operators_at_omega}, the subscripts of the $\mathsf{E}$ operators correspond to \Group{SU}{3} Dynkin labels. For instance, ``$-1,-1$'' means $-\alpha_{(1)}-\alpha_{(2)}$, where $\alpha_{(1)}$ and $\alpha_{(2)}$ are the two simple roots of \Group{SU}{3}. 
This means that the $[\Group{U}{1}\times\Group{U}{1}]^{(\omega)}$ charges of the $U_i^{(\omega)}$ are given by components of the simple roots of \Group{SU}{3}~\cite{Ibanez:1990ju,Beye:2014nxa}.

\begin{wraptable}[13]{r}{6cm}
\centering\vspace*{-1ex}\begin{tabular}{cccc}
  \toprule
  state & $[\Group{U}{1}\times\Group{U}{1}]^{(\omega)}$\\ 
  \midrule
  $U_1^{(\omega)}$ & $(\sqrt{2},0)$\\
  $U_2^{(\omega)}$ & $(-1/\sqrt{2},\sqrt{3/2})$\\
  $U_3^{(\omega)}$ & $(-1/\sqrt{2},-\sqrt{3/2})$\\
  \midrule
  $X^{(\omega)}$ & $(\sqrt{2}/3,0)$\\
  $Y^{(\omega)}$ & $(-1/\sqrt{18},1/\sqrt{6})$\\
  $Z^{(\omega)}$ & $(-1/\sqrt{18},-1/\sqrt{6})$\\
  \bottomrule
  \end{tabular}
  \caption{$[\Group{U}{1}\times\Group{U}{1}]^{(\omega)}$ charges of the untwisted and twisted matter fields.}
  \label{tab:U1timesU1_omega_1}
\end{wraptable}

The $U_i^{(\omega)}$ fields are charge eigenstates of the $[\Group{U}{1}\times\Group{U}{1}]^{(\omega)}$ symmetry the charges of which are given in \Cref{tab:U1timesU1_omega_1}.
The twisted states localized at the three fixed points are also charged under $[\Group{U}{1}\times\Group{U}{1}]^{(\omega)}$, and can be rotated into a basis of charge eigenstates~\cite[Equation~(14)]{Chun:1989se}\begin{equation}\label{eq:Phi_U_omega}
  \Phi_{-\nicefrac{2}{3}}^{(\omega)}\defeq\mathcal{U}_{(\omega)}\,\Phi_{-\nicefrac{2}{3}}
\end{equation}
with the localization eigenstates $\Phi_{-\nicefrac{2}{3}}$ from \eqref{eq:Phi_-2/3} and
\begin{equation}\label{eq:U_omega}
  \mathcal{U}_{(\omega)}= \frac{1}{\sqrt{3}} 
  \begin{pmatrix}
    \omega  & 1 & 1 \\
     1 & \omega  & 1 \\
    1 & 1 & \omega
  \end{pmatrix}\;.
\end{equation}
Notice that, in order to be able to describe the K\"ahler modulus $\varrho$ by a left-chiral superfield, our choice of $\mathcal{U}_{(\omega)}$ differs from \cite[Equation~(14)]{Chun:1989se} by Hermitean conjugation.\footnote{We denote the K\"ahler modulus by $\varrho$ whereas representation matrices are indicated by $\rho$.}   
The transformation \eqref{eq:U_omega} fulfills
\begin{equation}\label{eq:U_omega=T.Sinv.T}
  \mathcal{U}_{(\omega)}=-\ii\omega^2\,\rho(\mathsf{T}^{-1}\,\mathsf{S}\,\mathsf{T}^{-1})\;.
\end{equation}
This expression involves (cf.\ \cite[Equation~(7)]{Chun:1989se} and \cite[Equation~(4.36)]{Lauer:1990tm}\footnote{Note our expression for $\rho(\mathsf{T})$ differs from \cite{Chun:1989se,Lauer:1990tm} by an unphysical overall phase $\omega^2$.})
\begin{subequations}\label{eq:rho(S)_and_rho(T)}
\begin{align}
  \rho(\mathsf{S}) &= \frac{\ii}{\sqrt{3}}
  \begin{pmatrix}
    1 & 1 & 1 \\
    1 & \omega ^2 & \omega  \\
    1 & \omega  & \omega ^2 
  \end{pmatrix}\;,\label{eq:rho(S)}\\
  \rho(\mathsf{T})&=\diag (\omega^2,1,1) \;.\label{eq:rho(T)}
\end{align}  
\end{subequations}
Notice that we have fixed the phase of the $\mathsf{S}$ transformation in such a way that $\rho\bigl((\mathsf{S}\,\mathsf{T})^3\bigr)=\mathds{1}$, cf.\ \cite[Discussion around Equation~(11)]{Nilles:2020kgo}. In addition, the matrices $\mathsf{S}$ and $\mathsf{T}$ turn out to be reducible representation matrices of the finite modular group $T'$, which can be decomposed into the $\rep{2'}\oplus\rep{1}$ representation of $T'$. 
For later convenience let us note that 
\begin{equation}\label{eq:S^2=C}
  \rho(\mathsf{S})^2=-\begin{pmatrix}
    1 & 0 & 0\\
    0 & 0 & 1\\
    0 & 1 & 0
  \end{pmatrix}\eqdef \rho(\mathsf{C})\;.
\end{equation}
Here and in what follows, $\rho$ denotes an appropriate representation. 

The $[\Group{U}{1}\times\Group{U}{1}]^{(\omega)}$ charges of the three components of $\Phi_{-\nicefrac{2}{3}}^{(\omega)}$ are shown in \Cref{tab:U1timesU1_omega_1}.
One can determine the $[\Group{U}{1}\times\Group{U}{1}]^{(\omega)}$ generators in the basis of localization eigenstates \eqref{eq:Phi_-2/3},
\begin{subequations}\label{eq:ti_at_omega_localization_eigenstates}
\begin{align}
  \mathsf{t}^{(\omega)}_1 
  &= \bigl(\mathcal{U}_{(\omega)}\bigr)^\dagger\,\mathsf{t}_1\,\mathcal{U}_{(\omega)}=\frac{1}{2\sqrt{3}} \begin{pmatrix}
    0 & \omega ^2 & \omega ^2 \\
    \omega  & 0 & 1 \\
     \omega  & 1 & 0
  \end{pmatrix}\;,\\ 
  \mathsf{t}^{(\omega)}_2 &=\bigl(\mathcal{U}_{(\omega)}\bigr)^\dagger\,\mathsf{t}_2\,\mathcal{U}_{(\omega)}= \frac{\ii}{2\sqrt{3}} \begin{pmatrix}
    0 & -\omega ^2 & \omega ^2 \\
     \omega  & 0 & -1 \\
     -\omega  & 1 & 0 
  \end{pmatrix}\;.
\end{align}
\end{subequations}
Here, the $\mathsf{t}_i$ are diagonal and given by
\begin{subequations}\label{eq:diagonal_t_i}
\begin{align}
  \mathsf{t}_1&=\frac{1}{2\sqrt{3}}\diag(2,-1,-1)\;,\\
  \mathsf{t}_2&=\frac{1}{2}\diag(0,1,-1)\;.
\end{align}  
\end{subequations}
For later convenience, let us define 
\begin{equation}\label{eq:tomegapm}
  \mathsf{t}^{(\omega)}_\pm\defeq
  \frac{1}{\sqrt{2}}\bigl(\mathsf{t}^{(\omega)}_1\pm\ii\mathsf{t}^{(\omega)}_2\bigr)\;,
\end{equation}
and note that 
\begin{equation}\label{eq:Coupling_Phi_bar_Phi_at_omega}
 \begin{pmatrix}
  \Phi_{-\nicefrac{2}{3}}^\dagger\,
  \mathsf{t}^{(\omega)}_+
  \,\Phi_{-\nicefrac{2}{3}}\\
  \Phi_{-\nicefrac{2}{3}}^\dagger\,
  \mathsf{t}^{(\omega)}_-
  \,\Phi_{-\nicefrac{2}{3}}
  \end{pmatrix}  
  =\frac{1}{\sqrt{6}}\begin{pmatrix}
    \omega ^2\,\overline{X}\,Y+\overline{Y}\,Z+\omega\,\overline{Z}\, X\vphantom{\Bigl(\Bigr)}\\
    \omega\,\overline{Y}\, X+\overline{Z}\,Y+\omega ^2\,\overline{X}\,Z\vphantom{\Bigl(\Bigr)}
  \end{pmatrix}\;.
\end{equation}
Here, we put the expressions in a doublet because they get exchanged under the $\mathsf{S}^2$ transformation from \eqref{eq:S^2=C}.

Since there are two gauge bosons and three massless untwisted matter fields, we can form one $D$-flat combination of the $U_i^{(\omega)}$, which we associate with the $\varrho$ modulus.
In the fixed gauge \eqref{eq:untwisted_matter_fields}, 
\begin{equation}
  \varrho-\omega=\frac{1}{\sqrt{3}}\Bigl(U_1^{(\omega)}+U_2^{(\omega)}+U_3^{(\omega)}\Bigr)\;.
\end{equation}
This reveals that $\partial X^-_\mathrm{L} \partial X^+_\mathrm{R}$ is the vertex operator for $\varrho$, as expected.
Away from $\varrho=\omega$, the two orthogonal linear combinations of the $U_i^{(\omega)}$ get eaten by the $[\Group{U}{1}\times\Group{U}{1}]^{(\omega)}$ gauge bosons via the stringy Higgs mechanism \cite{Ibanez:1990ju}.

It is convenient to arrange the (massive) gauge fields in a doublet $W_{(\omega)}\defeq\Bigl(W^+_{(\omega)},W^-_{(\omega)}\Bigr)^{\transpose}$.
The contraction \eqref{eq:Coupling_Phi_bar_Phi_at_omega} can then be recast as
\begin{equation}\label{eq:W_pm_omega}
  \begin{pmatrix}
    \Phi_{-\nicefrac{2}{3}}^\dagger\,\mathsf{t}^{(\omega)}_+\,\Phi_{-\nicefrac{2}{3}}\vphantom{\Bigl(\Bigr)}\\
    \Phi_{-\nicefrac{2}{3}}^\dagger\,\mathsf{t}^{(\omega)}_-\,\Phi_{-\nicefrac{2}{3}}\vphantom{\Bigl(\Bigr)}
  \end{pmatrix}
  \cdot 
  \begin{pmatrix}
    0 & 1\vphantom{\Bigl(\Bigr)}\\ 1 &0\vphantom{\Bigl(\Bigr)}
  \end{pmatrix}
  \cdot 
  \begin{pmatrix}
    W^+_{(\omega)}\vphantom{\Bigl(\Bigr)}\\ W^-_{(\omega)}\vphantom{\Bigl(\Bigr)}
  \end{pmatrix}
  =
  \Phi_{-\nicefrac{2}{3}}^\dagger\,\mathsf{t}^{(\omega)}_-\,\Phi_{-\nicefrac{2}{3}}\,W^+_{(\omega)}
  +\Phi_{-\nicefrac{2}{3}}^\dagger\,\mathsf{t}^{(\omega)}_+\,\Phi_{-\nicefrac{2}{3}}\,W^-_{(\omega)}\;.
\end{equation}

\begin{wraptable}[11]{r}{4cm}
  \centering\begin{tabular}{cccc}
    \toprule
    state & $[\mathds{Z}_3\times\mathds{Z}_3]^{(\omega)}$\\ 
    \midrule
    $X^{(\omega)}$ & $(1,0)$\\
    $Y^{(\omega)}$ & $(1,2)$\\
    $Z^{(\omega)}$ & $(1,1)$\\
    \bottomrule
    \end{tabular}
    \caption{$[\mathds{Z}_3\times\mathds{Z}_3]^{(\omega)}$ charges of the twisted fields.}
    \label{tab:Z3timesZ3_omega_1}
\end{wraptable}

Once the $[\Group{U}{1}\times\Group{U}{1}]^{(\omega)}$ gauge symmetry is broken, there is a residual $[\mathds{Z}_3\times\mathds{Z}_3]^{(\omega)}$ symmetry seen by the twisted states. 
The $[\mathds{Z}_3\times\mathds{Z}_3]^{(\omega)}$ charges of the latter are not unique, possible choices can be found e.g.\ with the tools provided by \cite{Petersen:2009ip}. 
A convenient choice is shown in \Cref{tab:Z3timesZ3_omega_1}.
This choice differs from the one by \cite{Beye:2014nxa} but is physically equivalent.
As we shall see below in \Cref{sec:Delta(27)}, the space group rule arises from a different $\Group{U}{1}$ symmetry, which becomes linearly realized at a different critical point. 
The second $\mathds{Z}_3$ in \Cref{tab:Z3timesZ3_omega_1} corresponds, up to phases, to a permutation of the localization eigenstates of \eqref{eq:Phi_-2/3},
\begin{equation}\label{eq:permutation_Z_3_omega}
  \begin{pmatrix}
   X\\ Y\\ Z  
  \end{pmatrix}
  \xmapsto{~\mathds{Z}_3^{(\omega,2)}~}\begin{pmatrix}
    0 & \omega ^2 & 0 \\
     0 & 0 & 1 \\
     \omega  & 0 & 0
  \end{pmatrix}\cdot\begin{pmatrix}
    X\\ Y\\ Z  
     \end{pmatrix}\eqdef Z_{(\omega,2)}\cdot\Phi_{-\nicefrac{2}{3}}
  \;.
\end{equation} 
As we shall see in more detail below, the fact that the permutation symmetry \eqref{eq:permutation_Z_3_omega} derives from a continuous gauge theory is instrumental when showing explicitly that flavor symmetries in string models are gauged. 

This leaves us with the following picture: at $\varrho=\omega$ we have two massless gauge bosons corresponding to a $[\Group{U}{1}\times\Group{U}{1}]^{(\omega)}$ gauge symmetry. 
Away from $\varrho=\omega$, there is a residual $[\mathds{Z}_3\times\mathds{Z}_3]^{(\omega)}$ symmetry seen by the twisted fields. 
As we shall see next, at different critical values of $\varrho$ \eqref{eq:rho_crit1}, physically inequivalent $\Group{U}{1}\times\Group{U}{1}$ as well as $\mathds{Z}_3\times\mathds{Z}_3$ symmetries emerge.

\subsubsection{Massless untwisted fields at \texorpdfstring{$\varrho=\omega+1$}{rho=omega+1}}
\label{sec:massless_untwisted_fields_at_omega+1}

Consider the point $\varrho=\omega+1$. 
Now the massless gauge bosons correspond to the image of the stabilizer of $\omega$ under the $\mathsf{T}_\varrho$ transformation,
\begin{equation}\label{eq:massless_KK_and_winding_torusAtomega+1}
\begin{pmatrix*}[r]n_2 & -n_1\\ w^1& w^2\end{pmatrix*} 
\in\left\{  
\pm\begin{pmatrix*}[r]-1 & 0\\ -1 & -1\end{pmatrix*}, 
\pm\begin{pmatrix*}[r] 0 & -1\\ 1 & 0\end{pmatrix*}, 
\pm \begin{pmatrix*}[r]1 & 1\\ 0 & 1 \end{pmatrix*}\right\}\;.
\end{equation} 
That is, we get a new set of gauge bosons, which generate the symmetry $[\Group{U}{1}\times\Group{U}{1}]^{(\omega+1)}$. 

\begin{wrapfigure}[10]{r}{9cm}
  \centering\includegraphics{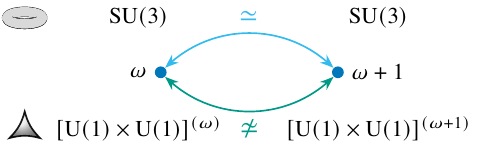}%
  \caption{$\varrho=\omega$ and $\varrho=\omega+1$ are physically equivalent for the torus but distinct for the orbifold.}
  \label{fig:torus_cs_orbifold_cartoon}
\end{wrapfigure}

A key message from our analysis is that these are \emph{physically different} gauge bosons. 
In order to see why this matters, let us first recall the situation on the torus (cf.\ \Cref{sec:Modular_transformations}). 
On the torus, for $\tau=\omega$ at a generic point in $\varrho$-moduli space we have a $\Group{U}{1}\times\Group{U}{1}$ symmetry, generated by the above-mentioned Cartan generators, which stay massless throughout the $\varrho$-moduli space. 
At both $\varrho=\omega$ and $\varrho=\omega+1$ this Abelian symmetry gets enhanced to $\Group{SU}{3}$. 
However, both $\Group{SU}{3}$ theories are physically equivalent, which is why the torus has an $\Group{SL}{2,\mathds{Z}}_\varrho$ symmetry. 
In other words, on the torus physics is the same at $\varrho=\omega$ and $\varrho=\omega+1$.
We will now establish that on the orbifold the points $\omega$ and $\omega+1$ are \emph{physically inequivalent}, see \Cref{fig:torus_cs_orbifold_cartoon} for a cartoon. 
This can be seen in two ways, 
\begin{enumerate}
  \item either by noting that the gauge bosons from \eqref{eq:W_pm_omega} from $\varrho=\omega$ are charged under $[\Group{U}{1}\times\Group{U}{1}]^{(\omega+1)}$, 
  \item or by realizing that the twisted eigenstates of a given $\Group{U}{1}\times\Group{U}{1}$ symmetry are not eigenstates of some other $\Group{U}{1}\times\Group{U}{1}$.
\end{enumerate}

It turns out that the second way is more direct. 
It is important to note that the twisted fields transform nontrivially under modular transformations of $\varrho$~\cite{Chun:1989se,Lauer:1990tm}. 
Specifically, under the transformation $\mathsf{T}_\varrho:\varrho\mapsto\varrho+1$
\begin{equation}\label{eq:rho(T)_twisted_fields}
 \Phi_{-\nicefrac{2}{3}}\xmapsto{~\mathsf{T}_\varrho~}  
 \rho(\mathsf{T})\,\Phi_{-\nicefrac{2}{3}}
 \;.
\end{equation}
In other words, 
\begin{equation}
  \Phi_{-\nicefrac{2}{3}}(\varrho+1)=\rho(\mathsf{T})\,\Phi_{-\nicefrac{2}{3}}(\varrho)\;.
\end{equation}
We therefore have to demand that 
\begin{equation}
  \mathcal{U}_{(\omega)}\,\Phi_{-\nicefrac{2}{3}}(\omega)\overset{!}{=}
  \mathcal{U}_{(\omega+1)}\,\Phi_{-\nicefrac{2}{3}}(\omega+1)
  =
  \mathcal{U}_{(\omega+1)}\,\rho(\mathsf{T})\,\Phi_{-\nicefrac{2}{3}}(\omega)\;.
\end{equation}
As a consequence, the matrix that relates the charge and position eigenstates is now given by 
\begin{equation}\label{eq:U_(omega+1)}
  \mathcal{U}_{(\omega+1)}=\mathcal{U}_{(\omega)}\,\rho\bigl(\mathsf{T}^{-1}\bigr)\;.
\end{equation}
Note that this matrix is not unique. 
We can add elements of the stabilizer of $\omega$ such as $\mathsf{S}\,\mathsf{T}$ to the transformation and still map $\omega$ to $\omega+1$. 
Similar statements apply to the transformation matrices discussed below. 
Hence, the generators of $[\Group{U}{1}\times\Group{U}{1}]^{(\omega+1)}$ become 
\begin{subequations}\label{eq:ti_at_omega+1_localization_eigenstates}
\begin{align}
  \mathsf{t}^{(\omega+1)}_1 
  &= \bigl(\mathcal{U}_{(\omega+1)}\bigr)^\dagger\,\mathsf{t}_1\,\mathcal{U}_{(\omega+1)}=\rho(\mathsf{T})\,\mathsf{t}^{(\omega)}_1\,[\rho(\mathsf{T})]^{-1}=\frac{1}{2\sqrt{3}} \begin{pmatrix}
    0 & \omega  & \omega  \\
     \omega ^2 & 0 & 1 \\
     \omega ^2 & 1 & 0
  \end{pmatrix}\;,\\ 
  \mathsf{t}^{(\omega+1)}_2 &=\bigl(\mathcal{U}_{(\omega+1)}\bigr)^\dagger\,\mathsf{t}_2\,\mathcal{U}_{(\omega+1)}=\rho(\mathsf{T})\,\mathsf{t}^{(\omega)}_2\,[\rho(\mathsf{T})]^{-1}= \frac{\ii}{2\sqrt{3}} 
    \begin{pmatrix*}[r]
    0 & -\omega  & \omega  \\
     \omega ^2 & 0 & -1 \\
     -\omega ^2 & 1 & 0
    \end{pmatrix*}\;.
\end{align}
\end{subequations}
Clearly, these generators do not commute with the previous ones, i.e.\ the $\mathsf{t}^{(\omega)}_i$ of \eqref{eq:ti_at_omega_localization_eigenstates}. 
This means that the corresponding gauge bosons couple to different linear combinations of the localization eigenstates than gauge bosons that are massless at $\varrho=\omega$ do.
Moreover, the mass equation \eqref{eq:mass_TD} reveals that the latter fields have now acquired a mass, $m^2=4/3$ in our convention in which $\alpha'=1$.

Similarly to the situation at $\varrho=\omega$, we obtain at $\varrho=\omega+1$ three massless untwisted fields $U_i^{(\omega+1)}$. 
These fields have the same charges as the $U_i^{(\omega)}$ in \Cref{tab:U1timesU1_omega_1}, yet with respect to \emph{different} $\Group{U}{1}$ generators.
As before, one combination of the $U_i^{(\omega+1)}$ corresponds to the $\varrho$ modulus whereas the other two combinations get eaten by the $[\Group{U}{1}\times\Group{U}{1}]^{(\omega+1)}$ gauge fields if $\varrho$ departs from $\omega+1$.
Analogously to what we have done before, we denote the massive multiplet by $W_{(\omega+1)}$.
The analogous statements apply to the charge eigenstates $\Phi_{-\nicefrac{2}{3}}^{(\omega+1)}\defeq\mathcal{U}_{(\omega+1)}\,\Phi_{-\nicefrac{2}{3}}$. 

At $\varrho=\omega+1$, the twisted fields still couple to the ``old'' $[\Group{U}{1}\times\Group{U}{1}]^{(\omega)}$ gauge bosons, which are now massive.
These gauge bosons get rotated under a linear combination of the $[\Group{U}{1}\times\Group{U}{1}]^{(\omega+1)}$ transformations. 
Likewise, the $[\Group{U}{1}\times\Group{U}{1}]^{(\omega+1)}$ get rotated under a combination of $[\Group{U}{1}\times\Group{U}{1}]^{(\omega)}$ transformations. 
To see this, consider the couplings of the localization eigenstates to the $[\Group{U}{1}\times\Group{U}{1}]^{(\omega+1)}$ gauge fields $W_{(\omega+1)}\defeq\Bigl(W^+_{(\omega+1)},W^-_{(\omega+1)}\Bigr)^{\transpose}$,
\begin{equation}\label{eq:Coupling_Phi_bar_Phi_at_omega+1}
  \begin{pmatrix}
    \Phi_{-\nicefrac{2}{3}}^\dagger\,\mathsf{t}^{(\omega+1)}_-\,\Phi_{-\nicefrac{2}{3}}\vphantom{\Bigl(\Bigr)}\\
    \Phi_{-\nicefrac{2}{3}}^\dagger\,\mathsf{t}^{(\omega+1)}_+\,\Phi_{-\nicefrac{2}{3}}\vphantom{\Bigl(\Bigr)}
  \end{pmatrix}\cdot\begin{pmatrix}
    W^+_{(\omega+1)}\vphantom{\Bigl(\Bigr)}\\ W^-_{(\omega+1)}\vphantom{\Bigl(\Bigr)}
  \end{pmatrix}
  =\frac{1}{\sqrt{6}}\begin{pmatrix}
    \omega\,\overline{X}\,Z+\omega^2\,\overline{Y}\,X+\overline{Z}\,Y\vphantom{\Bigl(\Bigr)}\\
    \omega^2\,\overline{Z}\,X+\omega\,\overline{X}\,Y+\overline{Y}\,Z\vphantom{\Bigl(\Bigr)}
  \end{pmatrix}\cdot\begin{pmatrix}
    W^+_{(\omega+1)}\vphantom{\Bigl(\Bigr)}\\ W^-_{(\omega+1)}\vphantom{\Bigl(\Bigr)}
  \end{pmatrix}\;,
\end{equation}
where $\mathsf{t}^{(\omega+1)}_{\pm}$ are defined in the same way as \Cref{eq:tomegapm}: $\mathsf{t}^{(\omega+1)}_{\pm}\defeq \frac{1}{\sqrt{2}}\left(\mathsf{t}^{(\omega+1)}_{1}\pm \ii \mathsf{t}^{(\omega+1)}_{2}\right)$. 
This coupling is analogous to \eqref{eq:W_pm_omega}.
The contractions of localization eigenstates pick up nontrivial phases under the $\mathds{Z}_3^{(\omega,2)}$ from \eqref{eq:permutation_Z_3_omega},
\begin{equation}\label{eq:Z_3_trafo_of_W_pm_omega+1}
  \begin{pmatrix}
    \omega\,\overline{X}\,Z+\omega^2\,\overline{Y}\,X+\overline{Z}\,Y\\
    \omega^2\,\overline{Z}\,X+\omega\,\overline{X}\,Y+\overline{Y}\,Z
  \end{pmatrix}
  \xmapsto{~\mathds{Z}_3^{(\omega,2)}~}\begin{pmatrix}
    \omega^2 & 0 \\ 0 & \omega
  \end{pmatrix}\cdot \begin{pmatrix}
    \omega\,\overline{X}\,Z+\omega^2\,\overline{Y}\,X+\overline{Z}\,Y\\
    \omega^2\,\overline{Z}\,X+\omega\,\overline{X}\,Y+\overline{Y}\,Z
  \end{pmatrix}\;.
\end{equation}
This means that the gauge fields transform as 
\begin{equation}
  \begin{pmatrix}
    W^+_{(\omega+1)}\\ W^-_{(\omega+1)}
  \end{pmatrix}\xmapsto{~\mathds{Z}_3^{(\omega,2)}~}\begin{pmatrix}
    \omega & 0 \vphantom{\Bigl(\Bigr)}\\ 0 & \omega^2\vphantom{\Bigl(\Bigr)}
  \end{pmatrix}\cdot 
  \begin{pmatrix}
    W^+_{(\omega+1)}\\ W^-_{(\omega+1)}
  \end{pmatrix}\;.
\end{equation}
As before (cf.\ our discussion around \Cref{eq:Coupling_Phi_bar_Phi_at_omega}), the $\mathsf{S}_\varrho^2$ transformation swaps the components of both the contractions of localization eigenstates and gauge fields, i.e.\  
\begin{equation}
  W^+_{(\omega+1)}\xleftrightarrow{~\mathsf{S}_\varrho^2~}W^-_{(\omega+1)}\;.
\end{equation} 
This operation does not commute with the $\mathds{Z}_3^{(\omega,2)}$ transformation \eqref{eq:Z_3_trafo_of_W_pm_omega+1}, and we thus see that $W_{(\omega+1)}$ forms a doublet under the non-Abelian group 
\begin{equation}
 S_3^{(\omega,\omega+1)}=\mathds{Z}_3^{(\omega,2)}\rtimes\mathds{Z}_2^{\mathsf{S}_\varrho^2}\;.   
\end{equation}
This shows, in particular, that the $\mathds{Z}_3^{(\omega,2)}$ symmetry does not commute with the $[\Group{U}{1}\times\Group{U}{1}]^{(\omega+1)}$ symmetries. 
Instead, the gauge symmetry from winding modes at $\varrho=\omega+1$ is given by 
\begin{equation}
  G_\mathrm{winding}^{(\omega+1)}=[\Group{U}{1}\times\Group{U}{1}]^{(\omega+1)}\rtimes\mathds{Z}_3^{(\omega,2)}\;. 
\end{equation}  
Analogous statements hold at $\varrho=\omega$ as well as other critical points discussed below.

The residual $\mathds{Z}_3$ symmetries away from $\varrho=\omega+1$ act on the twisted states via the matrices 
\begin{subequations}\label{eq:Z_3_omega+1}
\begin{align}
  Z_{(\omega+1,1)}&\defeq\omega\,\mathds{1}_3\;,\label{eq:Z_3_omega+1_1}\\
  Z_{(\omega+1,2)}&\defeq\begin{pmatrix}
    0 & \omega & 0  \\
    0 & 0 & 1 \\
     \omega^2 & 0 & 0
  \end{pmatrix}\;.\label{eq:Z_3_omega+1_2}
\end{align}  
\end{subequations}
Together with the residual $\mathds{Z}_{(\omega,2)}$ from \eqref{eq:permutation_Z_3_omega} the residual $\mathds{Z}_3$ symmetries \eqref{eq:Z_3_omega+1} give rise to a $\Delta(27)$ symmetry. 
In slightly more detail, 
\begin{subequations}
\begin{align}
  Z_{(\omega,2)}\cdot Z_{(\omega+1,2)}^2\cdot Z_{(\omega,2)}&=\rho(\mathsf{A})\;,\\
  Z_{(\omega+1,2)}^2\cdot Z_{(\omega,2)}^2\cdot Z_{(\omega+1,2)}^2&=\rho(\mathsf{B})
\end{align}  
\end{subequations}
with the representation matrices $\rho(\mathsf{A})$ and $\rho(\mathsf{B})$ from \Cref{tab:Delta(54)_representation_matrices} in \Cref{sec:Delta(54)}. 
The matrices $\rho(\mathsf{A})$ and $\rho(\mathsf{B})$ generate a $\Delta(27)$ symmetry. Conversely, these three $\mathds{Z}_3$ generators can be expressed by $\rho(\mathsf{A}),\rho(\mathsf{B})$ as
\begin{subequations}\label{eq:ThreeZ3}
\begin{align}
  Z_{(\omega+1,1)}&=\rho(\mathsf{A}^2\,\mathsf{B}^2\,\mathsf{A}\,\mathsf{B})\;,\\
  Z_{(\omega+1,2)}&=\rho(\mathsf{B}\,\mathsf{A}\,\mathsf{B})\;,\\
  Z_{(\omega,2)}&=\rho(\mathsf{B}^2\,\mathsf{A}\,\mathsf{B}^2)\;.
\end{align}
\end{subequations}
We have already seen in \eqref{eq:S^2=C} that $\rho(\mathsf{C})$ in \Cref{tab:Delta(54)_representation_matrices}, which enhances $\Delta(27)$ to $\Delta(54)$, emerges from the $\mathsf{S}_\varrho^2$ transformation.  
Our conventions for $\Delta(27)$ and $\Delta(54)$ are detailed in \Cref{sec:Delta(54)}.
In particular, $\rho$ without any further specification stands for representation matrices in the $\rep{3}_2$ representation of $\Delta(54)$. 
With the help of \Cref{eq:ThreeZ3}, we can now  express the structure of the residual gauge subgroup $\Delta(54)$ at $\omega+1$ as 
\begin{equation}\label{eq:Delta54atOmgea+1}
\Delta(54)=\left[\mathds{Z}^{\mathsf{A}^2\,\mathsf{B}^2\,\mathsf{A}\,\mathsf{B}}_3\times \mathds{Z}^{\mathsf{B}\,\mathsf{A}\,\mathsf{B}}_3\right]\rtimes \left[\mathds{Z}^{\mathsf{B}^2\,\mathsf{A}\,\mathsf{B}^2}_3\rtimes \mathds{Z}^{\mathsf{C}}_2\right]\;.
\end{equation}
Altogether we therefore have established that, away from the critical points, the twisted states see a $\Delta(54)$ symmetry. 
In this $\Delta(54)$, a $\Delta(27)$ subgroup emerges from the discrete remnants of relatively misaligned, i.e.\ non-commuting, $\Group{U}{1}$ symmetries which are linearly realized at different critical points in $\varrho$ moduli space. 
This allows us to interpret the contraction of \eqref{eq:Coupling_Phi_bar_Phi_at_omega+1} as a $\Delta(54)$ $\rep{2}_3$-plet, see \Cref{eq:CG_2_3} in \Cref{sec:Delta(54)}. 
This means that the generators of $[\Group{U}{1}\times\Group{U}{1}]^{(\omega+1)}$ serve as the \ac{CG} coefficients of the $\Delta(54)$ contraction of two triplets to a doublet. 
Likewise, the generators of $[\Group{U}{1}\times\Group{U}{1}]^{(\omega)}$ in \eqref{eq:W_pm_omega} yield the contraction to a $\rep{2}_4$-plet, cf.\ \Cref{eq:CG_2_4}. 
That is, \eqref{eq:Coupling_Phi_bar_Phi_at_omega} contains the corresponding \ac{CG} coefficients.
We shall return to this discussion in \Cref{sec:Delta(27)}.

\subsubsection{Massless untwisted fields at \texorpdfstring{$\varrho=\omega+2$}{rho=omega+2}}
\label{sec:massless_untwisted_fields_at_omega+2}

Once we move to $\varrho=\omega+2$, the twisted localization eigenstates pick up additional phases,
\begin{equation}
  \Phi_{-\nicefrac{2}{3}}\xmapsto{~\mathsf{T}^2_\varrho~}  \rho(\mathsf{T}^{2})\,\Phi_{-\nicefrac{2}{3}}
  \;.
\end{equation}
Hence,
\begin{equation}
     \mathcal{U}_{(\omega+2)}=\mathcal{U}_{(\omega)}\,\rho(\mathsf{T}^{-2})\;.
\end{equation}
Therefore, the generators of $[\Group{U}{1}\times\Group{U}{1}]^{(\omega+2)}$ become 
\begin{subequations}\label{eq:ti_at_omega+2_localization_eigenstates}
\begin{align}
  \mathsf{t}^{(\omega+2)}_1 
  &= \bigl(\mathcal{U}_{(\omega+2)}\bigr)^\dagger\,\mathsf{t}_1\,\mathcal{U}_{(\omega+1)}=\rho(\mathsf{T}^2)\,\mathsf{t}^{(\omega)}_1\,[\rho(\mathsf{T}^2)]^{-1}=\frac{1}{2\sqrt{3}} \begin{pmatrix}
    0 & 1 & 1 \\
     1 & 0 & 1 \\
     1 & 1 & 0 
  \end{pmatrix}\;,\\ 
  \mathsf{t}^{(\omega+2)}_2 &=\bigl(\mathcal{U}_{(\omega+2)}\bigr)^\dagger\,\mathsf{t}_2\,\mathcal{U}_{(\omega+2)}=\rho(\mathsf{T}^2)\,\mathsf{t}^{(\omega)}_2\,[\rho(\mathsf{T}^2)]^{-1}= \frac{\ii}{2\sqrt{3}} \begin{pmatrix*}[r]
    0 & -1 & 1 \\
     1 & 0 & -1 \\
     -1 & 1 & 0  \\
  \end{pmatrix*}\;.
\end{align}
\end{subequations}
The corresponding massless gauge bosons correspond to the image of the stabilizer of $\omega$ under the square of the $\mathsf{T}_\varrho$ transformation.
Analogously to our finding in \Cref{sec:massless_untwisted_fields_at_omega+1}, these generators do not commute with the $\mathsf{t}^{(\omega+1)}_i$, nor the $\mathsf{t}^{(\omega)}_i$. 
This means that we have found yet another set of inequivalent gauge fields, which can be combined to yet another doublet, $W_{(\omega+2)}\defeq\Bigl(W_{(\omega+2)}^+,W_{(\omega+2)}^-\Bigr)^{\transpose}$. 
As in \eqref{eq:Coupling_Phi_bar_Phi_at_omega} and \eqref{eq:Coupling_Phi_bar_Phi_at_omega+1}, the generators of $[\Group{U}{1}\times\Group{U}{1}]^{(\omega+2)}$ provide us with the \ac{CG} coefficients of the contraction of a triplet and antitriplet to a $\rep{2}_1$-plet of $\Delta(54)$ doublet (cf.\ \Cref{eq:CG_2_1}),
\begin{equation}\label{eq:Coupling_Phi_bar_Phi_at_omega+2}
  \begin{pmatrix}
    \Phi_{-\nicefrac{2}{3}}^\dagger\,\mathsf{t}^{(\omega+2)}_-\,\Phi_{-\nicefrac{2}{3}}\vphantom{\Bigl(\Bigr)}\\
    \Phi_{-\nicefrac{2}{3}}^\dagger\,\mathsf{t}^{(\omega+2)}_+\,\Phi_{-\nicefrac{2}{3}}\vphantom{\Bigl(\Bigr)}
  \end{pmatrix}\cdot\begin{pmatrix}
    W^+_{(\omega+2)}\vphantom{\Bigl(\Bigr)}\\ W^-_{(\omega+2)}\vphantom{\Bigl(\Bigr)}
  \end{pmatrix}
  =\frac{1}{\sqrt{6}}\begin{pmatrix}
    \overline{X}\, Z+\overline{Y}\, X+\overline{Z}\, Y\vphantom{\Bigl(\Bigr)}\\
    \overline{Z}\, X+\overline{X}\, Y+\overline{Y}\, Z\vphantom{\Bigl(\Bigr)}
  \end{pmatrix}\cdot\begin{pmatrix}
    W^+_{(\omega+2)}\vphantom{\Bigl(\Bigr)}\\ W^-_{(\omega+2)}\vphantom{\Bigl(\Bigr)}
  \end{pmatrix}\;,
\end{equation}
where $\mathsf{t}^{(\omega+2)}_{\pm}\defeq \frac{1}{\sqrt{2}}\left(\mathsf{t}^{(\omega+2)}_{1}\pm \ii \mathsf{t}^{(\omega+2)}_{2}\right)$. 
Similarly to what we have encountered before, the gauge fields that were massless at $\varrho=\omega+1$ have now acquire a mass of $m^2=4/3$. 
The gauge fields that were massless at $\varrho=\omega$ have an even larger mass, but this does not mean that there are no gauge fields with these quantum numbers at $m^2=4/3$. 
Rather, as we shall see next, there is another doublet of gauge fields which couple to the localization eigenstates in the same way as the members of $W_{(\omega)}$ do, and become massless at $\varrho=\omega+3$.

\subsubsection{Massless untwisted fields at \texorpdfstring{$\varrho=\omega+3$}{rho=omega+3}}
\label{sec:massless_untwisted_fields_at_omega+3}

Once we move on to $\varrho=\omega+3$, we find a $[\Group{U}{1}\times\Group{U}{1}]^{(\omega+3)}$ symmetry that is physically equivalent to the $[\Group{U}{1}\times\Group{U}{1}]^{(\omega)}$.
This is because $[\rho(\mathsf{T})]^3=\mathds{1}$. 
Therefore, $\varrho=\omega+3$ is physically indistinguishable from $\varrho=\omega$.
Notice, though, that the corresponding massless gauge bosons correspond to the image of the stabilizer of $\omega$ under the cube of the $\mathsf{T}_\varrho$ transformation.
That is, the $[\Group{U}{1}\times\Group{U}{1}]^{(\omega+3)}$ gauge symmetry gets mediated by gauge bosons with different winding numbers than those at $\varrho=\omega$.  

One can generalize these statements to become
\begin{equation}
\left[\Group{U}{1}\times\Group{U}{1}\right]^{(\omega+k)}\simeq[\Group{U}{1}\times\Group{U}{1}]^{(\omega+\ell)}\quad\text{if }\ell-k=0\bmod3\;.
\end{equation}
This means that physics repeats itself when shifting $\varrho$ by 3. 
That is, unlike on the torus, $\Group{SL}{2,\mathds{Z}}$ transformations do not relate physically equivalent critical points. 
However, physics repeats itself under $\mathds{T}_\varrho^3$ transformations. 
This suggests that the modular symmetry is $\Gamma(3)$, which we will discuss in some more detail in \Cref{sec:Gamma(3)}.

\begin{figure}[htb]
  \centering\includegraphics{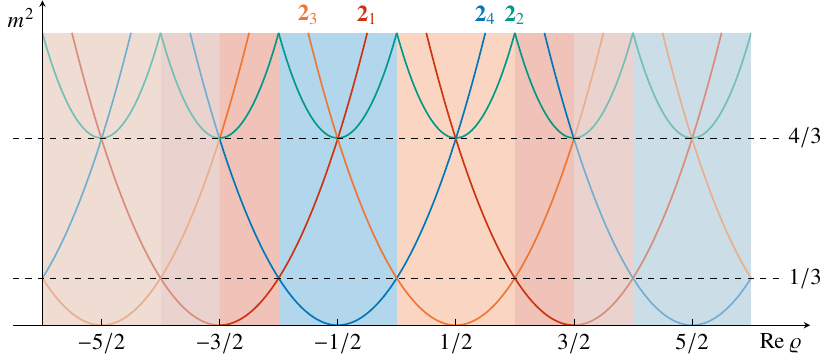}%
  \caption{Dependence of the masses of $N_\mathrm{L}=0$ doublets at $\im\varrho=\sqrt{3}/2$. The points $\omega$, $\omega+1$ and $\omega+2$ have $\re\varrho=-1/2$, $1/2$ and $3/2$, respectively, and are related by the $\mathsf{T}_\varrho$ transformation. 
  As indicated, at each of the three critical points, a \emph{different} $\Group{\Delta}{54}$ doublet becomes massless. 
  Only after a $\mathsf{T}_\varrho^3$ transformation a doublet with the same $\Group{\Delta}{54}$ quantum number becomes massless.
  This illustrates that physically equivalent situations are related by $\Gamma(3)$ (and not $\Group{SL}{2,\mathds{Z}}$) transformations.
  The regions outside our choice of the fundamental domain of $\Gamma(3)$ are grayed out.}
  \label{fig:cartoon_doublet_masses}
\end{figure}

We plot the masses of gauge fields in \Cref{fig:cartoon_doublet_masses}, where we label them by their $\Delta(54)$ representations.
For later convenience, we also include a pair of states $\rep{2}_2$, which becomes massless at different critical points (cf.\ \Cref{sec:massless_untwisted_fields_at_nu} below).

As one can see from \Cref{fig:cartoon_doublet_masses}, at the critical points, 3 doublets are mass degenerate. 
This is a consequence of the $\mathds{Z}_3$ stabilizer groups of these critical points. 
In other words, apart from the enhanced gauge symmetries mediated by the doublets that become massless at these points, there is a linearly realized $\mathds{Z}_3$ symmetry that relates the three massive doublets. 

We have argued that the discussion so far suggests that the modular symmetry is $\Gamma(3)$ since physics repeats itself only after a shift of $\varrho$ by 3. 
Before fully establishing the $\Gamma(3)$ symmetry in \Cref{sec:Gamma(3)}, we need to also allow the imaginary part of $\varrho$, i.e.\ the radius, to vary, and show that doublets come in infinite towers. 
We start by considering different radii.

\subsubsection{Massless untwisted fields at \texorpdfstring{$\varrho=\nu$}{rho=nu}}
\label{sec:massless_untwisted_fields_at_nu}

Starting from $\varrho=\omega+2$ (cf.\ \Cref{sec:massless_untwisted_fields_at_omega+2}), one can subject $\varrho$ to an $\mathsf{S}$ transformation,
\begin{equation}\label{eq:omega+2->nu}
  \omega+2\xmapsto{~\mathsf{S}~}\frac{-1}{\omega+2}\eqdef\nu=-1/2+\ii\sqrt{3}/6\;.
\end{equation} 
As before, we obtain two massless gauge bosons, which correspond to a $[\Group{U}{1}\times\Group{U}{1}]^{(\nu)}$ symmetry.
Importantly, because of \eqref{eq:U_omega=T.Sinv.T}, the transformation \eqref{eq:omega+2->nu} brings us back to the localization eigenstates up to a $\mathsf{T}_\varrho$ transformation.\footnote{As already mentioned below \eqref{eq:U_(omega+1)}, the transformation matrix $\mathcal{U}_{(\nu)}$ is not unique, however, the all viable transformation matrices lead to the conclusion that the localization eigenstates are charge eigenstates of $[\Group{U}{1}\times\Group{U}{1}]^{(\nu)}$.}  
Since the latter is diagonal in the localization eigenstates (cf.\ \eqref{eq:rho(T)}), the charge eigenstates at of $[\Group{U}{1}\times\Group{U}{1}]^{(\nu)}$ coincide with the localization eigenstates.  
That is, the generators of $[\Group{U}{1}\times\Group{U}{1}]^{(\nu)}$ are diagonal in the basis of localization eigenstates, and can be chosen to be given by the $\mathsf{t}_i$ of \eqref{eq:diagonal_t_i},
\begin{subequations}\label{eq:t_nu}
\begin{align}
  \mathsf{t}_1^{(\nu)}&= \mathsf{t}_1=\frac{1}{2\sqrt{3}}\diag(2,-1,-1)\;,\\
  \mathsf{t}_2^{(\nu)}&=\mathsf{t}_2=\frac{1}{2}\diag(0,1,-1)\;.
\end{align}  
\end{subequations}
Linear combinations of these generators provide us with the \ac{CG} coefficients of the contraction of a triplet and antitriplet to a $\rep{2}_2$-plet of $\Delta(54)$ doublet (cf.\ \Cref{eq:CG_2_1}),
\begin{equation}\label{eq:Coupling_Phi_bar_Phi_at_nu}
  \begin{pmatrix}
    \Phi_{-\nicefrac{2}{3}}^\dagger\,\mathsf{t}^{(\nu)}_+\,\Phi_{-\nicefrac{2}{3}}\vphantom{\Bigl(\Bigr)}\\
    \Phi_{-\nicefrac{2}{3}}^\dagger\,\mathsf{t}^{(\nu)}_-\,\Phi_{-\nicefrac{2}{3}}\vphantom{\Bigl(\Bigr)}
  \end{pmatrix}\cdot\begin{pmatrix}
    W^+_{(\nu)}\vphantom{\Bigl(\Bigr)}\\ W^-_{(\nu)}\vphantom{\Bigl(\Bigr)}
  \end{pmatrix}
  =\frac{1}{\sqrt{6}}\begin{pmatrix}
\overline{X}\, X+\omega\,\overline{Y}\, Y+\omega^2\,\overline{Z}\, Z\vphantom{\Bigl(\Bigr)}\\
\overline{X}\, X+\omega^2\,\overline{Y}\, Y+\omega\,\overline{Z}\, Z\vphantom{\Bigl(\Bigr)}
  \end{pmatrix}\cdot\begin{pmatrix}
    W^+_{(\nu)}\vphantom{\Bigl(\Bigr)}\\ W^-_{(\nu)}\vphantom{\Bigl(\Bigr)}
  \end{pmatrix}\;,
\end{equation}
where $\mathsf{t}^{(\nu)}_{\pm}\defeq \frac{1}{\sqrt{2}}\left(\mathsf{t}^{(\nu)}_{1}\pm \ii \mathsf{t}^{(\nu)}_{2}\right)$. 
This means that the massless gauge bosons at $\varrho=\nu$ form a doublet $\rep{2}_2$ of $\Delta(54)$.

\subsubsection{Massless untwisted fields at \texorpdfstring{$\varrho=\nu+1$}{rho=nu+1} and \texorpdfstring{$\varrho=\nu+2$}{rho=nu+2}}
\label{sec:massless_untwisted_fields_at_nu+1}

Both at $\varrho=\nu+1$ and $\varrho=\nu+2$, there are $\Group{U}{1}\times\Group{U}{1}$ symmetries the generators of which are diagonal in the basis of localization eigenstates.
This is because $\rho(\mathsf{T})$ is diagonal in this basis, cf.\ \Cref{eq:rho(T)_twisted_fields}. 
That is, similarly to the situation at $\varrho=\nu$, we can choose the generators to be given by \eqref{eq:diagonal_t_i}. 
Therefore, the massless gauge bosons at $\varrho=\nu+1$ and $\nu+2$ also all form the doublet $\rep{2}_2$ of $\Delta(54)$.

\begin{figure}[htb]
  \centering\includegraphics{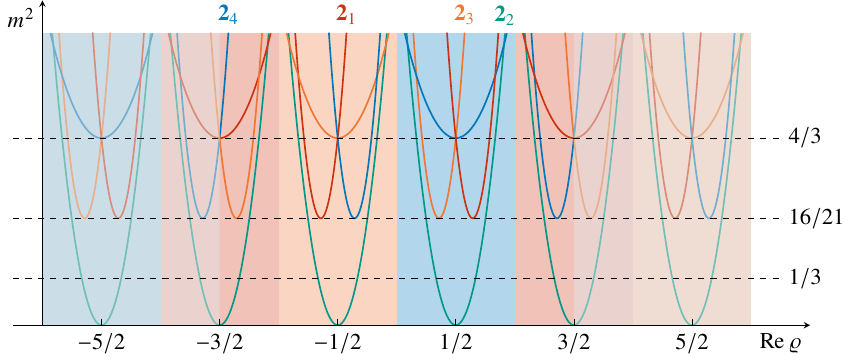}%
  \caption{Dependence of the masses of $N_\mathrm{L}=0$ doublets at $\im\varrho=1/2\sqrt{3}$. The points $\nu$, $\nu+1$ and $\nu+2$ have $\re\varrho=-1/2$, $1/2$ and $3/2$, respectively, and are related by the $\mathsf{T}_\varrho$ transformation. 
  As indicated, at each of the three critical points, \emph{only} the $\Group{\Delta}{54}$ doublet $\rep{2}_2$ becomes massless.
  Contrast this to the critical points $\omega$, $\omega+1$, and $\omega+2$ where a \emph{different} $\Group{\Delta}{54}$ doublet became massless.
  Notice that each symmetry enhanced point is now distinguished by the structure of doublets appearing at the first massive level, $m^2 = 4/3$.
  We color each region according to the doublet whose mass is minimized at $m^2=4/3$ at the symmetry enhanced point to distinguish each region in moduli space.
  Once again, this illustrates that physically equivalent situations are related by $\Gamma(3)$ (and not $\Group{SL}{2,\mathds{Z}}$) transformations.
  The regions outside our choice of the fundamental domain of $\Gamma(3)$ are grayed out.}
  \label{fig:cartoon_nu_doublet_masses}
\end{figure}

Previously we have used the fact that the generators of $[\Group{U}{1}\times\Group{U}{1}]^{(\omega)}$, $[\Group{U}{1}\times\Group{U}{1}]^{(\omega+1)}$ and $[\Group{U}{1}\times\Group{U}{1}]^{(\omega+2)}$ are different to infer that the corresponding critical points are physically inequivalent. 
Now see that the generators of $[\Group{U}{1}\times\Group{U}{1}]^{(\nu)}$, $[\Group{U}{1}\times\Group{U}{1}]^{(\nu+1)}$ and $[\Group{U}{1}\times\Group{U}{1}]^{(\nu+2)}$ can be chosen to be identical.
Does this mean that the three critical points $\nu$, $\nu+1$ and $\nu+2$ are equivalent?
As one can see from \Cref{fig:cartoon_nu_doublet_masses}, this is not the case. 
Define $a_{(\nu)}\defeq\re\varrho-1/2$ as the $\varrho$-axion at $\nu$, i.e.\ $\im\varrho=\sqrt{3}/6$. 
This field has quadratic couplings to $\rep{2}_2$ and $\rep{2}_3$, i.e.\ the gauge bosons which become massless at $\varrho=\nu$ and $\varrho=\omega$, respectively, whereas its couplings to $\rep{2}_1$ and $\rep{2}_4$ are linear. 
For the explicit form of these couplings see \Cref{eq:m^2_tower_axion} in \Cref{sec:Towers}.
Repeating this analysis at $\varrho=\nu+1$, we see that the local axion now has quadratic couplings to $\rep{2}_2$ and $\rep{2}_4$ whereas the couplings to $\rep{2}_1$ and $\rep{2}_3$ are linear. 
Likewise, the local axion at $\varrho=\nu+2$ has quadratic couplings to $\rep{2}_1$ and $\rep{2}_2$, and couplings linearly to $\rep{2}_3$ and $\rep{2}_4$.

Similarly to our discussion in \Cref{sec:massless_untwisted_fields_at_omega,sec:massless_untwisted_fields_at_omega+1,sec:massless_untwisted_fields_at_omega+2,sec:massless_untwisted_fields_at_omega+3}, physics at $\nu+k$ and $\nu+\ell$ is equivalent if $k-\ell=0\bmod3$. 
This means that we have a total of 6 inequivalent critical points $\varrho_\mathrm{crit}$ giving rise to linearly realized $[\Group{U}{1}\times\Group{U}{1}]^{(\varrho_\mathrm{crit})}$ gauge symmetries, where 
\begin{equation}
  \varrho_\mathrm{crit}\in\bigl\{\omega,\omega+1,\omega+2,\nu,\nu+1,\nu+2\bigr\}\;.
\end{equation} 
This is exactly what one expects to if the modular symmetry is $\Gamma(3)$, which we will discuss next, instead of \Group{SL}{2,\mathds{Z}}.

\subsubsection{\texorpdfstring{$\Gamma(3)$}{Gamma(3)}}
\label{sec:Gamma(3)}

Let us summarize the transformation of the $\Delta(54)$-doublets under modular generators,
\begin{subequations}\label{eq:2-plet-S_and_T}
\begin{align}
\begin{pmatrix}
\rep{2}_1 \\ \rep{2}_2 \\ \rep{2}_3 \\ \rep{2}_4 
\end{pmatrix}\xmapsto{~\mathsf{S}_\varrho~}
\begin{pmatrix}
\rep{2}_2 \\ \rep{2}_1 \\ \rep{2}_4 \\ \rep{2}_3 
\end{pmatrix}\;,\label{eq:2-plet-S}\\
\begin{pmatrix}
\rep{2}_1 \\ \rep{2}_2 \\ \rep{2}_3 \\ \rep{2}_4 
\end{pmatrix}\xmapsto{~\mathsf{T}_\varrho~}
\begin{pmatrix}
\rep{2}_4 \\ \rep{2}_2 \\ \rep{2}_1 \\ \rep{2}_3 
\end{pmatrix}\;.\label{eq:2-plet-T}
\end{align}  
\end{subequations}
Given that critical points related by $\Group{SL}{2,\mathds{Z}}_\varrho$ transformations are physically distinct, it is clear that $\Group{SL}{2,\mathds{Z}}_\varrho$ is not the symmetry of the low-energy \ac{EFT}. 
Instead, the modular symmetry is given by congruence subgroup 
\begin{equation}\label{eq:Gamma(3)}
  \Gamma(3)\defeq \left\{\begin{pmatrix}
    a & b\\ c & d
  \end{pmatrix}\in\Group{SL}{2,\mathds{Z}}_\varrho;\;\begin{pmatrix}
    a & b\\ c & d
  \end{pmatrix}=\begin{pmatrix}
    1 & 0\\ 0 & 1
  \end{pmatrix}\bmod3\right\}\;,
\end{equation}
which generated by three generators 
\begin{equation}\label{eq:Gamma3_generators}
  \Gamma(3) = \left\langle \mathsf{T}^3=\begin{pmatrix}
    1 & 3\\ 0 & 1
  \end{pmatrix} \;,\; \left(\mathsf{T}^3\,\mathsf{S}\right)^2\,\mathsf{S}^2=\begin{pmatrix}
    -8 & 3\\ -3 & 1
  \end{pmatrix}\;,\; \left(\mathsf{T}^2\,\mathsf{S}\right)^3=\begin{pmatrix}
    4 & -3\\ 3 & -2
  \end{pmatrix} \right\rangle\;,
\end{equation}
where the $\mathsf{S}$ and $\mathsf{T}$ are the common $2\times 2$ matrix generators of $\Group{SL}{2,\mathds{Z}}_\varrho$. 
It can be verified through \Cref{eq:2-plet-S_and_T} that these $\Delta(54)$-doublets remain invariant under the above three $\Gamma(3)$ generators~\eqref{eq:Gamma3_generators}.

\begin{figure}[htb!]
  \centering\includegraphics{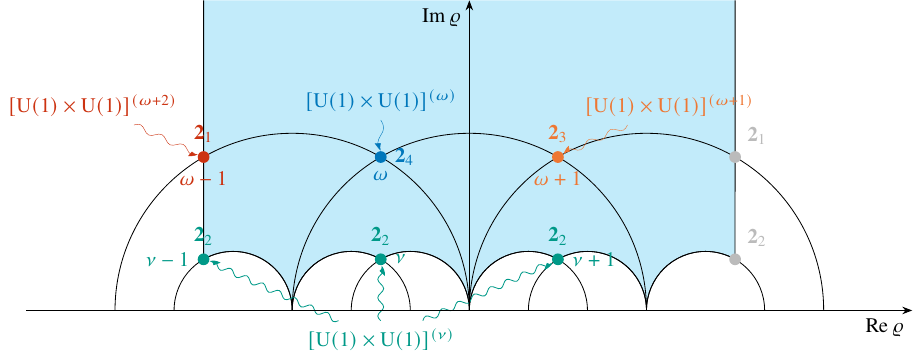}%
  \caption{Fundamental domain of $\Gamma(3)$.}
  \label{fig:Gamma(3)}
\end{figure}

We show the fundamental domain of $\Gamma(3)$ in \Cref{fig:Gamma(3)}, where we indicate the $\Group{U}{1}\times\Group{U}{1}$ symmetries that become exact that the critical points along with the representations of the gauge bosons under the $\Delta(54)$ symmetry.

\subsection{\texorpdfstring{$\Delta(27)$}{Delta(27)} from misaligned \texorpdfstring{\Group{U}{1}}{U(1)} symmetries and \texorpdfstring{$\Delta(54)$}{Delta(54)}}
\label{sec:Delta(27)}

\begin{wraptable}[11]{r}{4cm}
  \centering\begin{tabular}{cccc}
    \toprule
    state & $[\mathds{Z}_3\times\mathds{Z}_3]^{(\nu)}$\\ 
    \midrule
    $X$ & $(1,0)$\\
    $Y$ & $(1,2)$\\
    $Z$ & $(1,1)$\\
    \bottomrule
    \end{tabular}
    \caption{$[\mathds{Z}_3\times\mathds{Z}_3]^{(\nu)}$ charges of the localization eigenstates.}
    \label{tab:Z3timesZ3_nu_1}
\end{wraptable}

As we already have seen around \Cref{eq:Z_3_omega+1}, the relatively misaligned residual $[\mathds{Z}_3\times\mathds{Z}_3]^{(\omega)}$ and $[\mathds{Z}_3\times\mathds{Z}_3]^{(\omega+1)}$ symmetries give rise to a $\Delta(27)$ symmetry.   
Likewise, away from $\varrho=\nu$, there is a $[\mathds{Z}_3\times\mathds{Z}_3]^{(\nu)}$ symmetry, with one $\mathds{Z}_3$ acting on the charge eigenstates of $[\Group{U}{1}\times\Group{U}{1}]^{(\omega)}$ as permutation.
The $[\mathds{Z}_3\times\mathds{Z}_3]^{(\nu)}$ charges can be chosen to look like the $[\mathds{Z}_3\times\mathds{Z}_3]^{(\omega)}$ charges (cf.\ \Cref{tab:Z3timesZ3_omega_1}), yet crucially the charges in \Cref{tab:Z3timesZ3_nu_1} concern localization eigenstates. 
This means that the residual symmetries $[\mathds{Z}_3\times\mathds{Z}_3]^{(\omega)}$ and $[\mathds{Z}_3\times\mathds{Z}_3]^{(\nu)}$ again combine to $\Delta(27)$,
\begin{equation}
  [\mathds{Z}_3\times\mathds{Z}_3]^{(\omega)}\cup[\mathds{Z}_3\times\mathds{Z}_3]^{(\nu)}
  =
  [\mathds{Z}_3\times\mathds{Z}_3]\ltimes\mathds{Z}_3=\Delta(27)\;.
\end{equation}
We thus see that the $\Delta(27)$ arises from ``misaligned'' $\mathds{Z}_3\times\mathds{Z}_3$ symmetries which are discrete remnants of similarly ``misaligned'' $\Group{U}{1}\times\Group{U}{1}$ that become exact at different critical points in $\varrho$-moduli space. 
The different ways in which the residual $[\mathds{Z}_3\times\mathds{Z}_3]^{(\varrho_\mathrm{crit})}$ symmetries emerge at various critical point $\varrho_\mathrm{crit}$ are spelled out in \Cref{eq:Z3Z3Z3Z2} in \Cref{sec:Delta(54)}.

At this point it is instructive to discuss how these findings relate to earlier discussions of the flavor symmetry in the $\mathds{Z}_3$ orbifold. 

In \cite{Kobayashi:2006wq} a $\Delta(54)$ flavor symmetry was derived from string selection rules, which we review in \Cref{sec:String_selection_rules}. 
According to our discussion, the point-group and space-group selection rules $\mathds{Z}_3^{\mathrm{PG}}\times\mathds{Z}_3^{\mathrm{SG}}$ (cf.\ \eqref{eq:String_selection_Z_3_x_Z_3}) can also be understood as the $[\mathds{Z}_3\times\mathds{Z}_3]^{(\nu)}$ that remains unbroken in $\varrho$-moduli space. 
The cyclic permutation symmetry is then part of the $[\mathds{Z}_3\times\mathds{Z}_3]^{(\omega)}$ symmetry.
The $\Delta(27)$ symmetry can also be understood in terms of outer automorphisms of the Narain lattice \cite{Baur:2019iai}.

In \cite{Beye:2014nxa}, the gauge origin of two $\mathds{Z}_3$ factors contained in $[\mathds{Z}_3\times\mathds{Z}_3]^{(\omega)}$ was discussed, yet the gauge origin of the third, non-commuting $\mathds{Z}_3$ was not clarified. 
In \cite{Biermann:2019amx} it was pointed out that the ``missing'' $\mathds{Z}_3$ arises in an orbifold GUT picture as a discrete remnant of an \Group{SU}{3} symmetry of the upstairs theory.
However, despite these similarities, the stringy picture is quite different from the orbifold GUT story. 
One key difference is that in the orbifold GUT the ``missing'' $\mathds{Z}_3$ does not get enhanced to a continuous symmetry anywhere in the moduli space of the 4D \ac{EFT}.

The misalignment of the $[\mathds{Z}_3\times\mathds{Z}_3]^{(\varrho_\mathrm{crit})}$ symmetries is also instrumental for the understanding of the $\Gamma(3)$ modular symmetry: \Group{SL}{2,\mathds{Z}} transformations may lead to physically different situations and only $\Gamma(3)$ transformations map to equivalent physics. 
We will elaborate more on this in \Cref{sec:Interpretation_of_modular_flavor_symmetries}.

It is instructive to look at this from a slightly different angle. 
The gauge bosons have left-moving momenta the lengths of which coincide with the lengths of the non-Cartan roots of the adjoint representation of \Group{SU}{3}.
The weights of the twisted states are shorter by a factor 3. 
If we (artificially) rescale all weights by a factor $\sqrt{3}$ and rotate the axes of the weight system by $\pi/3$, the weights of the twisted states and gauge fields enjoy the same geometric relations as the $\rep{3}+\crep{3}$ and $\rep{10}+\crep{10}$ representations of \Group{SU}{3}, respectively. 
As is well known (cf.\ e.g.\ \cite{Luhn:2011ip}), \Group{SU}{3} can be broken spontaneously to $\Delta(27)$ by giving an appropriate \ac{VEV} to a \rep{10}-plet. 
In this regard the stringy mechanism of obtaining $\Delta(27)$ is somewhat similar to the 4D spontaneous breakdown of \Group{SU}{3} to $\Delta(27)$, while differing in many other important aspects.

Let us revisit \eqref{eq:Coupling_Phi_bar_Phi_at_omega}.
One can interpret this relation as a coupling between linear combinations of localization eigenstates to combinations of winding modes, e.g.\ 
\begin{align}
  \omega^2\,\overline{X}\,Y+\overline{Y}\,Z+\omega\,\overline{Z}\,X 
   &=\omega^2\,\CenterObject{\includegraphics[width=2cm]{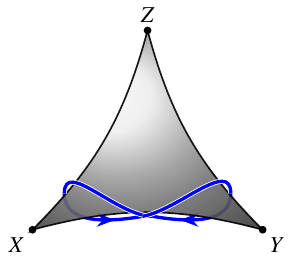}}
   +\CenterObject{\includegraphics[width=2cm]{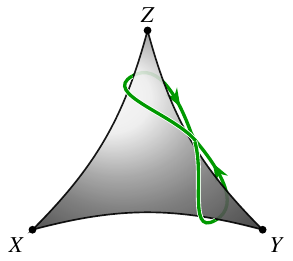}}
   +\omega\,\CenterObject{\includegraphics[width=2cm]{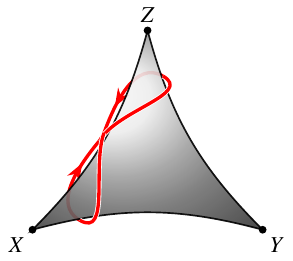}}
   \;.
\end{align}

\begin{figure}[htb]
  \centering\subcaptionbox{$\overline{X}Y$.\label{fig:DoubletXbarY}}
  {\includegraphics{DoubletXbarY.pdf}}\qquad
  \subcaptionbox{$\overline{Y}Z$.\label{fig:DoubletYbarZ}}
  {\includegraphics{DoubletYbarZ.pdf}}\par
  \subcaptionbox{$\overline{Z}X$.\label{fig:DoubletZbarX}}
  {\includegraphics{DoubletZbarX.pdf}}\qquad
  \subcaptionbox{$\rep{2}_2$.\label{fig:2_2}}
  {\includegraphics{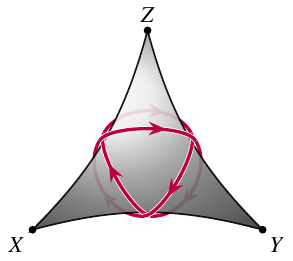}}\par
  \caption{Cartoons of the winding modes. Each winding string is accompanied by a version in which the string winds in the opposite direction, together with which it forms a doublet.}\label{fig:Doublet_winding_modes}
\end{figure}

In order to substantiate this statement, it is important to realize that the inequivalent $[\mathds{Z}_3\times\mathds{Z}_3]^{(\varrho_\mathrm{crit})}$ gauge fields build in fact infinite towers, which we will study in more detail in \Cref{sec:Towers}.

\subsection{Superpotential}
\label{sec:Superpotential} 

At each of the critical points, at the cubic level there is only one monomial of charge eigenstates allowed, 
\begin{equation}\label{eq:W_crit}
  \mathscr{W}_\mathrm{trilinear}=\mathcal{Y}(\varrho_\mathrm{crit})\,X^{(\varrho_\mathrm{crit})}\,Y^{(\varrho_\mathrm{crit})}\,Z^{(\varrho_\mathrm{crit})}\;.
\end{equation}
One can express the charge eigenstates $X_{(\varrho_\mathrm{crit})}$, $Y_{(\varrho_\mathrm{crit})}$ and $Z_{(\varrho_\mathrm{crit})}$ in terms of the localization eigenstates $X$, $Y$ and $Z$. 
The resulting expressions are shown in \Cref{tab:cubic_superpotential_terms_at_critical_points}. 

\begin{table}
 \centering\begin{tabular}{cc}
  \toprule 
  $\varrho$ & $\mathscr{W}_\mathrm{trilinear}$\\
  \midrule 
  $\omega$ & $\bigl(\omega\,X+Y+Z\bigr)\, \bigl(X+\omega\,Y+Z\bigr)\,\bigl(X+Y+\omega\,Z\bigr)$\\
  $\omega+1$ & $\bigl(\omega^2\,X+Y+Z\bigr)\, \bigl(\omega\,X+\omega\,Y+Z\bigr)\,\bigl(\omega\,X+Y+\omega\,Z\bigr)$\\
  $\omega+2$ & $\bigl(X+Y+Z\bigr)\, \bigl(\omega^2\,X+\omega\,Y+Z\bigr)\,\bigl(\omega^2\,X+Y+\omega\,Z\bigr)$\\
  $\nu,\nu+1,\nu+2$ & $X\,Y\,Z$\\
  \bottomrule
 \end{tabular}  
 \caption{Trilinear superpotential terms at the critical points. We use shorthands of the form \eqref{eq:XYZ_shorthands}.}
 \label{tab:cubic_superpotential_terms_at_critical_points}
\end{table}

It is instructive to compare the results in \Cref{tab:cubic_superpotential_terms_at_critical_points} with the complete superpotential that is valid at any point in K\"ahler moduli space. 
As pointed out in~\cite{Nilles:2020gvu}, the complete trilinear superpotential is given by
\begin{align}\label{eq:W_trilinear}
\mathscr{W}_\mathrm{trilinear}(\varrho; X,Y,Z)&= \mathcal{Y}_{2}(\varrho)\,\left(X^3+Y^3+Z^3\right)-3\sqrt{2}\,\mathcal{Y}_{1}(\varrho)\,X\,Y\,Z\;.
\end{align}
Here, we introduced two short-hand notations,
\begin{subequations}\label{eq:XYZ_shorthands}
\begin{align}
  X^3+Y^3+Z^3&\defeq X_1\,X_2\,X_3+Y_1\,Y_2\,Y_3+Z_1\,Z_2\,Z_3\;,\\
  X\,Y\,Z&\defeq \frac{1}{6}\bigl(X_1\,Y_2\,Z_3 +X_1\, Y_3\,Z_2+X_2\,Y_1\,Z_3+X_2\,Y_3\,Z_1+X_3\,Y_1\,Z_2+X_3\,Y_2\,Z_1 \bigr)\;,
\end{align}  
\end{subequations}
where the subscripts indicate additional quantum numbers such that these expressions are gauge invariant. 
For instance, in the case of standard embedding the twisted fields transform as $\rep{27}$-plets of $\mathrm{E}_6$.
Further, in \Cref{eq:W_trilinear} $\mathcal{Y}_{i}(\varrho)$ is the component of the modular form doublet, which can be constructed using the Dedekind eta-function $\eta$~\cite{Liu:2019khw},\footnote{The Dedekind eta-function here is defined as $\eta(\varrho)=q^{1/24}\,\prod_{\infty}^{n=1}\left(1-q^n\right)$ with $q=\ee^{2\pi\ii\varrho}$. This definition differs from the one used by \texttt{Mathematica} by a factor $\sqrt{2}$.}
\begin{equation}\label{eq:Y^(1)_2''}
\mathcal{Y}^{(1)}_{\rep{2''}}(\varrho) \defeq \begin{pmatrix}
\mathcal{Y}_{1}(\varrho)\vphantom{\dfrac{\eta(\varrho/3)^3}{\eta(\varrho)}} \\ \mathcal{Y}_{2}(\varrho)\vphantom{\dfrac{\eta(\varrho/3)^3}{\eta(\varrho)}}
\end{pmatrix} \defeq \mathcal{N}\,\begin{pmatrix}
-3\sqrt{2}\dfrac{\eta(3\varrho)^3}{\eta(\varrho)}\\ 3\dfrac{\eta(3\varrho)^3}{\eta(\varrho)}+\dfrac{\eta(\varrho/3)^3}{\eta(\varrho)}
\end{pmatrix}\;.
\end{equation}
The normalization factor is give by $\mathcal{N}=2^{-\nicefrac{1}{2}}3^{\nicefrac{1}{4}}\frac{\mathnormal{\Gamma}(2/3)^2}{\mathnormal{\Gamma}(1/3)}$~\cite{Chun:1989se,Lauer:1990tm}, where $\mathnormal{\Gamma}$ stands for the usual Gamma-function and should be distinguished from the $\Gamma$ in the congruence subgroup $\Gamma(3)$. 
Further, the upper index $(1)$ of $\mathcal{Y}^{(1)}_{\rep{2''}}(\varrho)$ indicates that the modular weight is $1$, while the lower index $\rep{2''}$ indicates that it forms the doublet $\rep{2''}$ of the finite modular group $T'$, and the corresponding representation matrices are given by 
\begin{subequations}\label{eq:rho_2''}
\begin{align}
  \rho_{\rep{2''}}(\mathsf{S})&=-\frac{\ii}{\sqrt{3}}\begin{pmatrix}
    1 & \sqrt{2} \\ \sqrt{2} & -1
    \end{pmatrix}\;,\\ 
  \rho_{\rep{2''}}(\mathsf{T})&=\begin{pmatrix}
    \omega & 0 \\ 0 & 1
  \end{pmatrix}\;,
\end{align}
\end{subequations}
respectively. The localization eigenstates transform with the representation matrices \eqref{eq:rho(S)_and_rho(T)}.

At the critical points in K\"ahler moduli space, the modular form doublet exhibits residual symmetries, i.e.\ it is an eigenvector of the corresponding residual symmetry group.
Specifically, at the critical points $\varrho=\omega+n$ for $n\in\mathds{Z}$, we have
\begin{equation}\label{eq:Y^(1)_2''@omega}
\mathcal{Y}^{(1)}_{\rep{2''}}(\omega+n)=2^{\nicefrac{1}{2}}3^{-\nicefrac{3}{4}}\, \begin{pmatrix}
\dfrac{\omega^{n+1}}{\sqrt{2}} \\ 1 \end{pmatrix} \;.
\end{equation}
At the other critical points $\varrho=\nu+n$ with $\nu=-1/2+\ii\sqrt{3}/6$ for $n\in\mathds{Z}$, we have
\begin{equation}\label{eq:Y^(1)_2''@nu}
  \mathcal{Y}^{(1)}_{\rep{2''}}(\nu+n)=3^{\nicefrac{1}{4}}\omega^{n+1}\,  \begin{pmatrix}
  1 \\ 0 \end{pmatrix}  \;.
\end{equation}
Therefore, at the various fixed points for $n\in\mathds{Z}$ the trilinear superpotential can be summarized as
\begin{subequations}
  \begin{align}
  \mathscr{W}_\mathrm{trilinear}(\varrho=\omega+n) &=2^{\nicefrac{1}{2}}3^{-\nicefrac{3}{4}}\,\left[ \left(X^3+Y^3+Z^3\right) - 3\omega^{n+1}\,X\,Y\,Z \right]\;,\\
  \mathscr{W}_\mathrm{trilinear}(\varrho=\nu+n) &=3^{\nicefrac{1}{4}}\omega^{n+1}\, X\,Y\,Z
  \end{align}
\end{subequations}
in agreement with~\Cref{eq:rho_2'',tab:cubic_superpotential_terms_at_critical_points}.

\section{Towers}
\label{sec:Towers}

Let us now discuss the winding modes in more detail.
As we shall see, they come in towers labeled by appropriate orbifold-invariant winding numbers. 
In order to construct orbifold-invariant states, it is convenient to define 
\begin{equation}\label{eq:orbifold_z_N_and_z_W}
 \begin{pmatrix}
  z_N\\ z_W
 \end{pmatrix}\defeq\begin{pmatrix}
  -\omega\, n_1 + n_2\\
   \omega\, w^2 + w^1 
 \end{pmatrix}\;.  
\end{equation}
The orbifold action and modular transformations of $\varrho$ act on these combinations as 
\begin{subequations}\label{eq:orbifold_and_modular_transformation_of_z_N_and_z_W}
  \begin{align}
    \rho_z(\theta)&=\begin{pmatrix}
      \omega &0 \\ 0 &\omega
    \end{pmatrix}\;,\label{eq:orbifold_transformation_of_z_N_and_z_W}\\
    \rho_z(\mathsf{S}_\varrho)&=\begin{pmatrix}
      0 &1 \\ -1 &0
    \end{pmatrix}\;,\\
    \rho_z(\mathsf{T}_\varrho)&=\begin{pmatrix}
      1 &1 \\ 0 &1
    \end{pmatrix}\;.
\end{align}  
\end{subequations}
Clearly, the modular transformations of $(z_N,z_W)$ coincide with the two generators $\mathsf{S}$ and $\mathsf{T}$ of $\Group{SL}{2,\mathds{Z}}$ matrices. 
Moreover, these matrices commute with $\theta$, and hence the corresponding transformations have a chance to survive orbifolding. 
As we have seen in \Cref{sec:Orbifolding}, $\Group{SL}{2,\mathds{Z}}_\varrho$ is broken by the orbifold action to its congruence subgroup $\Gamma(3)$.
Therefore, our aim is to build quantities that are invariant under the orbifold action and $\Gamma(3)$.

We can use $z_N$ and $z_W$ to construct various orbifold invariant quantities,
\begin{subequations}\label{eq:orbifold_tower_quantum_numbers}
  \begin{align}
    \TowerN^2 &\defeq |z_N|^2 = n_1^2 + n_2^2 + n_1 n_2\;,\\
    \TowerW^2 &\defeq |z_W|^2 = (w^1)^2 + (w^2)^2 - w^1 w^2\;,\\
    \TowerR^2 &\defeq 2 \re (z^*_N z_W) = w^1 (n_1 + 2 n_2) - w^2 (2 n_1 + n_2)\;,\\
    \TowerI^2 &\defeq \frac{4}{\sqrt{3}} \im (z_N^* z_W) = 2(n_1 w^1 + n_2 w^2)\;,\\
    \OrbN &\defeq -2\re z_N^3\bmod3 =(-n_1+n_2)\bmod 3\;,\label{eq:s_N_mod_def}\\
    \OrbW &\defeq -2\re z_W^3\bmod3=(w^1 + w^2)\bmod 3\;.\label{eq:s_W_mod_def}
\end{align}
\end{subequations}
We see that $\OrbN$ and $\OrbW$ result in expressions that have been used in \cite{Baur:2019iai}.\footnote{Note that in the conventions of \cite{Baur:2019iai}, $\OrbN$ and $\OrbW$ were denoted by $N$ and $W$, respectively. In our conventions, $N$ and $W$ are given by $|z_N|$ and $|z_W|$ up to a sign.}
$\TowerI^2$ is both orbifold and modular invariant. 
On the other hand, $\TowerN^2$, $\TowerW^2$ and $\TowerR^2$ follow the transformation laws
\begin{subequations}
\begin{align}
    \begin{pmatrix}
      \TowerW^2 \\ \TowerN^2 \\ \TowerR^2
    \end{pmatrix}
    &\xmapsto{~\mathsf{S}_\varrho~} \begin{pmatrix*}[r]
      \phantom{-}0 & \phantom{-}1 & 0 \\ 1 & 0 & 0 \\ 0 & 0 & -1
    \end{pmatrix*}\cdot\begin{pmatrix}
      \TowerW^2 \\ \TowerN^2 \\ \TowerR^2
    \end{pmatrix}\eqdef\rho_\mathrm{tower}(\mathsf{S}_\varrho) \,\begin{pmatrix}
      \TowerW^2 \\ \TowerN^2 \\ \TowerR^2
    \end{pmatrix}\;,\\
    \begin{pmatrix}
      \TowerW^2 \\ \TowerN^2 \\ \TowerR^2
    \end{pmatrix}
    &\xmapsto{~\mathsf{T}_\varrho~} \begin{pmatrix*}[r]
      1 & 0 & 0 \\ 1 & 1 & 1 \\ 2 & 0 & 1
    \end{pmatrix*}\cdot\begin{pmatrix}
      \TowerW^2 \\ \TowerN^2 \\ \TowerR^2
    \end{pmatrix}\eqdef\rho_\mathrm{tower}(\mathsf{T}_\varrho) \,\begin{pmatrix}
      \TowerW^2 \\ \TowerN^2 \\ \TowerR^2
    \end{pmatrix}\;.
\end{align}
\end{subequations}
The $s_N$ and $s_W$ follow transformation laws
\begin{subequations}
  \begin{align}
    \begin{pmatrix}
      \OrbN \\ \OrbW
    \end{pmatrix}
    &\xmapsto{~\mathsf{S}_\varrho~} \begin{pmatrix*}[r]
      \phantom{-}0 & \phantom{-}1 \\ -1 & 0
    \end{pmatrix*}\cdot\begin{pmatrix}
      \OrbN \\ \OrbW
    \end{pmatrix} \bmod 3 \eqdef\rho_\mathrm{orb}(\mathsf{S}_\varrho) \,\begin{pmatrix}
      \OrbN \\ \OrbW
    \end{pmatrix} \bmod 3\;,\\
    \begin{pmatrix}
      \OrbN \\ \OrbW
    \end{pmatrix}
    &\xmapsto{~\mathsf{T}_\varrho~} \begin{pmatrix*}[r]
      1 & 1 \\ 0 & 1
    \end{pmatrix*}\cdot\begin{pmatrix}
      \OrbN \\ \OrbW
    \end{pmatrix} \bmod 3 \eqdef\rho_\mathrm{orb}(\mathsf{T}_\varrho) \,\begin{pmatrix}
      \OrbN \\ \OrbW
    \end{pmatrix} \bmod 3\;.
  \end{align}
\end{subequations}
The corresponding matrices fulfill 
\begin{subequations}
\begin{align}
 \bigl[\rho_\mathrm{tower}(\mathsf{S}_\varrho)\bigr]^4&=\mathds{1}_3\;,
 &\bigl[\rho_\mathrm{tower}(\mathsf{T}_\varrho)\bigr]^3&=\mathds{1}_3\bmod 3\;,\\ \bigl[\rho_\mathrm{orb}(\mathsf{S}_\varrho)\bigr]^4&=\mathds{1}_2\;,&
 \bigl[\rho_\mathrm{orb}(\mathsf{T}_\varrho)\bigr]^3&=\mathds{1}_2\bmod 3\;.
\end{align}  
\end{subequations}
After orbifolding, these quantum numbers can be used to construct the mass equation, \cref{eq:mass_TD} and level matching, \cref{eq:level_matching}, as
\begin{equation}\label{eq:TowerMass}
  m^2 = \frac{2}{\sqrt{3} \im \varrho}\left( \TowerN^2 + |\varrho|^2 \TowerW^2 -\TowerR^2 \re \varrho \right)+ 2\left(N_\mathrm{L} + N_\mathrm{R} - \begin{cases} 1 & (\text{R})\\ \frac{3}{2} & (\text{NS}) \end{cases}\right)\;,
\end{equation}
and
\begin{equation}
  0 = \TowerI^2 + 2\left( N_\mathrm{L} - N_\mathrm{R} - \begin{cases} 1 & (\text{R}) \\ \frac{1}{2} & (\text{NS}) \end{cases} \right)\;.
\end{equation}
This demonstrates that $\TowerN^2$ and $\TowerW^2$ are squares of the \ac{KK} and winding numbers on the orbifold (rather than the torus).
This is consistent with the fact that, instead of the two duality groups $\Group{SL}{2,\mathds{Z}}_\tau$ and $\Group{SL}{2,\mathds{Z}}_\varrho$ of the upstairs theory, the orbifold has a $\Gamma(3)$ symmetry that fits into $\Group{SL}{2,\mathds{Z}}_\varrho$. 
In particular, the quantum numbers defined in \eqref{eq:orbifold_tower_quantum_numbers} can be used to label the massive gauge fields discussed in \Cref{sec:Orbifolding}. 
Examples of these labels for the doublets that form massless gauge bosons at $\varrho = \omega, \omega+1, \omega + 2,$ and $\nu$ are given in \Cref{tab:tower_labels}.

\begin{table}[t]
  \centering\begin{tabular}{ccccccc}
    \toprule
    $\Group{\Delta}{54}$ &$\Delta(27)$ & $(\OrbW, \OrbN)$ & $(\TowerW^2, \TowerN^2, \TowerR^2)$ & $\mathsf{A}$ & $\mathsf{B}$ & $\mathsf{C}$\\ 
    \midrule
    $\rep{2}_1$ & $\begin{pmatrix} \rep{1}_{0,2} \\ \rep{1}_{0,1}  \end{pmatrix}$ & $\begin{pmatrix} (2,0) \\ (1,0) \end{pmatrix}$ & $(1,3,3)$ & $\begin{pmatrix}  1 & 0 \\ 0 & 1\end{pmatrix}$ & $\begin{pmatrix} \omega^2 & 0 \\ 0 & \omega \end{pmatrix}$ & $\begin{pmatrix} 0 & 1 \\ 1 & 0 \end{pmatrix}$ \\
    \midrule
    $\rep{2}_2$  & $\begin{pmatrix} \rep{1}_{2,0} \\ \rep{1}_{1,0}  \end{pmatrix}$ & $\begin{pmatrix} (0,1) \\ (0,2) \end{pmatrix}$ & $(3,1,-3)$ & $\begin{pmatrix}  \omega^2 & 0 \\ 0 & \omega \end{pmatrix}$ & $\begin{pmatrix} 1 & 0 \\ 0 & 1 \end{pmatrix}$ & $\begin{pmatrix} 0 & 1 \\ 1 & 0 \end{pmatrix}$ \\
    \midrule
    $\rep{2}_3$  & $\begin{pmatrix} \rep{1}_{2,2} \\ \rep{1}_{1,1}  \end{pmatrix}$ & $\begin{pmatrix} (2,1) \\ (1,2) \end{pmatrix}$ & $(1,1,1)$ & $\begin{pmatrix}  \omega^2 & 0 \\ 0 & \omega \end{pmatrix}$ & $\begin{pmatrix} \omega^2 & 0 \\ 0 & \omega \end{pmatrix}$ & $\begin{pmatrix} 0 & 1 \\ 1 & 0 \end{pmatrix}$ \\
    \midrule
    $\rep{2}_4$  & $\begin{pmatrix} \rep{1}_{2,1} \\ \rep{1}_{1,2}  \end{pmatrix}$ & $\begin{pmatrix} (1,1) \\ (2,2) \end{pmatrix}$ & $(1,1,-1)$ & $\begin{pmatrix}  \omega^2 & 0 \\ 0 & \omega \end{pmatrix}$ & $\begin{pmatrix} \omega & 0 \\ 0 & \omega^2 \end{pmatrix}$ & $\begin{pmatrix} 0 & 1 \\ 1 & 0 \end{pmatrix}$ \\
    \bottomrule
  \end{tabular}
  \caption{\Group{\Delta}{54} doublets. $\mathsf{A}$, $\mathsf{B}$ and $\mathsf{C}$ are the generators, cf.\ \Cref{sec:Delta(54)}. Example values of the orbifold invariant quantum numbers $(\TowerW^2,\TowerN^2,\TowerR^2)$ are given for the massless doublets at different symmetry enhanced points, i.e. the $\rep{2}_4$ at $\varrho = \omega$, the $\rep{2}_3$ at $\varrho = \omega+1$, the $\rep{2}_1$ at $\varrho = \omega + 2$, and the $\rep{2}_2$ at $\varrho = \nu$.}
  \label{tab:tower_labels}
\end{table}
  
By expanding the mass equation, \Cref{eq:TowerMass}, in terms of perturbations in the axionic direction $\varrho_1$ about a critical point $\varrho_\mathrm{crit}$, we can see how the couplings of $\varrho-\varrho_\mathrm{crit}$ to the doublets differ depending on the value of $\varrho_\mathrm{crit}$. 
Sending $\varrho \rightarrow \varrho_\mathrm{crit}+\delta\varrho$ with $\delta\varrho \in \mathds{R}$, we find (neglecting contributions from the oscillators which are not relevant here)
\begin{equation}\label{eq:m^2_tower_axion}
  m^2 = \frac{2}{\sqrt{3}\im\varrho_\mathrm{crit}}\biggl\{ \left[ \TowerN^2 + \TowerW^2 \,|\varrho_\mathrm{crit}|^2 - \TowerR^2\, \re \varrho_\mathrm{crit}\right] + \left[ 2 \TowerW^2 \,\re \varrho_\mathrm{crit} - \TowerR^2 \right] (\delta\varrho) +\left[ \TowerW^2 \right] (\delta \varrho)^2\biggr\}\;.
\end{equation}

The quantum numbers presented in this section allow us to label states in towers. 
On the torus, $\Group{SL}{2,\mathds{Z}}_\varrho$ transformations relate different states on a tower of states. 
On the orbifold, only $\Gamma(3)$ transformations map a state from a given tower to another state on the same tower. 
The remaining $\Group{SL}{2,\mathds{Z}}_\varrho$ transformations relate states carrying different quantum numbers.
This means that orbifolding breaks the modular and gauge symmetries of the upstairs theory to a diagonal subgroup in which $\Group{SL}{2,\mathds{Z}}_\varrho$ transformations that do not belong to $\Gamma(3)$ have to be combined with additional transformations. 
As we shall discuss in more detail in the next sections, these additional transformations are the outer automorphisms of $\Delta(54)$.

\section{\texorpdfstring{$\mathcal{CP}$}{CP} transformation}
\label{sec:CP}

Let us revisit the mass equation~\eqref{eq:TowerMass} at $\tau=\omega$,
\begin{equation}\label{eq:MassEq_CP}
m^2 \propto \frac{2}{\sqrt{3}\,\im\varrho} \left|z_N-\overline{\varrho}\,z_W \right|^2 = \frac{2}{\sqrt{3}\,\im\varrho} \left|\overline{z}_N-\varrho\, \overline{z}_W\right|^2\;.
\end{equation}
The mass spectrum has, in addition to the modular $\Group{SL}{2,\mathds{Z}}_{\varrho}$ symmetry, a symmetry that can be related to $\mathcal{CP}$.
In slightly more detail, as we shall see, at special curves in $\varrho$ moduli space $\mathcal{CP}$ can be conserved as far as the $\mathds{T}^2/\mathds{Z}_3$ orbifold plane is concerned.

The canonical $\mathcal{CP}$ transformation of the K\"ahler modulus $\varrho$ is given by \cite{Dent:2001cc,Baur:2019kwi,Novichkov:2019sqv}
\begin{equation}
\varrho\xmapsto{~\mathcal{CP}~} -\overline{\varrho}\;.
\end{equation}
In order to be a symmetry of the mass equation \eqref{eq:MassEq_CP}, we have to demand that
\begin{subequations}\label{eq:zNzWCP}
\begin{align}
z_N &\xmapsto{~\mathcal{CP}~} \alpha\, \overline{z}_N \;,\\
z_W &\xmapsto{~\mathcal{CP}~} -\alpha\, \overline{z}_W \;.
\end{align}
\end{subequations}
Here, $\alpha$ is a phase and can only be $\alpha\in\pm\{1,\omega,\omega^2\}$. 
It originates from the universal residual symmetry $\mathds{Z}^{(\mathsf{S\,T})_\tau}_{3}$ and is therefore related to the $\mathds{Z}_3$ twist.
For convenience, let us choose $\alpha=\omega^2$. 
Then \eqref{eq:zNzWCP} becomes
\begin{subequations}
\begin{align}
z_N=n_2-n_1\,\omega& \eqmathbox[CP][r]{{}\xmapsto{~\mathcal{CP}~}{}} \omega^2\, \overline{z}_N\notag\\
&\eqmathbox[CP][r]{{}={}} n_2\,\omega^2-n_1\,\omega = -n_2\,(1+\omega)-n_1\,\omega=-n_2-(n_1+n_2)\,\omega \;,\\
z_W=w^1+w^2\,\omega &\eqmathbox[CP][r]{{}\xmapsto{~\mathcal{CP}~}{}} -\omega^2\, \overline{z}_W\notag\\
&\eqmathbox[CP][r]{{}={}}-w^1\,\omega^2-w^2\,\omega=w^1\,(1+\omega)-w^2\,\omega=w^1+(w^1-w^2)\,\omega \;.
\end{align}
\end{subequations}
That is, the canonical $\mathcal{CP}$ transformation acts on winding and \ac{KK} numbers as
\begin{equation}\label{eq:wwnnCP}
\left(w^1,w^2,n_1,n_2\right)\xmapsto{~\mathcal{CP}~} \left(w^1,w^1-w^2,n_1+n_2,-n_2\right)\;.
\end{equation}
The corresponding $\mathcal{CP}$ transformation matrix is
\begin{equation}
\label{eq:CPmat}
\mathcal{CP}=\begin{pmatrix}
1 & 0 & 0 & 0 \\
1 & -1 & 0 & 0 \\
0 & 0 & 1 & 1 \\
0 & 0 & 0 & -1 
\end{pmatrix}\;.
\end{equation}
This transformation law is consistent with $K_*$ in \cite[Equation~(48)]{Baur:2019iai}, where the $\mathcal{CP}$ symmetry is found from the outer automorphisms of the Narain lattice.
The $\mathcal{CP}$ transformation has order 2, $\mathcal{CP}^2=1$, so $\mathcal{CP}$ generates $\mathds{Z}^{\mathcal{CP}}_2$. 
On the other hand, $\mathcal{CP}$ can also be regarded as an outer automorphism of the $\mathds{Z}_3$ twist, because $\mathcal{CP} \,\theta\, \mathcal{CP}^{-1} = \theta^2$.
The $\mathcal{CP}$ symmetry extends the original modular group $\Group{SL}{2,\mathds{Z}}_{\varrho}$ to the so-called extended modular group $\Group{SL}{2,\mathds{Z}}_{\varrho}\rtimes \mathds{Z}^{\mathcal{CP}}_2\cong \Group{GL}{2,\mathds{Z}}_{\varrho}$~\cite{Baur:2019iai,Novichkov:2019sqv}.\footnote{In mathematics, the extended modular group was introduced in \cite{alma9951093201865}.} 

The above way of discussing the $\mathcal{CP}$ transformation may suggest that it is an accidental symmetry. 
However, this is not the case. 
Rather, orbifolding breaks the T-duality group from $\mathrm{O}(2,2;\mathds{Z})\cong \left(\Group{SL}{2,\mathds{Z}}_{\tau}\times \Group{SL}{2,\mathds{Z}}_{\varrho}\right)\rtimes \left(\mathds{Z}^{M}_2\times \mathds{Z}^{\mathcal{CP}}_2\right)$ to $\left(\mathds{Z}^{(\mathsf{ST})_\tau}_{3}\times\Group{SL}{2,\mathds{Z}}_{\varrho}\right)\rtimes \mathds{Z}^{\mathcal{CP}}_2$ after orbifolding.
Whether or not $\mathcal{CP}$ is conserved in the upstairs, i.e.\ torus, theory depends on the \acp{VEV} of $\tau$ and $\varrho$. 
In the downstairs theory, i.e.\ on the orbifold, $\tau$ is fixed, and the requirement of $\mathcal{CP}$ conservation restricts $\varrho$ to specific curves, as we discuss now in more detail.

According to the $\mathcal{CP}$ transformation~\eqref{eq:wwnnCP}, we can infer the $\mathcal{CP}$ transformation of orbifold-invariant winding and \ac{KK} numbers $(s_W,s_N)$ to be
\begin{equation}
\mathcal{CP}~\colon ~ (s_W,s_N) \xmapsto{~\mathcal{CP}~} (-s_W, s_N)\;.
\end{equation}
Therefore, the transformation behavior of the $\Delta(54)$ doublets, i.e.\ gauge fields, under $\mathcal{CP}$ can be summarized as
\begin{subequations}
\label{eq:doubletsCP}
\begin{align}
\rep{2}_1 &= \begin{pmatrix} (2,0) \\ (1,0)\end{pmatrix} \xmapsto{~\mathcal{CP}~} \begin{pmatrix} (1,0) \\ (2,0)\end{pmatrix} = \rep{2}_1  \;,&
\rep{2}_2 &= \begin{pmatrix} (0,1) \\ (0,2)\end{pmatrix} \xmapsto{~\mathcal{CP}~} \begin{pmatrix} (0,1) \\ (0,2)\end{pmatrix} = \rep{2}_2  \;,\\
\rep{2}_3 &= \begin{pmatrix} (2,1) \\ (1,2)\end{pmatrix} \xmapsto{~\mathcal{CP}~} \begin{pmatrix} (1,1) \\ (2,2)\end{pmatrix} = \rep{2}_4  \;,&
\rep{2}_4 &= \begin{pmatrix} (1,1) \\ (2,2)\end{pmatrix} \xmapsto{~\mathcal{CP}~}
\begin{pmatrix} (2,1) \\ (1,2)\end{pmatrix} = \rep{2}_3 \;.
\end{align}
\end{subequations}
In short,
\begin{equation}\label{eq:2-plet-CP}
\begin{pmatrix}
\rep{2}_1 \\ \rep{2}_2 \\ \rep{2}_3 \\ \rep{2}_4 
\end{pmatrix}\xmapsto{~\mathcal{CP}~}
\begin{pmatrix}
\rep{2}_1 \\ \rep{2}_2 \\ \rep{2}_4 \\ \rep{2}_3 
\end{pmatrix}\;.
\end{equation}
This already shows that, in our setup, the $\mathcal{CP}$ transformation is more subtle than one may naively expect: a ``normal'' $\mathcal{CP}$ transformation does not swap gauge bosons of different $\Group{U}{1}$ factors. 
It also reveals that at a generic point in moduli space $\mathcal{CP}$ is broken since the masses of the gauge bosons transforming as $\rep{2}_3$ and $\rep{2}_4$ do not coincide. 

Let us discuss the twisted sector states.
We know that the canonical $\mathcal{CP}$ transformation of the twisted fields is given by
\begin{equation}
(X, Y, Z)^{\transpose} \xmapsto{~ \mathcal{CP}~} (\overline{X}, \overline{Y}, \overline{Z})^{\transpose}\;.
\end{equation}
That is, the $\mathcal{CP}$ transformation exchanges the twisted fields of $\theta$-twisted sector with their complex conjugate partners from the $\theta^2$-twisted sector.
By using the \acp{OPE} between twisted states, we can infer the $\mathcal{CP}$ transformation of the untwisted states from the $\mathcal{CP}$ transformation of twisted states,
\begin{subequations}
\label{eq:OpeCP}
\begin{align}
\rep{2}_1 & =\frac{1}{\sqrt{3}}\binom{\overline{X}\, Z+\overline{Y}\, X+\overline{Z}\, Y}{\overline{Z}\, X+\overline{X}\, Y+\overline{Y}\, Z} \xmapsto{~\mathcal{CP}~} \frac{1}{\sqrt{3}}\binom{\overline{Z}\, X+\overline{X}\, Y+\overline{Y}\, Z}{\overline{X}\, Z+\overline{Y}\, X+\overline{Z}\, Y} = \rep{2}_1 \;, \\
\rep{2}_2 & =\frac{1}{\sqrt{3}}\binom{\overline{X}\, X+\omega\,\overline{Y}\, Y+\omega^2\,\overline{Z}\, Z}{\overline{X}\, X+\omega^2\,\overline{Y}\, Y+\omega\,\overline{Z}\, Z} \xmapsto{~\mathcal{CP}~} \frac{1}{\sqrt{3}}\binom{\overline{X}\, X+\omega\,\overline{Y}\, Y+\omega^2\,\overline{Z}\, Z}{\overline{X}\, X+\omega^2\,\overline{Y}\, Y+\omega\,\overline{Z}\, Z} = \rep{2}_2 \;, \\
\rep{2}_3 & =\frac{1}{\sqrt{3}}\binom{\omega\,\overline{X}\,Z+\omega^2\,\overline{Y}\,X+\overline{Z}\,Y}{\omega^2\,\overline{Z}\,X+\omega\,\overline{X}\,Y+\overline{Y}\,Z} \xmapsto{~\mathcal{CP}~} \frac{1}{\sqrt{3}}\binom{\omega\,\overline{Z}\, X+\omega ^2\,\overline{X}\,Y+\overline{Y}\,Z}{\omega ^2\,\overline{X}\,Z+\omega\,\overline{Y}\, X+\overline{Z}\,Y}=\rep{2}_4 \;,\\
\rep{2}_4 & =\frac{1}{\sqrt{3}}\binom{\omega ^2\,\overline{X}\,Y+\overline{Y}\,Z+\omega\,\overline{Z}\, X}{\omega\,\overline{Y}\, X+\overline{Z}\,Y+\omega ^2\,\overline{X}\,Z} \xmapsto{~\mathcal{CP}~} \frac{1}{\sqrt{3}}\binom{\omega^2\,\overline{Y}\,X+\overline{Z}\,Y+\omega\,\overline{X}\,Z}{\omega\,\overline{X}\,Y+\overline{Y}\,Z+\omega^2\,\overline{Z}\,X}=\rep{2}_3\;.
\end{align}
\end{subequations}
As one can see, this is exactly the same $\mathcal{CP}$ transformation of the untwisted doublets as~\Cref{eq:doubletsCP}, as it should be. 

It is noteworthy that the above $\mathcal{CP}$ transformation is actually reflected in \Cref{fig:cartoon_doublet_masses,fig:cartoon_nu_doublet_masses}: 
when we change $\varrho$ to $-\overline{\varrho}$, i.e.\ $\re\varrho\mapsto -\re\varrho$, the figure gets reflected at the imaginary axis, which precisely interchanges $\rep{2}_3$ and $\rep{2}_4$ while $\rep{2}_1$ and $\rep{2}_2$ remain unchanged.

\begin{figure}[htb]
  \centering\includegraphics{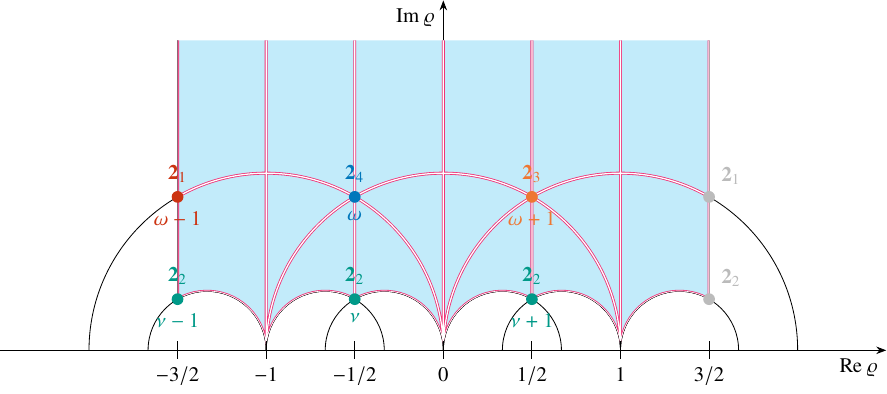}%
  \caption{$\mathcal{CP}$-conserving curves (hollow) in the fundamental domain of $\Gamma(3)$.}
  \label{fig:CP_curves}
\end{figure}

This raises the question whether $\rep{2}_3$ and $\rep{2}_4$ play a more special role for $\mathcal{CP}$ than the other doublets. 
It turns out that they do not, and the doublets are, as far as $\mathcal{CP}$ is concerned, on similar footing. 
This is because the $\mathcal{CP}$ transformation in \Cref{eq:doubletsCP,eq:OpeCP} is not unique.  
Instead, the $\mathcal{CP}$-conserving loci in $\varrho$-moduli space are determined by the condition that $\varrho$ equals $-\overline\varrho$ up to an $\Group{SL}{2,\mathds{Z}}$ transformation, i.e.\ 
\begin{equation}\label{eq:varrho_CP}
-\overline{\varrho}_{\mathcal{CP}} = \frac{a\, \varrho_{\mathcal{CP}} + b}{c\,\varrho_{\mathcal{CP}} +d} 
\end{equation}
for some $\begin{psmallmatrix} 
    a & b\\ c & d
\end{psmallmatrix}\in\Group{SL}{2,\mathds{Z}}$.\footnote{One may refer to the map $\varrho\mapsto-\bigl((a\,\varrho+b)/(c\,\varrho+d)\bigr)^*$ as ``generalized'' $\mathcal{CP}$ transformation \cite{Holthausen:2012dk,Feruglio:2012cw}. However, this term has been used rather generously. We stress that our $\mathcal{CP}$ transformation is a proper class-inverting automorphism of the enhanced symmetry, see \cite{Chen:2014tpa} for a detailed discussion.}   
This defines $\mathcal{CP}$-conserving curves in the $\varrho$-moduli space. 
They are given by the vertical lines of $\re\varrho_{\mathcal{CP}}=0$, $\pm 1/2$, $\pm 1$, \dots and the $\Group{SL}{2,\mathds{Z}}$ images of the arcs $|\varrho_{\mathcal{CP}}|=1$ in the upper half plane $\mathcal{H}_{\varrho}$, see \Cref{fig:CP_curves}. 
When the K\"ahler modulus $\varrho$ is fixed away from these curves, the $\mathcal{CP}$ symmetry is broken. 
As we shall see, the way $\mathcal{CP}$ is broken is somewhat subtle.

The enhanced (residual) symmetry of this family of $\mathds{Z}^{\mathcal{CP}}_2$ transformations is also reflected in \Cref{fig:cartoon_doublet_masses}, in which we can see the degeneracy of the masses of certain untwisted doublets on the vertical lines where $\mathcal{CP}$ is conserved.
For example, at $\re\varrho_{\mathcal{CP}}=0$, the masses of the doublets $\rep{2}_3$ and $\rep{2}_4$ are the same. 
The action of the residual $\mathds{Z}^{\mathcal{CP}}_2$ consists in the exchange $\rep{2}_3\leftrightarrow\rep{2}_4$. 
Likewise, if we reflect at the axis $\re\varrho_{\mathcal{CP}}=-1$, we see that this swaps $\rep{2}_1$ and $\rep{2}_3$ while leaving the other doublets unchanged. 
A reflection around the circle of radius 1 and with center at $-1$ exchanges $\rep{2}_2$ and $\rep{2}_3$ while not affecting $\rep{2}_1$ and $\rep{2}_4$.
This is different from the reflection at the circle with center at 0, which swaps $\rep{2}_1$ and $\rep{2}_2$, and geometrically corresponds to the $\mathsf{S}_\varrho$ transformation.
One may verify that, given that $\varrho$ attains a $\mathcal{CP}$ conserving value \eqref{eq:varrho_CP}, the masses of the respective doublets coincide so that swapping them is indeed a symmetry.

\begin{wrapfigure}[14]{r}{6cm}
\centering\includegraphics{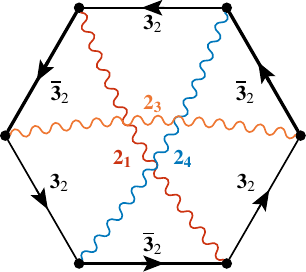}
\caption{$\mathcal{CP}$ violation appears at the loop level.}
\label{fig:CP_loop}
\end{wrapfigure}

It is instructive to discuss the above findings in the context of flavor symmetries. 
It is known that both $\Delta(27)$ and $\Delta(54)$ are so-called type-I groups, which do not allow for a physical $\mathcal{CP}$ transformation for a generic matter content \cite{Chen:2014tpa}.
This question has been studied in the context of the $\mathds{Z}_3$ orbifold in \cite{Nilles:2018wex}, where it was found that, provided there are at least 3 $\Delta(54)$ doublets, $\mathcal{CP}$ is violated in decays of these doublets. 
The loop diagram used in \cite[Figure~3]{Nilles:2018wex} has, according to our findings, a rather peculiar interpretation: the doublets are gauge bosons and the \ac{CG} coefficients are \Group{U}{1} generators, cf.\ \Cref{fig:CP_loop}.
In other words, $\mathcal{CP}$ violation arises from gauge interactions due to the ``misalignment'' of \Group{U}{1} generators. 
We also see that at the $\mathcal{CP}$ conserving points in moduli space the discrete symmetry gets enhanced to a group allowing for a $\mathcal{CP}$ transformation, in agreement with the discussion in \cite{Baur:2019iai}.
This symmetry enhancement leads to degeneracies of the doublet masses (cf.\ \Cref{fig:cartoon_doublet_masses,fig:cartoon_nu_doublet_masses}), such that the loop shown in \Cref{fig:CP_loop} no longer has a nontrivial imaginary part. 

How can one characterize this kind of $\mathcal{CP}$ violation?
It is possible to attribute $\mathcal{CP}$ violation to the departure of $\varrho$ from $\mathcal{CP}$-conserving regions in moduli space. 
However, it is important to note that $\mathcal{CP}$ violation does not (only) come about through the presence of phases in $\varrho$-dependent (Yukawa) couplings but also from the \ac{CG} coefficients \cite{Chen:2009gf,Chen:2014tpa}. 
That is, even if one sets the $\varrho$-dependent Yukawa couplings to zero, $\mathcal{CP}$ is still violated in diagrams like the one shown in \Cref{fig:CP_loop}. 
In this regard, the situation is similar to the scenario discussed in \cite{Ratz:2016scn}.
Altogether we thus see that the way $\mathcal{CP}$ is violated is more subtle than usually appreciated. 

This raises the question of why $\varrho$ settles at a $\mathcal{CP}$-violating point. 
In the framework of duality invariant gaugino condensation \cite{Nilles:1990jv,Font:1990nt} it has been conjectured that $\varrho$ always gets stabilized on the boundary of the fundamental domain of $\Group{SL}{2,\mathds{Z}}$ \cite{Cvetic:1991qm}. 
More detailed analyses \cite{Dent:2001ut,Novichkov:2022wvg,Leedom:2022zdm} revealed that, using the most general expressions allowed by modular invariance, there are local minima away from this boundary. 
These minima violate $\mathcal{CP}$. 
In \cite{Knapp-Perez:2023nty} a different route has been suggested: the dynamics of matter fields can not only move $\varrho$ away from the critical points by a parametrically small amount but also leads to vacua in which $\varrho$ settles at $\mathcal{CP}$-violating values. 
A necessary condition for this is that there is $\mathcal{CP}$ violation in the matter interactions. 
As we have seen, the \ac{CG} coefficients derived from the misaligned $[\Group{U}{1}\times\Group{U}{1}]^{(\varrho_\mathrm{crit})}$ symmetries do have this property. 
This means that the subtle ways in which the $\mathcal{CP}$ transformation is realized in this scheme provide us with all the ingredients needed to explain why $\mathcal{CP}$ is violated in the ground state. 
It will be interesting to explore these questions in explicit string models that come close to the \ac{SM}.

\section{Discussion \& applications}
\label{sec:Applications}

\subsection{Dissecting the eclectic symmetry of the $\mathds{Z}_3$ orbifold}
\label{sec:dissecting_eclectic_symmetry}

The combination of modular, traditional, $R$ and $\mathcal{CP}$ symmetries has been dubbed ``eclectic'' \cite{Nilles:2020nnc}. 
Many important insights into the structure of the eclectic flavor group have been obtained previously using outer automorphisms of Narain space group \cite{Baur:2019kwi,Baur:2019iai}.
Our analysis has led to a more precise understanding of these ingredients, and how they interact with each other. 
Specifically, we find that the so-called eclectic symmetry of the $\mathds{Z}_3$ orbifold plane can be written as
\begin{equation}\label{eq:ecl}
  G_\mathrm{eclectic}^{\mathds{Z}_3}=
  \overunderbraces{&\br{3}{=\,\protect\Delta(54)}& &\br{2}{=\,S_4}}%
  {&\Delta(27)&\rtimes&\bigl(\mathds{Z}_2^{\mathsf{C}=\mathsf{S}_\varrho^2}&\mathbin{\boldsymbol{.}}&A_4\bigr)&\rtimes\mathds{Z}_2^{\mathcal{CP}}&}%
  {&&&\br{3}{=\,T'}&&}
  =
  \Delta(54)\rtimes S_4\;.
\end{equation}
Here, the ``$\mathbin{\boldsymbol{.}}$'' between $\mathds{Z}_2^{\mathsf{C}=\mathsf{S}_\varrho^2}$ and $A_4$ indicates a so-called non-split extension (cf.\ \Cref{sec:Group_extensions}). 
We further suppressed the $R$ symmetries coming from stabilizer group of the complex structure modulus $\tau$, cf.\ \Cref{app:Modular_transformations,app:vertex_op_SL2Z_trans}. 
At a generic point in $\varrho$ moduli space, only the $\Delta(54)$ symmetry is linearly realized. 
The outer automorphism group of this $\Delta(54)$ is $S_4$, which is generically broken by the \ac{VEV} of $\varrho$. 
$\Delta(54)$ gets extended ``maximally'' by the $S_4$ symmetry as the modulus $\varrho$ transforms faithfully under it. 
Notice, though, that in the jargon of modular flavor symmetries, the finite modular group is $T'$ (and not $S_4$).
This is because $T'$ comprises the holomorphic transformations of $\varrho$, even though $\varrho$ is not faithful under $T'$ but only $A_4$. 

At special points in $\varrho$ moduli space, subgroups of $S_4$ can be linearly realized. 
In particular, the residual subgroup of $S_4$ at $\varrho=\omega$ is $\mathds{Z}_3^{(\mathsf{S}\,\mathsf{T})_\varrho}\rtimes \mathds{Z}_2^{\mathcal{CP}\circ \mathsf{T}_\varrho} \cong S_3$. 
It corresponds to the permutation symmetry among three massive doublets with mass degeneracy. 
One can see this symmetry in \Cref{fig:cartoon_doublet_masses}, where the masses of $\rep{2}_1$, $\rep{2}_2$ and $\rep{2}_3$ coincide at $\re\varrho=-1/2$.
The residual subgroup of $S_4$ is $\mathds{Z}^{S_\varrho}_2\times \mathds{Z}_2^{\mathcal{CP}}$ at $\varrho=\ii$, which corresponds to the interchange symmetry between two mass-degenerate doublets. 
The $\mathds{Z}_2^{\mathcal{CP}}$ can be seen in \Cref{fig:cartoon_doublet_masses}, which shows that the masses of $\rep{2}_3$ and $\rep{2}_4$ are equal.

\subsection{Interpretation of modular flavor symmetries}
\label{sec:Interpretation_of_modular_flavor_symmetries}

As we have seen, different critical points in the fundamental domain of $\Gamma(3)$ come with massless states that are distinguished by their orbifold-invariant quantum numbers defined in \Cref{sec:Towers}. 
These quantum numbers can be mapped to $\Delta(54)$ quantum numbers. 
Crucially, we have established that the quantum numbers are different at different fixed points. 
This means that the fundamental domain of $\Group{SL}{2,\mathds{Z}}_\varrho$ is insufficient to describe the moduli space of $\varrho$. 
Rather, the appropriate moduli space is the fundamental domain of $\Gamma(3)$. 
Since the modular transformations are gauged, it is impossible for an observer to tell whether they live in universe in which $\varrho$ is close to, say, $\omega$ or $\omega+1$. 
However, the relative misalignment between the symmetries is physical, and gives rise to $\Delta(27)$. 
This means that, while it might be sufficient to scan over the fundamental domain of $\Group{SL}{2,\mathds{Z}}_\varrho$ to see whether a model provides us with a good fit, the physical fundamental domain of $\varrho$ is really the one of $\Gamma(3)$. 
In short, at least in \ac{TD} models, modular flavor symmetries come with additional quantum numbers that distinguish members of multiplets under the finite modular group.

\subsection{Implications for the K\"ahler potential and near-critical points}
\label{sec:Kaehler}

It has been pointed out that values of the modulus close to critical points $\varrho_\mathrm{crit}$ can allow us to address mass hierarchies \cite{Feruglio:2022koo,Feruglio:2023mii,Petcov:2023vws,Chen:2023mwt,Ding:2024ozt}. 
The entries of mass matrices are then proportional to powers of the departure of the modulus from $\varrho_\mathrm{crit}$, somewhat analogous to \ac{FN} scheme~\cite{Froggatt:1978nt}.
Our analysis shows that, in the \ac{TD} approach, at the critical points additional $\Group{U}{1}$ symmetries appear, mediated by gauge fields that become massless at $\varrho=\varrho_\mathrm{crit}$.
Additional massless fields provide us with the degrees of freedom that get eaten by the gauge fields for $\varrho\ne\varrho_\mathrm{crit}$.

The charges of the matter fields transforming in multiplets of the finite modular group have different $\Group{U}{1}$ charges, cf.\ e.g.\ \Cref{tab:U1timesU1_omega_1}.
Therefore, at $\varrho=\varrho_\mathrm{crit}$, the K\"ahler metric of the matter fields is diagonal,\footnote{Even though the superpotential discussed in \Cref{sec:Superpotential} is ``off-diagonal'', \ac{RG} corrections to the K\"ahler potential due to these interactions are diagonal.} schematically
\begin{equation}
 K\supset\frac{1}{(-\ii\varrho+\ii\bar\varrho)^{\nicefrac{2}{3}}} \Phi_{(\varrho_\mathrm{crit})}^\dagger\,\ee^{2g\,V}\,\Phi_{(\varrho_\mathrm{crit})}\;.   
\end{equation}
Here, $g$ denotes the gauge coupling and $V$ the gauge fields. 
This gives us more control over the K\"ahler potential, which is otherwise less constrained~\cite{Chen:2019ewa}. 

This protection mechanism is different from other proposals in the literature in that no additional flavon is needed. 
In order to understand this, let us briefly review the protection mechanisms proposed so far. 
In \cite[Section~3.2]{Nilles:2020kgo} it has been shown that the $\Delta(54)$ symmetry greatly restricts the K\"ahler potential. 
However, one needs to break the $\Delta(54)$ with appropriate flavons. 
This reintroduces the problem as higher-order terms involving those flavons and matter fields lead to additional terms \cite{Leurer:1992wg,Leurer:1993gy,Dudas:1995yu,Chen:2012ha,Chen:2013aya}. 
In \cite{Baur:2022hma} these terms and the corresponding additional parameters have been used to obtain realistic fermion masses. 
In \cite{Chen:2021prl} the extra off-diagonal terms in the K\"ahler potential have been suppressed by adding a traditional flavor symmetry and breaking it to a diagonal subgroup. 
This is at the expense of introducing an additional flavon. 

On the other hand, our observation here only relies on the $\Group{U}{1}$ symmetries that become exact at the critical points. 
The breaking of these symmetries is parametrized by the departure of $\varrho$ from $\varrho_\mathrm{crit}$, and no additional flavons are required. 
All the ingredients for this mechanism are local in the sense that we do not rely on features stemming from the other critical points. 
Therefore, even if we switch on a Wilson line in the $\mathds{Z}_3$ plane, thus breaking $\Delta(27)$, we expect these statements to continue holding. 

Altogether we see that the \Group{U}{1} symmetries greatly enhance the predictive power of near-critical analyses. 
The same hierarchy that appears in the expansion of the modular forms around $\varrho_\mathrm{crit}$ also controls extra terms in the K\"ahler metrics. 
These facts are expected to increase the precision in models in which the modulus is parametrically close to one of the critical points. 

Notice also that the near-critical analysis allows us to parametrize $\mathcal{CP}$ violation. 
Define $u\defeq\varrho-\varrho_{\mathcal{CP}}$ as the departure of $\varrho$ from some $\mathcal{CP}$ conserving point $\varrho_{\mathcal{CP}}$. 
Then $u$ is an order parameter for not only the amount of $\mathcal{CP}$ breaking in superpotential couplings but also the mass splitting of the doublets in diagrams like the one shown in \Cref{fig:CP_loop}. 
In the latter, the phases are predicted by group theory. 

Let us also note that the appearance of $\Group{U}{1}\times\Group{U}{1}$ gauge bosons is a firm prediction of the scheme. 
That is, if a fit of fermion masses fixes $\varrho$ very close to a critical point, in the \ac{TD} scheme this comes with an explicit prediction for possibly not too heavy gauge bosons along with the $\Group{U}{1}\times\Group{U}{1}$ charges of the matter fields. 
If $\varrho$ is sufficiently close to one of the critical points, in principle it is conceivable that the model could be tested by searching for these gauge bosons directly or indirectly.

\section{Summary}
\label{sec:Summary}

In this work, we have studied the stringy origin of modular flavor symmetries. 
As we have seen, a detailed discussion of the massive states is instrumental for a precise understanding of how congruence subgroups of $\Group{SL}{2,\mathds{Z}}$ come about, and what they mean. 
The massive states arrange themselves in towers carrying different quantum numbers of the residual symmetries of the model. 
We have defined orbifold-invariant winding and \ac{KK} numbers that distinguish these towers. 
There is a direct correspondence between these numbers and $\Delta(54)$ quantum numbers. 
At six different critical points in $\varrho$-moduli space, a member of a given set of towers becomes massless while the members of the other towers are massive with mass relations reflecting increased discrete symmetries.  
Modular transformations relate these towers, and therefore act as outer automorphisms on the corresponding symmetries. 

It is worthwhile to stress that we cannot restrict $\varrho$ to the fundamental domain of $\Group{SL}{2,\mathds{Z}}_\varrho$ to describe the model.
Rather, we need the full fundamental domain of $\Gamma(3)$ to have a complete description of the $\varrho$-moduli space. 
The fact that different critical points in the fundamental domain of $\Gamma(3)$ are distinguished by the quantum numbers of the states that get massless at these points is crucial for the interpretation of congruence subgroups.

The massive states are also key in showing that the traditional flavor symmetries are gauged. 
This is because, as we have seen, the different subgroups of the traditional flavor symmetry $\Delta(27)$ get enhanced to different, relatively misaligned $\Group{U}{1}\times\Group{U}{1}$ symmetries at different critical points in the fundamental domain of $\Gamma(3)$. 
These gauge fields of the $\Group{U}{1}\times\Group{U}{1}$ symmetries, as well as the scalar fields required to break the $\Group{U}{1}\times\Group{U}{1}$ symmetries to $\mathds{Z}_3\times\mathds{Z}_3$, are members of different towers. 
At finite $\varrho$, at most one such combination of states becomes massless. 
We have argued that these symmetries constrain the K\"ahler potential without the need of introducing, or using, additional flavons. 
The detailed analysis of this question is left for future work. 
As we have mentioned, these gauge fields are a firm prediction of the scheme, and can possibly be searched for at future experiments. 
That is, the \ac{TD} scheme of modular flavor symmetries makes explicit, and in principle testable, predictions.  

Away from the critical points the discrete $\mathds{Z}_3\times\mathds{Z}_3$ remnants of the misaligned $\Group{U}{1}\times\Group{U}{1}$ symmetries combine to $\Delta(27)=(\mathds{Z}_3\times\mathds{Z}_3)\rtimes\mathds{Z}_3$. 
Two of these $\mathds{Z}_3$ symmetries correspond to the so-called point- and space-group string selection rules.  
By deriving the $\mathds{Z}_3$ factors from $\Group{U}{1}\times\Group{U}{1}$ we were able to show explicitly that these selection rules derive from continuous gauge symmetries. 
$\Delta(27)$ gets enhanced to $\Delta(54)$ by the $\mathsf{S}_\varrho^2$ transformation, which, despite being modular, leaves $\varrho$ invariant. 
The generators of the misaligned $\Group{U}{1}\times\Group{U}{1}$ symmetries provide us with \ac{CG} coefficients of $\Delta(54)$ governing the couplings of the twisted states to the various $\Group{U}{1}\times\Group{U}{1}$ gauge fields. 
At generic points in $\varrho$-moduli space these \ac{CG} coefficients lead to $\mathcal{CP}$ violation. 

\begin{wrapfigure}[9]{r}{9cm}
  \centering\includegraphics{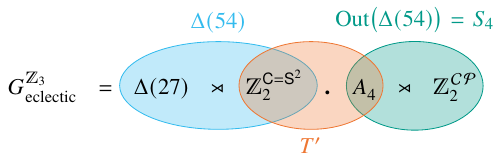}%
   \caption{Eclectic symmetry of the $\mathds{Z}_3$ orbifold plane.}
   \label{fig:eclectic_Venn}
\end{wrapfigure}

The ``full'', or, more precisely, extended (finite) modular symmetry includes the $\mathcal{CP}$ transformation, which is also referred to as a modular symmetry.
In slightly more detail, $\varrho$ is invariant under $\Delta(54)$. 
On the other hand, generic values of $\varrho$ are not invariant under the maximal outer automorphism group of $\Delta(54)$, $S_4=A_4\rtimes\mathds{Z}_2^{\mathcal{CP}}$. 
This $A_4$ is the intersection of $S_4$ with the finite modular group $T'$, cf.\ \Cref{fig:eclectic_Venn}. 
$T'$ can be thought of as the non-split extension of $A_4$. 
In addition to these symmetries there are linearly realized $R$-symmetries associated with the stabilizer group of the complex structure modulus $\tau$ at $\omega$.

\begin{wrapfigure}[23]{r}{7cm}
  \centering\includegraphics{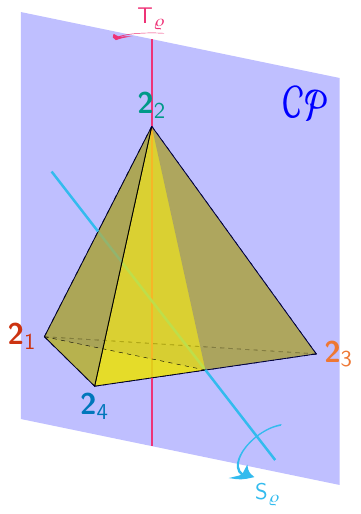}
  \caption{Extended modular $S_4$.}
  \label{fig:Full_modular_S_4}
\end{wrapfigure}

\mbox{}The extended finite modular group $S_4$ acts on $\Delta(54)$ as an outer automorphism. 
It is noteworthy that it acts on the four doublets of $\Delta(54)$ via the most common $S_4$ permutation, as demonstrated by \Cref{eq:2-plet-S}, \Cref{eq:2-plet-T} and \Cref{eq:2-plet-CP}. 
Since $S_4$ can be regarded as the full symmetry of the tetrahedron, we visualize its permutation on the four gauge bosons doublets in \Cref{fig:Full_modular_S_4}.  
This symmetry includes, apart from the tetrahedral symmetry $A_4$, also the reflection at a ``$\mathcal{CP}$ plane''. 
As usual, the $\mathsf{S}_\varrho$ transformation can be visualized as a $180^\circ$ rotation about an axis that runs through the centers of two non-adjacent edges. 
$\mathsf{T}_\varrho$ acts as a $120^\circ$ rotation about an axis that goes through one vertex and the center of the opposite face. 
This picture clearly illustrates that certain modular transformations are maps between states with different quantum numbers. 
At special locations in $\varrho$-moduli space portions of this $S_4$ are linearly realized. 
This includes the $\mathcal{CP}$-conserving curves shown in \Cref{fig:CP_curves}. 
Away from these curves, $\mathcal{CP}$ is broken both by the \ac{VEV} of $\varrho$ and the \ac{CG} coefficients of $\Delta(54)$. 
That is, even if one drops all $\varrho$-dependent interactions in the superpotential, $\mathcal{CP}$ is still broken by the relatively misaligned gauge interactions with the heavy $\Delta(54)$ doublets.  
Along the $\mathcal{CP}$-conserving curves, symmetries become linearly realized, and doublet masses become degenerate thus eliminating this kind of $\mathcal{CP}$ violation. 

As we have seen, the properties of the massive towers play an important role in the understanding of modular flavor symmetries. 
It will be interesting to explore their role in other applications.

\subsection*{Acknowledgments}

We would like to thank Mu-Chun Chen, V.~Knapp-Perez, Arvind Rajaraman, and Saul Ramos-Sanchez for useful discussions. 
XQL, XGL, MR and AS are supported by U.S.\ National Science Foundation under Grant No.\ PHY-2210283. AS is supported by the National Science Foundation Graduate Research Fellowship Program. This material is based upon work supported by the National Science Foundation Graduate Research Fellowship Program under Grant No.\  DGE-2235784. Any opinions, findings, and conclusions or recommendations expressed in this material are those of the authors and do not necessarily reflect the views of the National Science Foundation.

\appendix
\section{Modular transformations} 
\label{app:Modular_transformations}

A heterotic string state is characterized by its oscillator numbers, $N_\mathrm{L}$ and $N_\mathrm{R}$, $\mathrm{E}_8\times\mathrm{E}_8$ (or $\Group{Spin}{32}/\mathds{Z}_2$) momenta $\pi$, and \ac{KK} and winding numbers, $n_i$ and $w^i$ which all determine the state's left- and right-moving momenta, $p_\mathrm{L}$ and $p_\mathrm{R}$.
In this appendix, we discuss how $n_i$ and $w^i$ behave under target-space modular transformations.

In the absence of Wilson lines and momenta on the $\mathrm{E}_8 \times \mathrm{E}_8$ lattice, the mass equation and level-matching condition for the heterotic string compactified on $\mathds{T}^2$ are given by~\cite{Fraiman:2018ebo}
\begin{align}
  m^2 &= p_\mathrm{L}^2 + p_\mathrm{R}^2 + 2\left(N_\mathrm{L} + N_\mathrm{R} - \begin{cases} 1 & (\text{R})\\ \frac{3}{2} & (\text{NS}) \end{cases}\right)
  \;,\label{eq:mass_TD}\\
  0 &= p_\mathrm{L}^2 - p_\mathrm{R}^2 + 2\left( N_\mathrm{L} - N_\mathrm{R} - \begin{cases} 1 & (\text{R}) \\ \frac{1}{2} & (\text{NS}) \end{cases} \right)
  \;,\label{eq:level_matching}
\end{align}
respectively. 
The difference of these two equations yields a particularly simple form for the mass squared,
\begin{equation}\label{app:mSq_subtract}
  m^2 = 2 p_\mathrm{R}^2 + 2\left( 2N_\mathrm{R} - \begin{cases} 0 & (\text{R}) \\ 1 & (\text{NS}) \end{cases}  \right)\;.
\end{equation}
Here,
\begin{subequations}
\begin{align}
  p^2_\mathrm{L}&=\frac{\alpha '}{2\im\tau\,\im\varrho}\left|n_2-n_1\,\tau-\varrho\,(w^1+w^2\,\tau)\right|^2 \;,\label{eq:p_L_squared} \\
  p^2_\mathrm{R}&=\frac{\alpha '}{2\im\tau\,\im\varrho}\left|n_2-n_1\,\tau-\overline{\varrho}\,(w^1+w^2\,\tau)\right|^2\;,\label{eq:p_R_squared}
\end{align}  
\end{subequations}
where $\varrho$ is the K\"ahler modulus and $\tau$ is the complex structure modulus.
In what follows we will set $\alpha' = 1$. 
The components of the left- and right-moving momenta are explicitly given in terms of the \ac{KK} and winding numbers as well as the background fields as
\begin{subequations}
\begin{align}
  p_{\mathrm{L},a} &=  \sqrt{\frac{1}{2}}\widehat{e}^m_a\,\left[ n_m - (G_{mn} + B_{mn})\,w^n \right]\;, \label{eq:pL_formula}\\
  p_{\mathrm{R},a} &=  \sqrt{\frac{1}{2}}\widehat{e}^m_a\,\left[ n_m + (G_{mn} - B_{mn})\,w^n \right]\;, \label{eq:pR_formula}
\end{align}
\end{subequations}
where $G_{mn}$ is the background metric, $B_{mn}$ is the background Kalb--Ramond field, and $\widehat{e}^m_a$ is the torus dual vielbien (i.e.\ $\widehat{e}^{\transpose} \widehat{e} = G^{-1}$, cf.\ e.g.\ \cite{Fraiman:2018ebo}). 
In the $\mathds{T}^2$ compactification, one can write these explicitly in terms of the K\"ahler modulus $\varrho = \varrho_1 + \ii \varrho_2$ and complex structure $\tau = \tau_1 + \ii \tau_2$ as
\begin{subequations}
\begin{align}
  G_{mn} &= \frac{\varrho_2}{\tau_2}\begin{pmatrix} 1 & \tau_1 \\ \tau_1 & |\tau|^2 \end{pmatrix}\;,\\
  B_{mn} &= \begin{pmatrix} 0 & \varrho_1 \\ -\varrho_1 & 0 \end{pmatrix}\;,\\
  \widehat{e}_a^m &= \sqrt{\frac{\tau_2}{\varrho_2}}\begin{pmatrix} -1 & 0 \\ \frac{\tau_1}{\tau_2} & -\frac{1}{\tau_2} \end{pmatrix}\;.
\end{align}
\end{subequations}
Of course, the choice of the vielbein and metric is not strictly unique due to the symmetries of the metric.
Using these results, the mass equation \Cref{app:mSq_subtract} can be written as
\begin{equation}\label{app:mSq_simp}
  m^2 = \frac{1}{\im\tau\,\im\varrho}\left| n_2 - n_1\, \tau - \overline{\varrho}\, (w^1 + w^2\, \tau) \right|^2 + 2\left(2 N_\mathrm{R} - \begin{cases} 0 & (\text{R}) \\ 1 & (\text{NS}) \end{cases} \right)\,.
\end{equation}
Likewise, we can write the level matching equation explicitly as
\begin{equation}\label{app:levelMatchExplicit}
  0 = 2(n_1\, w^1 + n_2\, w^2)  + 2\left(N_\mathrm{L} - N_\mathrm{R} - \begin{cases} 1 & (\text{R}) \\ \frac{1}{2} & (\text{NS}) \end{cases} \right)\,.
\end{equation}

Modular transformations leave the mass equation invariant. 
Additionally they will be full symmetries of the underlying \ac{CFT}, which we will investigate in \Cref{app:OPEs,app:vertex_op_SL2Z_trans}.
We subject the moduli $\tau$ and $\varrho$ to the modular transformations 
\begin{subequations}
\begin{empheq}[left={(\varrho,\tau)\empheqlbrace}]{align}
  &\xmapsto{~\mathsf{S}_\tau~}\bigl(\varrho,-1/\tau\bigr)\;,\\   
  &\xmapsto{~\mathsf{T}_\tau~}(\varrho,\tau+1)\;,\\   
  &\xmapsto{~\mathsf{S}_\varrho~}\bigl(-1/\varrho,\tau\bigr)\;,\\   
  &\xmapsto{~\mathsf{T}_\varrho~}(\varrho+1,\tau)\;.
\end{empheq}  
\end{subequations}
We apply these transformations to the mass squared, \Cref{app:mSq_simp}, dropping the oscillator contribution which is not relevant for the transformations,
\begin{subequations}
  \begin{empheq}[left={ \dfrac{\left| n_2 - n_1 \tau -\overline{\varrho} (w^1 + w^2 \tau) \right|^2}{\im\tau\im\varrho}\empheqlbrace}]{align}
    &\xmapsto{~\mathsf{S}_\tau~}\frac{1}{\im\tau\im\varrho}\left|n_{1}+n_{2}\, \tau -\overline{\varrho}\, (-w^2 + w^1 \, \tau)\right|^2 \;,\\   
    &\xmapsto{~\mathsf{T}_\tau~}\frac{1}{\im\tau\im\varrho}\left|n_{2}-n_{1}-n_{1}\, \tau - \overline{\varrho}\, (w^1 + w^2 + w^2 \, \tau)\right|^2 \;,\\   
    &\xmapsto{~\mathsf{S}_\varrho~}\frac{1}{\im\tau\im\varrho}\left|-w^1-w^2 \, \tau -\overline{\varrho}\, (n_2 - n_1 \, \tau)\right|^2 \;,\\   
    &\xmapsto{~\mathsf{T}_\varrho~}\frac{1}{\im\tau\im\varrho}\left|n_{2}-w^1 - (n_{1}+w^2) \, \tau -\overline{\varrho}\, (w^1 + w^2 \, \tau)\right|^2  \;.
\end{empheq}  
\end{subequations}
Next, we extract coefficients of $1$, $\tau$, $\varrho$, and $\varrho\,\tau$ and demand that they remain invariant under the transformation on $\tau$ or $\varrho$. 
This yields the transformation rules for the momentum and winding numbers
\begin{subequations}
  \label{eq:nwModularTrans}
\begin{empheq}[left={\bigl(w^1,w^2,n_1,n_2\bigr)\empheqlbrace}]{align}
    &\xmapsto{~\mathsf{S}_\tau~}\bigl(-w^2, w^1, -n_2, n_1\bigr)\;,\\
    &\xmapsto{~\mathsf{T}_\tau~}\bigl(w^1-w^2, w^2, n_1, n_2+n_1\bigr)\;,\\
  &\xmapsto{~\mathsf{S}_\varrho~}\bigl(-n_2, n_1, -w^2, w^1\bigr)\;,\\
  &\xmapsto{~\mathsf{T}_\varrho~}\bigl(w^1, w^2, n_1-w^2, n_2+w^1\bigr)\;.
\end{empheq}
\end{subequations} 
The transformations of the momentum and winding numbers are unique up to an overall minus sign since the mass squared is quadratic in all of the momentum and winding numbers. This overall minus sign determines the presentation of \Group{SL}{2,\mathds{Z}}. 
We have chosen the overall sign such that we are using the presentation $(\mathsf{S\,T})^3 = -\mathsf{S}^2 = \mathds{1}_4$ for both $\Group{SL}{2,\mathds{Z}}_\tau$ and $\Group{SL}{2,\mathds{Z}}_\varrho$.

These linear transformations are described by the matrices \eqref{eq:STmatrices}.
The transformation matrices $(\mathsf{S}\, \mathsf{T})_\tau$ and $(\mathsf{S}\, \mathsf{T})_\varrho$ are given by
\begin{subequations}\label{app:ST_tau_mat}
\begin{align}
(\mathsf{S}\,\mathsf{T})_\tau&=\begin{pmatrix*}[r]
0 & -1 & \phantom{-}0 & \phantom{-}0 \\
 1 & -1 & 0 & 0 \\
 0 & 0 & -1 & -1 \\
 0 & 0 & 1 & 0 
\end{pmatrix*}\;, \\
(\mathsf{S}\,\mathsf{T})_\varrho&=\begin{pmatrix*}[r]
 -1 & \phantom{-}0 & \phantom{-}0 & -1 \\
 0 & -1 & 1 & 0 \\
0 & -1 & 0 & 0 \\
 1 & 0 & 0 & 0 \\
\end{pmatrix*}
 \;.
\end{align}  
\end{subequations}
In addition to \Cref{eq:S_T_relations_torus}, the matrices \eqref{eq:STmatrices} satisfy
\begin{subequations}\label{eq:additional_S_T_relations_torus}
\begin{align}  
(\mathsf{S}^2\, \mathsf{T})_\tau &=(\mathsf{T}\, \mathsf{S}^2)_\tau\;,& (\mathsf{S}^2 \,\mathsf{T})_\varrho &=(\mathsf{T}\, \mathsf{S}^2)_\varrho\;, &
\mathsf{S}_\tau\, \mathsf{S}_\varrho &=\mathsf{S}_\varrho\, \mathsf{S}_\tau\;,\\
\mathsf{T}_\tau\, \mathsf{T}_\varrho &=\mathsf{T}_\varrho\, \mathsf{T}_\tau\;,&\mathsf{S}_\tau\, \mathsf{T}_\varrho &=\mathsf{T}_\varrho\, \mathsf{S}_\tau\;,&\mathsf{T}_\tau\, \mathsf{S}_\varrho &=\mathsf{S}_\varrho\, \mathsf{T}_\tau\;.
\end{align}
\end{subequations}
That is, the transformations \eqref{eq:STmatrices} provide us with a representation of the T-duality group $\Group{SL}{2,\mathds{Z}}_{\tau}\times \Group{SL}{2,\mathds{Z}}_{\varrho}$, as expected.
If we further consider the mirror symmetry $\mathds{Z}^{M}_{2}$ and $\mathcal{CP}$ symmetry $\mathds{Z}^{\mathcal{CP}}_{2}$, then we get the complete T-duality group $\bigl(\Group{SL}{2,\mathds{Z}}_{\tau}\times \Group{SL}{2,\mathds{Z}}_{\varrho}\bigr)\rtimes\bigl(\mathds{Z}^{M}_{2}\times\mathds{Z}^{\mathcal{CP}}_{2}\bigr)\cong \Group{O}{2,2;\mathds{Z}}$.

At generic points in moduli space, $\mathsf{S}_\tau^2$ and $\mathsf{S}_\varrho^2$ leave the moduli invariant. 
However, these transformations do not leave the \ac{KK} and winding numbers unchanged, rather 
\begin{equation}\label{eq:S_squared_KK_winding}
  \mathsf{S}_\tau^2\,\begin{pmatrix}
    w^1\\ w^2\\ n_1\\ n_2  
  \end{pmatrix}
  =
  \mathsf{S}_\varrho^2\,\begin{pmatrix}
    w^1\\ w^2\\ n_1\\ n_2  
    \end{pmatrix}
  =\begin{pmatrix}
    -w^1\\ -w^2\\ -n_1\\ -n_2  
    \end{pmatrix}\;.
\end{equation}
Orbifolding fixes $\tau=\omega\defeq\ee^{2\pi\ii/3}$ which is invariant under $(\mathsf{S}\,\mathsf{T})_\tau$, an order 3 transformation in $\Group{SL}{2,\mathds{Z}}_\tau$ with our presentation. Together with $\mathsf{S}^2$, we can form the order 6 transformation $(\mathsf{S}^3\,\mathsf{T})_\tau$ which acts as an $R$-symmetry. 
Notice that, while the representation matrix of the $(\mathsf{S}^3\,\mathsf{T})_\tau$ transformation has order 6, it actually has order 18 when acting on twisted states as the modular weights are fractional with denominator 3.
Likewise, the special point $\varrho=\omega$ is invariant under $(\mathsf{S}^3\, \mathsf{T})_\varrho$, which is again an order 6 transformation in $\Group{SL}{2,\mathds{Z}}_\varrho$ (it is order 3 on the moduli which see $\Group{PSL}{2,\mathds{Z}}$). 
The actions of these combined transformations on the \ac{KK} and winding numbers are 
\begin{subequations}
  \begin{empheq}[left={\bigl(w^1,w^2,n_1,n_2\bigr)\empheqlbrace}]{align}
    &\xmapsto{~(\mathsf{S}^3\,\mathsf{T})_\tau~}\bigl(w^2,-w^1+w^2,n_1+n_2,-n_1\bigr)\;,\\
    &\xmapsto{~(\mathsf{S}^3\,\mathsf{T})_\varrho~}\bigl(w^1+n_2,w^2-n_1,w^2,-w^1\bigr)\;,\\
    &\xmapsto{~(\mathsf{S}^3\, \mathsf{T})_\tau^6~}\bigl(w^1,w^2,n_1,n_2\bigr)\;,\\
    &\xmapsto{~(\mathsf{S}^3\,\mathsf{T})_\varrho^6~}\bigl(w^1,w^2,n_1,n_2\bigr)\;.
\end{empheq}
\end{subequations}
Evidently, both $(\mathsf{S}^3\,\mathsf{T})_\tau$ and $(\mathsf{S}^3\, \mathsf{T})_\varrho$ are order 6 transformations when acting on the \ac{KK} and winding numbers. 
However, their action on twisted states has a higher order because of the factional modular weights of the latter. 
Notice also that, while an additional $R$-symmetry becomes linearly realized if $\varrho$ takes a critical value such as $\omega$ with nontrivial stabilizer symmetry, it is possible to rewrite the product of two $R$-symmetries as one $R$-symmetry and a non-$R$-symmetry (see e.g.\ \cite{Chen:2014gua}).
Discussing this in detail is beyond the scope of this work.

\section{Some group theory}

\subsection{\texorpdfstring{$\Delta(54)$}{Delta(54)}}
\label{sec:Delta(54)}

The finite group $\Delta(54)$ has the presentation 
\begin{equation}\label{eq:presenation_Delta(54)}
  \Delta(54)=\Braket{\mathsf{A},\mathsf{B},\mathsf{C} | \mathsf{A}^3=\mathsf{B}^3=\mathsf{C}^2=(\mathsf{A}\,\mathsf{B})^3=(\mathsf{A}\,\mathsf{B}^2)^3=(\mathsf{A}\,\mathsf{C})^2=(\mathsf{B}\,\mathsf{C})^2=\mathds{1}}\;.
\end{equation}
There are two inequivalent 3-dimensional complex representation, $\rep{3}_1$ and $\rep{3}_2$. 
The other two 3-dimensional representations are their complex conjugates, namely $\overline{\rep{3}}_1$ and $\overline{\rep{3}}_2$. 
The relevant representation matrices of $\rep{3}_1$ and $\overline{\rep{3}}_1$ are shown in \Cref{tab:Delta(54)_representation_matrices}.
Further, $\Delta(54)$ has two 1-dimensional representations, $\rep{1}$ and $\rep{1'}$, as well as four doublets. 
The $\Delta(54)$ doublets can be obtained by contracting a $\crep{3}_2$ with $\rep{3}_2$,
\begin{align}
  \overline{\Phi}_{\crep{3}_2}\otimes\Phi_{\rep{3}_2}
  &=
  \begin{pmatrix}
    \overline{X}\\ \overline{Y}\\ \overline{Z}
  \end{pmatrix}_{\crep{3}_2}  
  \otimes 
  \begin{pmatrix}
    X\\ Y\\ Z
  \end{pmatrix}_{\rep{3}_2}
  \\
  &=\bigl(\overline{\Phi}_{\crep{3}_2}\otimes\Phi_{\rep{3}_2}\bigr)_{\rep{1}}
  +\bigl(\overline{\Phi}_{\crep{3}_2}\otimes\Phi_{\rep{3}_2}\bigr)_{\rep{2}_1}
  +\bigl(\overline{\Phi}_{\crep{3}_2}\otimes\Phi_{\rep{3}_2}\bigr)_{\rep{2}_2}
  +\bigl(\overline{\Phi}_{\crep{3}_2}\otimes\Phi_{\rep{3}_2}\bigr)_{\rep{2}_3}
  +\bigl(\overline{\Phi}_{\crep{3}_2}\otimes\Phi_{\rep{3}_2}\bigr)_{\rep{2}_4}\;,\notag
\end{align}
where
\begin{subequations}\label{eq:CG_doublets}
\begin{align}
  \bigl(\overline{\Phi}_{\crep{3}_2}\otimes\Phi_{\rep{3}_2}\bigr)_{\rep{2}_1}
  &=\frac{1}{\sqrt{3}}\begin{pmatrix}
    \overline{X}\, Z+\overline{Y}\, X+\overline{Z}\, Y \\
    \overline{Z}\, X+\overline{X}\, Y+\overline{Y}\, Z
  \end{pmatrix}\;,\label{eq:CG_2_1}\\    
  \bigl(\overline{\Phi}_{\crep{3}_2}\otimes\Phi_{\rep{3}_2}\bigr)_{\rep{2}_2}
  &=\frac{1}{\sqrt{3}}\begin{pmatrix}
    \overline{X}\, X+\omega\,\overline{Y}\, Y+\omega^2\,\overline{Z}\, Z\\
      \overline{X}\, X+\omega^2\,\overline{Y}\, Y+\omega\,\overline{Z}\, Z
    \end{pmatrix}\;,\label{eq:CG_2_2}\\  
  \bigl(\overline{\Phi}_{\crep{3}_2}\otimes\Phi_{\rep{3}_2}\bigr)_{\rep{2}_3}
  &=\frac{1}{\sqrt{3}}\begin{pmatrix}
    \omega\,\overline{X}\,Z+\omega^2\,\overline{Y}\,X+\overline{Z}\,Y\\
    \omega^2\,\overline{Z}\,X+\omega\,\overline{X}\,Y+\overline{Y}\,Z
  \end{pmatrix}\;,\label{eq:CG_2_3}\\
  \bigl(\overline{\Phi}_{\crep{3}_2}\otimes\Phi_{\rep{3}_2}\bigr)_{\rep{2}_4}
  &=\frac{1}{\sqrt{3}}\begin{pmatrix}
    \omega ^2\,\overline{X}\,Y+\overline{Y}\,Z+\omega\,\overline{Z}\, X\\
    \omega\,\overline{Y}\, X+\overline{Z}\,Y+\omega ^2\,\overline{X}\,Z
  \end{pmatrix}\;.\label{eq:CG_2_4}
\end{align}   
\end{subequations}

\begin{table}[t]
\centering\begin{tabular}{ccccccc}
    \toprule
    irrep & components & $\mathsf{A}$ & $\mathsf{B}$ & $\mathsf{C}$\\ 
    \midrule
    $\rep{3}_2$ & $\begin{pmatrix} X \\ Y \\ Z \end{pmatrix}$ & $\begin{pmatrix} 0 & 1 & 0 \\ 0 & 0 & 1 \\ 1 & 0 & 0 \end{pmatrix}$ & $\begin{pmatrix} 1 & 0 & 0 \\ 0 & \omega & 0 \\ 0 & 0 & \omega^2 \end{pmatrix}$ & $\begin{pmatrix*}[r] -1 & 0 & 0 \\ 0 & 0 & -1 \\ 0 & -1 & 0\end{pmatrix*}$ \\
    $\overline{\rep{3}}_2$ & $\begin{pmatrix} \Xb \\ \Yb \\ \Zb \end{pmatrix}$ & $\begin{pmatrix} 0 & 1 & 0 \\ 0 & 0 & 1 \\ 1 & 0 & 0 \end{pmatrix}$ & $\begin{pmatrix} 1 & 0 & 0 \\ 0 & \omega^2 & 0 \\ 0 & 0 & \omega \end{pmatrix}$ & $\begin{pmatrix*}[r] -1 & 0 & 0 \\ 0 & 0 & -1 \\ 0 & -1 & 0\end{pmatrix*}$ \\
    \midrule 
    $\rep{2}_1$  & $\begin{pmatrix} W^+_{(\omega+2)}\\ W^-_{(\omega+2)} \end{pmatrix}$ & $\begin{pmatrix}  1 & 0 \\ 0 & 1\end{pmatrix}$ & $\begin{pmatrix} \omega^2 & 0 \\ 0 & \omega \end{pmatrix}$ & $\begin{pmatrix} 0 & 1 \\ 1 & 0 \end{pmatrix}$ \\
    $\rep{2}_2$  & $\begin{pmatrix} W^+_{(\nu)}\\ W^-_{(\nu)} \end{pmatrix}$ & $\begin{pmatrix}  \omega^2 & 0 \\ 0 & \omega \end{pmatrix}$ & $\begin{pmatrix} 1 & 0 \\ 0 & 1 \end{pmatrix}$ & $\begin{pmatrix} 0 & 1 \\ 1 & 0 \end{pmatrix}$ \\
    $\rep{2}_3$  & $\begin{pmatrix} W^+_{(\omega+1)}\\ W^-_{(\omega+1)} \end{pmatrix}$ & $\begin{pmatrix}  \omega^2 & 0 \\ 0 & \omega \end{pmatrix}$ & $\begin{pmatrix} \omega^2 & 0 \\ 0 & \omega \end{pmatrix}$ & $\begin{pmatrix} 0 & 1 \\ 1 & 0 \end{pmatrix}$ \\
    $\rep{2}_4$  & $\begin{pmatrix} W^+_{(\omega)}\\ W^-_{(\omega)} \end{pmatrix}$ & $\begin{pmatrix}  \omega^2 & 0 \\ 0 & \omega \end{pmatrix}$ & $\begin{pmatrix} \omega & 0 \\ 0 & \omega^2 \end{pmatrix}$ & $\begin{pmatrix} 0 & 1 \\ 1 & 0 \end{pmatrix}$\\
    \bottomrule
  \end{tabular}
  \caption{Various \Group{\Delta}{54} representation matrices.}
  \label{tab:Delta(54)_representation_matrices}
\end{table}

The group structure of $\Delta(54)$ is $\left[\mathds{Z}_3\times \mathds{Z}_3\right]\rtimes \left[\mathds{Z}_3\rtimes \mathds{Z}_2\right]$. 
However, the $\mathds{Z}_3$ subgroups among them are not unique.  
Our discussion in \Cref{sec:MasslesWinding} illustrates this point.
The group structure $\Delta(54)$ has the following different implementations according to the different enhanced gauge symmetry $\left[\Group{U}{1}\times\Group{U}{1}\right] \rtimes \left[\mathds{Z}_3\rtimes\mathds{Z}_2\right]$ it inherits at various critical points:
\begin{subequations}\label{eq:Z3Z3Z3Z2}
\begin{align}
\varrho=\omega:~\Delta(54)&=\left[\mathds{Z}^{\mathsf{A}^2\,\mathsf{B}^2\,\mathsf{A}\,\mathsf{B}}_3\times \mathds{Z}^{\mathsf{B}^2\,\mathsf{A}\,\mathsf{B}^2}_3\right]\rtimes \left[\mathds{Z}^{\mathsf{A}}_3\rtimes \mathds{Z}^{\mathsf{C}}_2\right] \;,\\ 
\varrho=\omega+1:~\Delta(54)&=\left[\mathds{Z}^{\mathsf{A}^2\,\mathsf{B}^2\,\mathsf{A}\,\mathsf{B}}_3\times \mathds{Z}^{\mathsf{B}\,\mathsf{A}\,\mathsf{B}}_3\right]\rtimes \left[\mathds{Z}^{\mathsf{B}^2\,\mathsf{A}\,\mathsf{B}^2}_3\rtimes \mathds{Z}^{\mathsf{C}}_2\right] \label{eq:B.4c}\;,\\ 
\varrho=\omega+2:~\Delta(54)&=\left[\mathds{Z}^{\mathsf{A}^2\,\mathsf{B}^2\,\mathsf{A}\,\mathsf{B}}_3\times \mathds{Z}^{\mathsf{A}}_3\right]\rtimes \left[\mathds{Z}^{\mathsf{B}\,\mathsf{A}\,\mathsf{B}}_3\rtimes \mathds{Z}^{\mathsf{C}}_2\right] \;,\\ 
\varrho=\nu:~\Delta(54)&=\left[\mathds{Z}^{\mathsf{A}^2\,\mathsf{B}^2\,\mathsf{A}\,\mathsf{B}}_3\times \mathds{Z}^{\mathsf{B}}_3\right]\rtimes \left[\mathds{Z}^{\mathsf{B}^2\,\mathsf{A}\,\mathsf{B}^2}_3\rtimes \mathds{Z}^{\mathsf{C}}_2\right] \;,\\
\varrho=\nu+1:~\Delta(54)&=\left[\mathds{Z}^{\mathsf{A}^2\,\mathsf{B}^2\,\mathsf{A}\,\mathsf{B}}_3\times \mathds{Z}^{\mathsf{B}}_3\right]\rtimes \left[\mathds{Z}^{\mathsf{B}\,\mathsf{A}\,\mathsf{B}}_3\rtimes \mathds{Z}^{\mathsf{C}}_2\right] \;,\\ 
\varrho=\nu+2:~\Delta(54)&=\left[\mathds{Z}^{\mathsf{A}^2\,\mathsf{B}^2\,\mathsf{A}\,\mathsf{B}}_3\times \mathds{Z}^{\mathsf{B}}_3\right]\rtimes \left[\mathds{Z}^{\mathsf{A}}_3\rtimes \mathds{Z}^{\mathsf{C}}_2\right] \;.
\end{align}
\end{subequations}
Note that \Cref{eq:B.4c} has already appeared in \Cref{eq:Delta54atOmgea+1}. 
We see that different $\mathds{Z}_3$ subgroups have indeed undergone swapping or mixing at different points. 
Because these different critical points are linked by modular transformations, modular transformations also act on residual gauge subgroup $\Delta(54)$ as an outer automorphism, resulting in the exchange or mixture of different $\mathds{Z}_3$ factors among them.
One can write down the specific outer automorphism relations among the generators of $T'$ and $\Delta(54)$ through \Cref{eq:rho(S)_and_rho(T)} and \Cref{tab:Delta(54)_representation_matrices},
\begin{subequations}
\begin{align}
\rho(\mathsf{S})\,\rho(\mathsf{A})\, \rho(\mathsf{S})^{-1} &= \rho(\mathsf{B})\;,& \rho(\mathsf{T})\, \rho(\mathsf{A})\, \rho(\mathsf{T})^{-1} &= \rho(\mathsf{B}^2\,\mathsf{A}\,\mathsf{B}^2)\;, \\
\rho(\mathsf{S})\, \rho(\mathsf{B})\, \rho(\mathsf{S})^{-1} &= \rho(\mathsf{A}^2)\;,& \rho(\mathsf{T})\, \rho(\mathsf{A})\, \rho(\mathsf{T})^{-1} &= \rho(\mathsf{B})\;. 
\end{align}
\end{subequations}
Both $\rho(\mathsf{S}) $ and $\rho(\mathsf{T})$ commute with $\rho(\mathsf{C})$ because $\mathsf{C}=\mathsf{S}^2$ is the center of the finite modular group $T'$. 
In other words, this means that truly nontrivial outer automorphisms only act on the normal subgroup $\Delta(27)$ generated by the generators $\mathsf{A}$ and $\mathsf{B}$.
The entire symmetry group can be written as $\Delta(27)\rtimes T'$, and it is actually a part of the so-called eclectic flavor group.

On the other hand, for the massless gauge bosons at each critical point, one can verify that they only undergo nontrivial transformations under the subgroup $S_3=\mathds{Z}_3\rtimes\mathds{Z}_2\in \Delta(54)$ in the right parentheses of \Cref{eq:Z3Z3Z3Z2}. 
At each critical point, they remain unchanged under the two $\mathds{Z}_3$ factors in the respective left parentheses of \Cref{eq:Z3Z3Z3Z2}, as they should.
This is consistent with the fact that they are gauge bosons of $\Group{U}{1}\times\Group{U}{1}$. 
Therefore, each of the massless gauge bosons forms a doublet representation of $S_3$, and when combined together, they form the group $\left(\mathds{Z}_3\times\mathds{Z}_3\right)\rtimes \mathds{Z}_2\cong \mathds{Z}_3\rtimes S_3$, which is a subgroup of $\Delta(54)$. 
In other words, the fields in the untwisted sector only see a part of the residual gauge group $\Delta(54)$.

\subsection{Group extensions}
\label{sec:Group_extensions}

Let $G$ and $H$ be groups.
A group $K$ is called an extension of $G$ by $H$ if there exists a short exact sequence of the type 
\begin{equation}
  1\xrightarrow{~\eqmathbox[pi]{}~} G\xrightarrow{~\eqmathbox[pi]{\iota}~} K \xrightarrow{~\eqmathbox[pi]{\pi}~} H\xrightarrow{~\eqmathbox[pi]{}~} 1\;.
\end{equation}
Here, the image of the injective map $\iota$ is equal to the kernel of the surjective map $\pi$,
\begin{equation}
  \operatorname{im}\iota=\ker\pi\cong G\;.
\end{equation} 
This implies that $G$ is a normal subgroup of $K$, and the quotient group $K/G=K/\ker\pi$ is isomorphic to $H$. 
Group extensions provide a framework for constructing larger groups from smaller ones, capturing the interplay between $G$ and $H$ within $K$. 
If there exist a homomorphism 
\begin{equation}
  s\colon H\xrightarrow{~s~}K\text{ such that }\pi\circ s =\operatorname{id}_H\;,  
\end{equation}
then we say the extension is split; otherwise, it is non-split.

In practical terms, a split extension occurs when $K$ contains a subgroup
$Q\cong H$ such that $K = G\,Q$ and $G\cap Q = \{\mathds{1}\}$. 
Here, $Q$ complements $G$, allowing $K$ to be decomposed (or ``split'') into $G$ and $Q$ in a well-defined manner.
Equivalently, $K$ can be written as the semi-direct product $K=G\rtimes H$ where $H$ acts on $G$ via automorphisms.
Conversely, for non-split extensions, there is no subgroup $Q\subseteq K$ isomorphic
to $H$ that complements $G$. In this case, we write $K$ as $K=G\mathbin{\boldsymbol{.}} H$.

The two group $S_4$ and $T'$ appearing in the eclectic group~\eqref{eq:ecl} are respectively examples of the split extension group and the non-split extension group. 
Their group presentations are given by
\begin{subequations}
\begin{align}
S_4&=\Braket{\mathsf{S},\,\mathsf{T},\,\mathsf{CP} ~|~  \mathsf{S}^2=\left(\mathsf{S}\,\mathsf{T}\right)^3=\mathsf{T}^3=\mathsf{CP}^2=\mathds{1}\,,\, (\mathsf{CP})\,\mathsf{S}\,(\mathsf{CP})^{-1}=\mathsf{S}^{-1}\,,(\mathsf{CP})\,\mathsf{T}\,(\mathsf{CP})^{-1}=\mathsf{T}^{-1}}\;, \\
T'&=\Braket{\mathsf{S},\,\mathsf{T}~|~  \mathsf{S}^4=\left(\mathsf{S}\,\mathsf{T}\right)^3=\mathsf{T}^3=\mathsf{S}^2\,\mathsf{T}\,\mathsf{S}^{-2}\,\mathsf{T}^{-1}=\mathds{1}\,}\;.
\end{align}  
\end{subequations}
It is not difficult to find that the generators $\mathsf{S}$ and $\mathsf{T}$ generate the normal subgroup $A_4$ of $S_4$, while $\mathsf{CP}$ generates the $\mathds{Z}^{\mathcal{CP}}_2$ subgroup which does not intersect with $A_4$, and it acts on the $A_4$ by the automorphism. 
The quotient $S_4/A_4\cong \mathds{Z}^{\mathcal{CP}}_2$ distinguishes between even and odd permutations. 
Therefore, $S_4$ here is just a semi-direct product of the $A_4$ subgroup and $\mathds{Z}^{\mathcal{CP}}_2 \colon S_4=A_4\rtimes\mathds{Z}^{\mathcal{CP}}_2$, or equivalently, $S_4$ is the split expansion group of $A_4$ by $\mathds{Z}^{\mathcal{CP}}_2$. 
The corresponding short exact sequence is given by
\begin{equation}
  1\xrightarrow{~\eqmathbox[pi]{}~} A_4\xrightarrow{~\eqmathbox[pi]{\iota}~} S_4 \xrightarrow{~\eqmathbox[pi]{\pi}~} \mathds{Z}^{\mathcal{CP}}_2\xrightarrow{~\eqmathbox[pi]{}~} 1\;.
\end{equation}
On the other hand, in $T'$, we can find that there is no subgroup isomorphic to $A_4$, but only a $\mathds{Z}^{\mathsf{S}^2}_2$ center. 
And the quotient $T'/\mathds{Z}^{\mathsf{S}^2}_2\cong A_4$ gives the usual non-homogeneous finite modular group, which is exactly the same as the normal subgroup $A_4$ in the previous example. 
In this case, $T'$ is just a non-split extension of $\mathds{Z}^{\mathsf{S}^2}_2$ by $A_4$: $T'=\mathds{Z}^{\mathsf{S}^2}_2\mathbin{\boldsymbol{.}} A_4$, the corresponding short exact sequence reads
\begin{equation}
  1\xrightarrow{~\eqmathbox[pi]{}~} \mathds{Z}^{\mathsf{S}^2}_2\xrightarrow{~\eqmathbox[pi]{\iota}~} T' \xrightarrow{~\eqmathbox[pi]{\pi}~}  A_4\xrightarrow{~\eqmathbox[pi]{}~} 1\;.
\end{equation}
However, it is more common to say that $T'$ is the double cover of $A_4$. 
Note also that, although $T'$ is not a semi-direct product of $A_4$ and $\mathds{Z}_2$, it is a semi-direct product of $Q_8$ and $\mathds{Z}_3\colon T'=Q_8\rtimes \mathds{Z}_3$.

\section{Untwisted vertex operators}

\subsection{Vertex operators on \texorpdfstring{$\mathds{T}^2$}{T^2}}
\label{sec:untwisted_vertex_operators_on_torus}

Here we collect the definitions of untwisted vertex operators for states relevant in our discussions. 
These states have $N_\mathrm{R} = 1/2$ and \ac{NS} boundary conditions. 
Define
\begin{subequations}
\begin{align}
  X_{\mathrm{L/R}}^{\pm} &\defeq \frac{1}{\sqrt{2}}\left(X^1_{\mathrm{L/R}}\pm\ii X^2_{\mathrm{L/R}} \right)\;,\\
  p_{\mathrm{L/R}}^{\pm} &\defeq \frac{1}{\sqrt{2}}\left(p^1_{\mathrm{L/R}}\pm\ii p^2_{\mathrm{L/R}} \right)\;.
\end{align}
\end{subequations}
The bosonic oscillator content of a state is described by the vertex operators
\begin{align}
  \ii \overline{\partial} X_\mathrm{L}^\pm &\defeq \frac{1}{\sqrt{2}}\left( \ii \overline{\partial} X_\mathrm{L}^1 \pm \ii (\ii \overline{\partial} X_\mathrm{L}^2) \right)\;,\\
  \ii \partial X_\mathrm{R}^\pm &\defeq \frac{1}{\sqrt{2}}\left( \ii \partial X_\mathrm{R}^1 \pm \ii (\ii \partial X_\mathrm{R}^2) \right)\;.
\end{align}
The bar denotes that the left movers are anti-holomorphic while the right movers are holomorphic with respect to the worldsheet coordinates, however we will not denote this distinction in our notation moving forward. 
A state with momentum $p^{\pm}_{\mathrm{L/R}}$ in the compact dimension is described by exponentiation of these momentum operators as
\begin{equation}
  \exp\left[ \ii \left(p_\mathrm{L} \cdot X_\mathrm{L} + p_\mathrm{R} \cdot X_\mathrm{R} \right)\right]\;,
\end{equation}
where $p \cdot X \defeq p^+ \cdot X^- + p^- \cdot X^+$. 
Given the \ac{KK} numbers $n_i$, winding numbers $w^i$, K\"ahler modulus $\varrho$, and complex structure modulus $\tau$, one can compute the $p_\mathrm{L/R}$ as stated in \Cref{eq:pL_formula,eq:pR_formula}. 
So, provided the moduli have been specified, we can represent the winding and \ac{KK} part of a vertex operator by providing the winding and \ac{KK} numbers $n_i$ and $w^i$. 
On the orbifold, $\tau$ is fixed at $\omega = \exp(2\pi \ii /3)$ so we need to only specify the K\"ahler modulus $\varrho$ and the $n_i$, $w^i$. 
As we are interested in the orbifold, we only consider vertex operators always taken at $\tau = \omega$. 
We will denote this component of the vertex operator as
\begin{equation}
  V(\widehat{N})^{(\varrho)} \defeq \exp\left( p(w^i,n_i,\varrho)\cdot X \right)\;,
\end{equation}
where $p(w^i,n_i,\varrho)$ denotes evaluating the momenta in \Cref{eq:pL_formula,eq:pR_formula} with the given $\widehat{N}=(w^1,w^2,n_1,n_2)$, and $\varrho$. 
Given the $N_\mathrm{L}$ and $N_\mathrm{R}$ oscillator numbers, we can then write down the full vertex operator for the untwisted state. 
The relevant untwisted states in our case have $(N_\mathrm{L},N_\mathrm{R}) = (0,1/2)$. 
The massless gauge bosons have $(N_\mathrm{L},N_\mathrm{R})=(0,1/2)$ and their vertex operators have the form
\begin{equation}
  V(\widehat{N})^{(\varrho)} \otimes (\ii \partial X_\mathrm{R}^\mu)\;.
\end{equation}
In addition, there are complex scalars which arise when the right oscillator is in the compact dimension. 
In this case, two independent components of the complex scalar can be written as
\begin{equation}
  V(\widehat{N})^{(\varrho)} \otimes (\ii \partial X_\mathrm{R}^+)
  \quad\text{and}\quad
  V(\widehat{N})^{(\varrho)} \otimes (\ii \partial X_\mathrm{R}^-)\;.
\end{equation}

\subsection{Untwisted vertex operators on \texorpdfstring{$\mathds{T}^2 / \mathds{Z}_3$}{T^2/Z_3}}
\label{sec:untwisted_vertex_operators_on_orbifold}

To orbifold, we introduce the $\mathds{Z}_3$ twist $\theta$ which acts on the torus coordinates as
\begin{equation}\label{eq:orbifold_action_on_X_pm}
  X^\pm \xmapsto{~\theta~} \omega^{\mp 1}X^\pm\;.
\end{equation}
This acts similarly on oscillators in the compact dimension
\begin{equation}
  (\ii \partial X_{\mathrm{L/R}}^\pm) \xmapsto{~\theta~} \omega^{\mp 1}(\ii \partial X_{\mathrm{L/R}}^\pm)\;,
\end{equation}
while it does not affect oscillators in the non-compact dimensions. 
The exponential piece of the vertex operator is also affected by the orbifold twist. Using $p \cdot X \defeq p^+ \cdot X^- + p^- \cdot X^+$ along with the explicit formulae for the $p_\mathrm{L}$ and $p_\mathrm{R}$, \Cref{eq:pL_formula,eq:pR_formula}, one can verify that the action of the orbifold twist $\theta$ on the exponential pieces of untwisted vertex operators can be represented as
\begin{equation}
  \begin{pmatrix} w^1 \\ w^2 \\ n_1 \\ n_2 \end{pmatrix} \xmapsto{~\theta~}
  \begin{pmatrix} 0 & -1 & 0 & 0 \\
    1 & -1 & 0 & 0 \\
    0 & 0 & -1 & -1 \\
    0 & 0 & 1 & 0  \end{pmatrix}\cdot
  \begin{pmatrix} w^1 \\ w^2 \\ n_1 \\ n_2 \end{pmatrix}\eqdef \theta\, \begin{pmatrix} w^1 \\ w^2 \\ n_1 \\ n_2 \end{pmatrix}\;.
\end{equation}
We stress that the meaning of the above equation is that the effect of transforming the string coordinates $X^\pm \xmapsto{\theta} \omega^{\mp 1}X^{\pm}$ in the exponential piece of the vertex operator can be equivalently achieved by transforming the $w^i$ and $n_i$ as specified above. 
It is important to note that the orbifold action consists in \eqref{eq:orbifold_action_on_X_pm} and not in the transformation of the $w^i$ and $n_i$. 
Lastly, one must remember to also transform any oscillators in the compact dimensions as detailed before.
For brevity of notation, we denote the action of the twist on the $\widehat{N} = (w^1,w^2,n_1,n_2)$ as $\widehat{N} \xmapsto{\theta} \theta\, \widehat{N}$. 
As expected, the matrix form of $\theta$ acting on the \ac{KK} and winding numbers is order 3.

Constructing states which are invariant under the orbifold action depends on how many oscillators states have in the compact dimension. 
The exponential part of a vertex operator can be used to construct three types of relevant quantities on the orbifold,
\begin{subequations}\label{app:orb_inv_vertex_ops}
\begin{align}
  V_1(\widehat{N})^{(\varrho)} &\defeq \frac{1}{\sqrt{3}}\sum_{i=1}^{3}V(\theta^i \widehat{N})\;,\\
  V_\omega (\widehat{N})^{(\varrho)} &\defeq \frac{1}{\sqrt{3}}\sum_{i=1}^{3}\omega^i\, V(\theta^i \widehat{N})\;,\\
  V_{\omega^{-1}}(\widehat{N})^{(\varrho)} &\defeq \frac{1}{\sqrt{3}}\sum_{i=1}^{3}\omega^{-i}\,V(\theta^i \widehat{N})\;.
\end{align}
\end{subequations}
One can verify that the action of the orbifold twist on each of these exponential pieces of vertex operators is
\begin{subequations}
\begin{align}
  V_1(\widehat{N})^{(\varrho)} &\xmapsto{~\theta~} V_1(\widehat{N})^{(\varrho)}\;,\\
  V_\omega(\widehat{N})^{(\varrho)} &\xmapsto{~\theta~} \omega\, V_\omega (\widehat{N})^{(\varrho)}\;,\\
  V_{\omega^{-1}}(\widehat{N})^{(\varrho)} &\xmapsto{~\theta~} \omega^{-1}\, V_{\omega^{-1}} (\widehat{N})^{(\varrho)}\;.
\end{align}
\end{subequations}
Recalling the oscillators in compact dimensions transform as $(\ii \partial X^\pm) \xmapsto{~\theta~} \omega^{\mp 1}(\ii \partial X^\pm)$ (while oscillators in non-compact dimensions are not affected by the twist), we can build orbifold-invariant states. 
For $(N_\mathrm{L},N_\mathrm{R})=(0,1/2)$ with the oscillator in the non-compact dimension, the orbifold invariant state is
\begin{equation}
  V_1 (\widehat{N})^{(\varrho)} \otimes (\ii \partial X_\mathrm{R}^\mu)\;.
\end{equation}
At a symmetry enhanced point, these would build \Group{SU}{3} gauge bosons on $\mathds{T}^2$, and on the orbifold they correspond to \Group{U}{1} gauge bosons. 
The scalars with $(N_\mathrm{L},N_\mathrm{R})=(0,1/2)$ have oscillators in the compact dimension, so the exponential and oscillator components of the operator must transform in such a way to leave the state invariant overall. 
This is easy to construct, the orbifold invariant states are
\begin{equation}
  V_\omega (\widehat{N})^{(\varrho)} \otimes (\ii \partial X_\mathrm{R}^+)
  \quad\text{and}\quad
  V_{\omega^{-1}}(\widehat{N})^{(\varrho)} \otimes (\ii \partial X_\mathrm{R}^-)\;.
\end{equation}

\subsection{Relevant untwisted states at \texorpdfstring{$\varrho = \omega$}{rho=omega}}
\label{sec:untwisted_vertex_operators_at_omega}

So far, the construction of these untwisted states has been rather general. 
Now, we fix $\varrho = \omega$ and construct the explicit states which are relevant to our discussions. We will simply state the definitions of various vertex operators here; proof that these states do indeed correspond to \Group{U}{1} gauge bosons or charge eigenstates of these gauge bosons via \acp{OPE} are given in \cref{app:OPEs}. 
We start by constructing the massless $\Group{U}{1} \times \Group{U}{1}$ gauge bosons at $\varrho = \omega$. 
We denote this gauge symmetry as $[\Group{U}{1}\times\Group{U}{1}]^{(\omega)}$. 
These states have $(N_\mathrm{L},N_\mathrm{R})=(0,1/2)$ with the right moving oscillator in the non-compact dimension. 
The relevant \ac{KK} and winding numbers for these gauge bosons are denoted as
\begin{equation}
  \widehat{N}_4 = (-1,0,-1,1)^{\transpose}\;,
\end{equation}
where the $4$ is denoting that at $\varrho = \omega$, the massless doublets that appear are in the $\rep{2}_4$ representation of \Group{\Delta}{54}. 
The vertex operators for the $[\Group{U}{1}\times\Group{U}{1}]^{(\omega)}$ gauge bosons are given by 
\begin{subequations}\label{app:gauge_bosons_omega}
\begin{align}
  \mathcal{V}^{(\omega)}_1 &\defeq \frac{\ii}{\sqrt{2}}\left(V_1 (-\widehat{N}_4)^{(\omega)} - V_1 (\widehat{N}_4)^{(\omega)}\right)\otimes (\ii \partial X_\mathrm{R}^\mu)\;,\\
  \mathcal{V}^{(\omega)}_2 &\defeq \frac{1}{\sqrt{2}}\left(V_1 (\widehat{N}_4)^{(\omega)} + V_1 (-\widehat{N}_4)^{(\omega)}\right)\otimes (\ii \partial X_\mathrm{R}^\mu)\;.
\end{align}
\end{subequations}
In addition, there are massless untwisted complex scalars at $\varrho = \omega$ which are charged under the $[\Group{U}{1}\times\Group{U}{1}]^{(\omega)}$ gauge bosons. These states have vertex operators
\begin{subequations}\label{app:Ui_omega}
\begin{align}
  U_1^{(\omega)} &\defeq \frac{1}{\sqrt{3}}\left( -\ii (\ii \partial X_\mathrm{L}^-) + V_\omega (\widehat{N}_4)^{(\omega)} + V_\omega (-\widehat{N}_4)^{(\omega)} \right) \otimes (\ii \partial X_\mathrm{R}^+)\;,\\
  U_2^{(\omega)} &\defeq \frac{1}{\sqrt{3}}\left( -\ii (\ii \partial X_\mathrm{L}^-) + \omega V_\omega (\widehat{N}_4)^{(\omega)} + \omega^2\, V_\omega (-\widehat{N}_4)^{(\omega)} \right) \otimes (\ii \partial X_\mathrm{R}^+)\;,\\
  U_3^{(\omega)} &\defeq \frac{1}{\sqrt{3}}\left( -\ii (\ii \partial X_\mathrm{L}^-) + \omega^2 V_\omega (\widehat{N}_4)^{(\omega)} + \omega\, V_\omega (-\widehat{N}_4)^{(\omega)} \right) \otimes (\ii \partial X_\mathrm{R}^+)\;.
\end{align}
\end{subequations}
The charges of the $U_1^{(\omega)}$, $U_2^{(\omega)}$, and $U_3^{(\omega)}$ under the $[\Group{U}{1}\times\Group{U}{1}]^{(\omega)}$ gauge symmetry can be expressed as combinations of simple roots of \Group{SU}{3}, $\alpha_{(1)} = (\sqrt{2},0)$, $\alpha_{(2)} = (-1/\sqrt{2},\sqrt{3/2})$, and $-\alpha_{(1)} - \alpha_{(1)} = (-1/\sqrt{2},-\sqrt{3/2})$, respectivley. 
Their $\mathcal{CPT}$ conjugates are given by
\begin{subequations}\label{app:UiBar_omega}
\begin{align}
  \overline{U}_1^{(\omega)} &\defeq \frac{1}{\sqrt{3}}\left( \ii (\ii \partial X_\mathrm{L}^+) + V_{\omega^{-1}} (\widehat{N}_4)^{(\omega)} + V_{\omega^{-1}} (-\widehat{N}_4)^{(\omega)} \right) \otimes (\ii \partial X_\mathrm{R}^-)\;,\\
  \overline{U}_2^{(\omega)} &\defeq \frac{1}{\sqrt{3}}\left( \ii (\ii \partial X_\mathrm{L}^+) + \omega\, V_{\omega^{-1}} (\widehat{N}_4)^{(\omega)} + \omega^2\, V_{\omega^{-1}} (-\widehat{N}_4)^{(\omega)} \right) \otimes (\ii \partial X_\mathrm{R}^-)\;,\\
  \overline{U}_3^{(\omega)} &\defeq \frac{1}{\sqrt{3}}\left( \ii (\ii \partial X_\mathrm{L}^+) + \omega^2\, V_{\omega^{-1}} (\widehat{N}_4)^{(\omega)} + \omega\, V_{\omega^{-1}} (-\widehat{N}_4)^{(\omega)} \right) \otimes (\ii \partial X_\mathrm{R}^-)\;.
\end{align}
\end{subequations}
The charges of the $\overline{U}_1^{(\omega)}$, $\overline{U}_2^{(\omega)}$, and $\overline{U}_3^{(\omega)}$ under the $[\Group{U}{1}\times\Group{U}{1}]^{(\omega)}$ gauge symmetry are the negatives of the charges of the $U_i^{(\omega)}$ in \eqref{app:Ui_omega}, $-\alpha_{(1)} = (-\sqrt{2},0)$, $-\alpha_{(1)} = (1/\sqrt{2},-\sqrt{3/2})$, and $\alpha_{(1)} + \alpha_{(1)} = (1/\sqrt{2},\sqrt{3/2})$, respectively.

\section{\acp{OPE}}\label{app:OPEs}
 
\subsection{\texorpdfstring{$\Group{U}{1}\times\Group{U}{1}$}{U(1)xU(1)} charges of untwisted states}

Given \Group{U}{1} gauge bosons $V_i$ and a \Group{U}{1} primary state $U$, we can find the charge $q_i$ of $U$ under the \Group{U}{1} generated by $V_i$.
This can be accomplished by looking at the $1/(z-w)$ singular piece of the \ac{OPE}, i.e.\ 
 \begin{equation}
   V_i\bigl(X(z)\bigr)\, U\bigl(X(w)\bigr) \sim \frac{q_i}{z-w}\,U\bigl(X(w)\bigr) + \dots \,,
\end{equation}
where ``$\dots$" denotes terms regular at $z=w$. 
The standard \acp{OPE} which will be used are
\begin{align}
  \ii \partial X^{j}(z)\,\ee^{\ii k\cdot X(w)} &\sim \frac{k_j}{z-w}\,\ee^{\ii k\cdot X(w)} + \dots\;,\label{eq:OPEs_partial}
   \\
  \ee^{\ii k_1 \cdot X(z)}\,\ee^{\ii k_2 \cdot X(w)} &\sim (z-w)^{k_1 \cdot k_2}\bigl(1 + (z-w)\ii k_1 \cdot \partial X(w) \bigr)\,\ee^{\ii (k_1+k_2)\cdot X(w)} \epsilon(k_1,k_2)+\dots\;.\label{eq:OPEs_exp}
\end{align}
These \acp{OPE} arise from the Frenkel--Ka\u c--Segal construction of affine Lie algebras applied to the heterotic string~\cite{Blumenhagen:2013fgp}. 
Here, we will focus on the enhanced \Group{SU}{3} which arises on the $\mathds{T}^2$ compactification of the heterotic string (the string actually provides a realization of some level $n$ Kac-Moody algebra, but this distinction will not be relevant in computing the \Group{U}{1} charges) and the eventual $\Group{U}{1} \times \Group{U}{1}$ gauge group that is left over after orbifolding to $\mathds{T}^2 / \mathds{Z}_3$. 
The phase $\epsilon(k_1,k_2)$ arises from cocycles \cite{Chun:1991js}, but can be computed directly from group-theoretical considerations. 
In particular, cocycle operators are needed to ensure both locality of the \acp{OPE} and that the group algebra (in our case \Group{SU}{3} Lie algebra) is satisfied. 
We work in the Cartan--Weyl basis of \Group{SU}{3} with the unconventional normalization of the generators being
\begin{align}
     \operatorname{Tr}\left( \mathsf{T}^a \mathsf{T}^b \right) = \delta^{ab}\;.
\end{align}
This will be useful later to match the normalization of the \Group{SU}{3} weights of various string states. 
The generators of the Cartan subalgebra of \Group{SU}{3}, $\mathsf{T}^3$ and $\mathsf{T}^8$, have the forms
\begin{align}
    \mathsf{T}^3 = \frac{1}{\sqrt{2}}\begin{pmatrix} 1 & 0 & 0 \\ 0 & -1 & 0 \\ 0 & 0 & 0 \end{pmatrix}\;,\quad \mathsf{T}^8 = \frac{1}{\sqrt{6}}\begin{pmatrix} 1 & 0 & 0 \\ 0 & 1 & 0 \\ 0 & 0 & -2 \end{pmatrix}\;.
\end{align}
The raising and lowering operators $\mathsf{E}_{\alpha}$ will generically be all zeroes except for a single $1$ in an off-diagonal entry. 
The raising and lowering operators are labeled by their associated root $\alpha = (\alpha_{(i)}^1,\alpha_{(i)}^2)$ which is found through the commutation relations
\begin{subequations}
\begin{align}
     [\mathsf{T}^3,\mathsf{E}_{\alpha}] &= \alpha^1\, \mathsf{E}_{\alpha}\;,\\ 
     [\mathsf{T}^8,\mathsf{E}_{\alpha}] &= \alpha^2\, \mathsf{E}_{\alpha}\;.
\end{align}    
\end{subequations}
For example, we have the simple roots of \Group{SU}{3} given by $\alpha_{(1)} = (\sqrt{2},0)$, $\alpha_{(2)} = (-1/\sqrt{2},\sqrt{3/2})$, and find
\begin{equation}
    \mathsf{E}_{\alpha_{(1)}} = \begin{pmatrix} 0 & 1 & 0 \\ 0 & 0 & 0 \\ 0 & 0 & 0 \end{pmatrix}\;,\quad \mathsf{E}_{\alpha_{(2)}} = \begin{pmatrix} 0 & 0 & 0 \\ 0 & 0 & 1 \\ 0 & 0 & 0 \end{pmatrix}\;.
\end{equation}

On the string, the Cartan subalgebra of \Group{SU}{3} on $\mathds{T}^2$ is generated by the operators $\ii \partial X_\mathrm{L}^1(z)\otimes \partial X_\mathrm{R}^\mu (z)$ and $\ii \partial X_\mathrm{L}^2(z) \otimes \partial X_\mathrm{R}^\mu(z)$, or linear combinations of these operators. 
The raising and lowering operators are associated with the string states that have $N_\mathrm{L} = 0$, $N_\mathrm{R} = 1/2$ acting on the \ac{NS} vacuum. 
Their vertex operators (without cocycle operators attached) have the form
 \begin{equation}
     \mathsf{E}_{n,m}(z) = \exp\bigl( \ii (n\, \alpha_{(1)} + m\, \alpha_{(2)})\cdot X_\mathrm{L}(z) \bigr) \otimes \partial X_\mathrm{R}^\mu (z)\;.
   \end{equation}
Using \eqref{eq:OPEs_partial}, one can see that the root associated with the raising or lowering operator $\mathsf{E}_{n,m}$ is $n\,\alpha_{(1)} + m\,\alpha_{(2)}$. 
Note that the right-moving oscillators carry a Lorentz vector index in the non-compact dimensions, and will not be expanded in the \acp{OPE} we will need to compute.

Cocycle factors will arise in the \acp{OPE} between $\mathsf{E}_{n_1,m_1}$ and $\mathsf{E}_{n_2,m_2}$. 
Let $\alpha = n_1\, \alpha_{(1)} + m_1\,\alpha_{(2)}$, $\beta = n_2\, \alpha_{(1)} + m_2\, \alpha_{(2)}$. 
In our convention, when $\alpha = - \beta$ (which means $n_1 = -n_2$ and $m_1 = -m_2$, and $\alpha\cdot \beta = -2$), the cocycle factor is $1$ i.e.\ $\epsilon(\alpha,-\alpha) = 1$. 
Otherwise, the cocycle factor $\epsilon(\alpha,\beta)$ can be determined through the \Group{SU}{3} commutation relations
 \begin{equation}
     \left[\mathsf{E}_{\alpha},\mathsf{E}_{\beta}\right] = \begin{dcases}
         N_{\alpha,\beta}\mathsf{E}_{\alpha+\beta} &\text{for } \alpha\cdot \beta = -1\;,\\
         \alpha\cdot \mathsf{H} &\text{for } \alpha\cdot \beta = -2\;,\\
         0 &\text{else}\;,
       \end{dcases}
\end{equation}
where $\mathsf{H} = (\mathsf{T}^3,\mathsf{T}^8)$ are the generators of the Cartan subalgebra of \Group{SU}{3}.
In this basis, the $N_{\alpha,\beta}$ will take on values $(-1,0,1)$. 
If the commutator vanishes, one can check that the corresponding \ac{OPE} will be non-singular and so no adjustment is needed from cocycles. 
Otherwise, provided $\alpha\neq -\beta$, we take $\epsilon(\alpha,\beta)=N_{\alpha,\beta}$ which will take values in $(-1,1)$. 
The \ac{OPE} between states associated with roots of \Group{SU}{3} then has the form
\begin{equation}
     \mathsf{E}_{n_1,m_1}(z)\,\mathsf{E}_{n_2,m_2}(w) \sim \frac{N_{\alpha,\beta}}{z-w}\,\mathsf{E}_{n_1+n_2,m_1+m_2}(w)\;.
\end{equation}
One can verify that the \acp{OPE} with $\epsilon(\alpha,\beta)=-1$ are
\begin{align}
\nonumber
     \epsilon(\alpha_{(1)},-\alpha_{(1)} - \alpha_{(2)}) &= \epsilon(-\alpha_{(1)},- \alpha_{(2)}) = \epsilon(\alpha_{(2)},\alpha_{(1)})\\
      &= \epsilon(-\alpha_{(2)},\alpha_{(1)}+\alpha_{(2)}) = \epsilon(\alpha_{(1)} + \alpha_{(2)},-\alpha_{(1)}) = \epsilon(-\alpha_{(1)} - \alpha_{(2)},\alpha_{(2)}) = -1\;.
   \end{align}
The rest of the cocycle factors can be taken to be equal to $1$.

Now, we turn to investigate the \Group{U}{1} charges of the states which appear at the symmetry enhanced point $\varrho = \tau = \omega$ after orbifolding.
We note that the \Group{U}{1} charge eigenstates at other symmetry enhanced points $\varrho'$ related to $\omega$ by an $\Group{SL}{2,\mathds{Z}}_{\varrho}$ transformation are constructed in an identical manner once the \ac{KK} and winding numbers corresponding to the various raising and lowering operators of the \Group{SU}{3} at $\varrho'$ are determined.
The right moving oscillators will not play a role in the \acp{OPE} we compute, so we will not write them here. 
Once again, let $\widehat{N}_4 = (-1,0,-1,1)^{\transpose}$. 
Then, the orbifold invariant untwisted vertex operators as defined in \Cref{app:orb_inv_vertex_ops} at $\varrho = \omega$ can be written in terms of these operators associated with \Group{SU}{3} raising and lowering operators as
\begin{subequations}   
\begin{align}
     V_1(\widehat{N}_4)^{(\omega)} &= \frac{1}{\sqrt{3}}\left( \mathsf{E}_{1,0}+\mathsf{E}_{0,1}+\mathsf{E}_{-1,-1} \right)\;,\\
     V_\omega (\widehat{N}_4)^{(\omega)} &= \frac{1}{\sqrt{3}}\left( \mathsf{E}_{1,0}+\omega^2\, \mathsf{E}_{0,1} + \omega\, \mathsf{E}_{-1,-1} \right)\;,\\
     V_1(-\widehat{N}_4)^{(\omega)} &= \frac{1}{\sqrt{3}}\left( \mathsf{E}_{-1,0}+\mathsf{E}_{0,-1}+\mathsf{E}_{1,1} \right)\;,\\
     V_\omega(-\widehat{N}_4)^{(\omega)} &= \frac{1}{\sqrt{3}}\left( \mathsf{E}_{-1,0}+\omega^2\, \mathsf{E}_{0,-1} + \omega\, \mathsf{E}_{1,1} \right)\;.
\end{align}
\end{subequations}
One can derive the following useful \acp{OPE}:
\begin{subequations}
\begin{align}
     V_1(\widehat{N}_4)^{(\omega)}\, V_\omega (\widehat{N}_4)^{(\omega)} &\sim \frac{\ii V_\omega (-\widehat{N}_4)^{(\omega)}}{z-w}\;,\\
     V_1(\widehat{N}_4)^{(\omega)}\, V_\omega (-\widehat{N}_4)^{(\omega)} &\sim \frac{\ii\partial X_\mathrm{L}^-}{z-w}\;,\\
     V_1(\widehat{N}_4)^{(\omega)}\, \ii\partial X_\mathrm{L}^- &\sim -\frac{V_\omega (\widehat{N}_4)^{(\omega)}}{z-w}\;,\\
     V_1(-\widehat{N}_4)^{(\omega)}\, V_\omega (\widehat{N}_4)^{(\omega)} &\sim -\frac{\ii\partial X_\mathrm{L}^-}{z-w}\;,\\
     V_1(-\widehat{N}_4)^{(\omega)}\, V_\omega (-\widehat{N}_4)^{(\omega)} &\sim -\frac{\ii V_\omega (\widehat{N}_4)^{(\omega)}}{z-w}\;,\\
     V_1(-\widehat{N}_4)^{(\omega)}\, \ii\partial X_\mathrm{L}^- &\sim \frac{V_\omega (-\widehat{N}_4)^{(\omega)}}{z-w}\;,\\
     V_1(\widehat{N}_4)^{(\omega)}\, V_1(\widehat{N}_4)^{(\omega)} &\sim 0\;,\\
     V_1(-\widehat{N}_4)^{(\omega)}\, V_1(-\widehat{N}_4)^{(\omega)} &\sim 0\;,\\
     V_1(\widehat{N}_4)^{(\omega)}\, V_1(-\widehat{N}_4)^{(\omega)} &\sim \frac{1}{(z-w)^2}\;,\\
     V_1(-\widehat{N}_4)^{(\omega)}\, V_1(\widehat{N}_4)^{(\omega)} &\sim \frac{1}{(z-w)^2}\;.
\end{align}
\end{subequations}
The two \Group{U}{1} gauge bosons on the orbifold at $\varrho = \tau = \omega$ as defined in \Cref{app:gauge_bosons_omega} have vertex operators
\begin{align}
     \mathcal{V}_1^{(\omega)} &= \frac{\ii}{\sqrt{2}}\left( V_1(-\widehat{N}_4)^{(\omega)} - V_1(\widehat{N}_4)^{(\omega)} \right) \otimes \ii \partial X_\mathrm{R}^\mu\;,\\
     \mathcal{V}_2^{(\omega)} &= \frac{1}{\sqrt{2}}\left( V_1(\widehat{N}_4)^{(\omega)} + V_1(-\widehat{N}_4)^{(\omega)} \right)\otimes \ii \partial X_\mathrm{R}^\mu\;.
\end{align}
To start, we need to ensure these vectors satisfy the current algebra for two \Group{U}{1} gauge bosons \cite{Harvey:2017rko}. 
This means we need
\begin{align}
     \mathcal{V}_i^{(\omega)}(z)\,\mathcal{V}_j^{(\omega)}(w) \sim \frac{\delta_{ij}}{(z-w)^2}+\dots\;.
\end{align}
This leads to the \acp{OPE},
\begin{subequations}
\begin{align}
   \mathcal{V}_1^{(\omega)}\,\mathcal{V}_1^{(\omega)} &\sim \frac{1}{(z-w)^2}\;,
     &
     \mathcal{V}_1^{(\omega)}\, \mathcal{V}_2^{(\omega)} &\sim 0\;,\\
   \mathcal{V}_2^{(\omega)}\,\mathcal{V}_2^{(\omega)} &\sim \frac{1}{(z-w)^2}\;,
     &
   \mathcal{V}_2^{(\omega)}\, \mathcal{V}_1^{(\omega)} &\sim 0\;,
\end{align}
\end{subequations}
as required. 
We can then look for states with well-defined \Group{U}{1} charges under the $\mathcal{V}_i^{(\omega)}$ gauge bosons. 
The $U_{i}^{(\omega)}$ states, defined in \Cref{app:Ui_omega}, have charges given by the simple roots of \Group{SU}{3}. 
Their charges can be read off from the \acp{OPE}
\begin{subequations}
\begin{align}
     \mathcal{V}_1^{(\omega)}\,U_1^{(\omega)} & \sim \frac{\sqrt{2}}{z-w}U_1^{(\omega)}\;,
        &
     \mathcal{V}_2^{(\omega)}\,U_1^{(\omega)} & \sim 0\;,\\
     \mathcal{V}_1^{(\omega)}\,U_2^{(\omega)} & \sim -\frac{1}{\sqrt{2}}\frac{1}{z-w}U_2^{(\omega)} \;,
        &
     \mathcal{V}_2^{(\omega)}\,U_2^{(\omega)} & \sim \sqrt{\frac{3}{2}}\frac{1}{z-w}U_2^{(\omega)} \;,\\
     \mathcal{V}_1^{(\omega)}\,U_3^{(\omega)} & \sim -\frac{1}{\sqrt{2}}\frac{1}{z-w}U_3^{(\omega)} \;,
        &
     \mathcal{V}_2^{(\omega)}\,U_3^{(\omega)} & \sim -\sqrt{\frac{3}{2}}\frac{1}{z-w}U_3^{(\omega)} \;.
\end{align}
\end{subequations}
This means that $U_1^{(\omega)}$ has charge $\alpha_{(1)} = (\sqrt{2},0)$, $U_2^{(\omega)}$ has charge $\alpha_{(2)} = (-1/\sqrt{2},\sqrt{3/2})$, and $U_3^{(\omega)}$ has charge $-\alpha_{(1)}-\alpha_{(2)} = (-1/\sqrt{2},-\sqrt{3/2})$. 
A similar computation shows that the $\overline{U}_1^{(\omega)}$, $\overline{U}_2^{(\omega)}$, and $\overline{U}_3^{(\omega)}$ states defined in \Cref{app:UiBar_omega} have charges $-\alpha_{(1)}$, $-\alpha_{(2)}$, and $\alpha_{(1)} + \alpha_{(2)}$, respectively.
The charges of the massless untwisted \Group{U}{1} primary fields at $\varrho = \tau = \omega$ are given in \Cref{tab:U1timesU1_omega_1} of the main text.
The $U_i^{(\omega)}$ states are built solely out of winding modes with no left-moving oscillators ($N_\mathrm{L} = 0$).

\subsection{\ac{OPE}s between two twisted localization eigenstates}

As detailed in \cite[Equation~(3.24)]{Lauer:1990tm}, the \ac{OPE} between two twisted localization eigenstates yields a sum over an infinite number of untwisted states with momenta determined by the orbifold twist and two localization eigenstates. In the case of the $\mathds{Z}_3$ orbifold, one can solve these \ac{OPE}s to show that untwisted states labeled by the mod $3$ quantum numbers $(\OrbW,\OrbN)$, defined in \Cref{eq:orbifold_tower_quantum_numbers}, are associated with linear combinations of twisted localization eigenstates \cite{Nilles:2018wex}. We label these sums over untwisted states with mod $3$ quantum numbers $(\OrbW,\OrbN)$ as $V^{(\OrbW,\OrbN)}$. We will not need the explicit form of this sum. The result is
\begin{subequations}
\begin{align}
  \begin{pmatrix}V^{(2,0)} \\ V^{(1,0)} \end{pmatrix} &= \frac{1}{\sqrt{3}}\begin{pmatrix}
    \overline{X}\, Z+\overline{Y}\, X+\overline{Z}\, Y \\
    \overline{Z}\, X+\overline{X}\, Y+\overline{Y}\, Z
  \end{pmatrix}\;,\\
  \begin{pmatrix}V^{(0,1)} \\ V^{(0,2)} \end{pmatrix} &= \frac{1}{\sqrt{3}}\begin{pmatrix}
    \overline{X}\, X+\omega\,\overline{Y}\, Y+\omega^2\,\overline{Z}\, Z\\
      \overline{X}\, X+\omega^2\,\overline{Y}\, Y+\omega\,\overline{Z}\, Z
    \end{pmatrix}\;,\\
  \begin{pmatrix}V^{(2,1)} \\ V^{(1,2)} \end{pmatrix} &= \frac{1}{\sqrt{3}}\begin{pmatrix}
    \omega\,\overline{X}\,Z+\omega^2\,\overline{Y}\,X+\overline{Z}\,Y\\
    \omega^2\,\overline{Z}\,X+\omega\,\overline{X}\,Y+\overline{Y}\,Z
  \end{pmatrix}\;,\\
  \begin{pmatrix}V^{(1,1)} \\ V^{(2,2)} \end{pmatrix} &= \frac{1}{\sqrt{3}}\begin{pmatrix}
    \omega ^2\,\overline{X}\,Y+\overline{Y}\,Z+\omega\,\overline{Z}\, X\\
    \omega\,\overline{Y}\, X+\overline{Z}\,Y+\omega ^2\,\overline{X}\,Z
  \end{pmatrix}\;.
\end{align}
\end{subequations}

\section{Vertex operator transformations under \texorpdfstring{$\Group{SL}{2,\mathds{Z}}_\varrho \times \Group{SL}{2,\mathds{Z}}_\tau$}{SL(2,Z)_rho x SL(2,Z)_tau}}\label{app:vertex_op_SL2Z_trans}

\subsection{\texorpdfstring{$\Group{SL}{2,\mathds{Z}}_\varrho$}{SL(2,Z)_rho} Transformations}

In order for the modular transformations to be a symmetry of our theory, we should ensure that the underlying CFT obeys this symmetry as well. 
These conditions are detailed in \cite{Ibanez:1990ju}, and we repeat the relevant information below. 
Recall the quantities
\begin{align}
     X_{\mathrm{L}/\mathrm{R}}^{\pm} &\defeq \frac{1}{\sqrt{2}}\left(X^1_{\mathrm{L}/\mathrm{R}}\pm\ii X^2_{\mathrm{L}/\mathrm{R}} \right)\;,\\
     p_{\mathrm{L}/\mathrm{R}}^{\pm} &\defeq \frac{1}{\sqrt{2}}\left(p^1_{\mathrm{L}/\mathrm{R}}\pm\ii p^2_{\mathrm{L}/\mathrm{R}} \right)\;.
\end{align}
Modular transformations in general do not leave the $p_{\mathrm{L}/\mathrm{R}}^{\pm}$ invariant, but cause background dependent rotations on the momentum vectors. 
In the $\mathds{T}^2$ compactified theory, in order for $\Group{SL}{2,\mathds{Z}}_\varrho$ transformations to be a true symmetry of the string theory we must have that these transformations are not only symmetries of the mass spectrum but also preserve the S-matrix. 
This requirement means we need the winding and \ac{KK} mode vertex operators, $V \sim \exp(X^+ \cdot p^- + X^- \cdot p^+)$, to remain invariant under modular transformations; the $X^+$ and $X^-$ must rotate opposite the $p^-$ and $p^+$, respectively. 
Under a transformation $\gamma \in \Group{SL}{2,\mathds{Z}}_{\varrho}$ of the K\"ahler modulus $\varrho \xmapsto{~\gamma~} \frac{a\,\varrho + b}{c\,\varrho + d}$, theses fields transform as
\begin{subequations}
\begin{align}
     (p_\mathrm{L}^+)\big|_{\varrho} &\xmapsto{~\gamma~} \lambda\, \left( \frac{c\,\overline{\varrho}+d}{c\,\varrho+d} \right)^{1/2}(p_\mathrm{L}^+)\big|_{\gamma\cdot \varrho}\;,&
     (p_\mathrm{R}^+)\big|_{\varrho} &\xmapsto{~\gamma~} \lambda\, \left( \frac{c\,\varrho + d}{c\,\overline{\varrho}+d} \right)^{1/2}(p_\mathrm{R}^+)\big|_{\gamma \cdot \varrho}\;,\\
     X_\mathrm{L}^+ &\xmapsto{~\gamma~} \lambda\, \left( \frac{c\,\varrho+d}{c\,\overline{\varrho}+d} \right)^{1/2}X_\mathrm{L}^+\;,&
     X_\mathrm{R}^+ &\xmapsto{~\gamma~} \lambda\, \left( \frac{c\,\overline{\varrho} + d}{c\,\varrho+d} \right)^{1/2}X_\mathrm{R}^+\;,
\end{align}
\end{subequations}
with the $X^-$ and $p^-$ transformations given by conjugation, and $(\cdots)\big|_{\varrho}$ denoting the evaluation of the quantity in parenthesis at the point where the K\"ahler modulus takes the value $\varrho$ (and similarly for the point $\gamma \cdot \varrho$). 
The $p_{\mathrm{L}/\mathrm{R}}$ depend explicitly on the value of the K\"ahler modulus, while the string coordinates $X_{\mathrm{L}/\mathrm{R}}$ do not which is why we neglect to give them any labelling for the value of the K\"ahler modulus. 
$\lambda$ is an additional phase which depends only on the parameters of the modular transformation, and on the presentation of \Group{SL}{2,\mathds{Z}} used. 
These transformations ensure the $X\cdot p = X^+ \cdot p^- + X^- \cdot p^+$ remains invariant so that the winding modes which build the \Group{SU}{3} gauge bosons fulfill the correct \acp{OPE} to build an \Group{SU}{3} gauge symmetry at another enhanced symmetry point under modular transformations. 
Note that these transformations do not leave oscillators in the compact dimensions invariant. 
In particular, under the same transformation of the K\"ahler modulus $\varrho \xmapsto{~\gamma~} \frac{a\varrho + b}{c\varrho + d}$ we have
\begin{subequations}
\begin{align}
     \ii\partial X_\mathrm{L}^+ &\mapsto \overline{\lambda}\, \left( \frac{c\,\varrho+d}{c\,\overline{\varrho}+d} \right)^{1/2}\ii \partial X_\mathrm{L}^+\;,\\
     \ii \partial X_\mathrm{R}^+ &\mapsto \overline{\lambda}\, \left( \frac{c\,\overline{\varrho} + d}{c\,\varrho+d} \right)^{1/2}\ii \partial X_\mathrm{R}^+\;,
\end{align}
\end{subequations}
with the $\ii \partial X^-$ transformation rules given by conjugation. 
The field associated with the degree of freedom corresponding to the K\"ahler modulus, $\widetilde{\varrho} = \partial X_\mathrm{L}^- \partial X_\mathrm{R}^+$, then transforms as
 \begin{equation}\label{eq:stringRhoTrans}
     (\widetilde{\varrho})\big|_{\varrho} \mapsto \left( \frac{c\,\overline{\varrho}+d}{c\,\varrho+d} \right)(\widetilde{\varrho})\big|_{\gamma \cdot \varrho}\;.
\end{equation}

\subsection{Unbroken \texorpdfstring{$\Group{SL}{2,\mathds{Z}}_{\tau}$}{SL(2,Z)_tau} transformations}

After orbifolding, the complex structure modulus $\tau$ is frozen at $\tau = \omega$. 
 Generically, $\Group{SL}{2,\mathds{Z}}_\tau$ transformations are then not allowed as they change the complex structure modulus. 
However, there are two transformations of which $\tau = \omega$ is a fixed point. These are $\mathsf{S}^2_\tau$ and $(\mathsf{S\,T})_\tau$ which enact $\omega \mapsto \omega$ and $\omega \mapsto -1/(1+\omega) = \omega$, respectively. 
These act on the momentum and winding numbers (see \cref{app:ST_tau_mat}) as
\begin{subequations}\label{app:S2_and_ST_tau}
\begin{align}
     \mathsf{S}^2_\tau &= \begin{pmatrix}
                   -1 & 0 & 0 & 0 \\
                   0 & -1 & 0 & 0 \\
                   0 & 0 & -1 & 0 \\
                   0 & 0 & 0 & -1
                   \end{pmatrix}\;,\\ 
     \quad (\mathsf{S\,T})_\tau &=  \begin{pmatrix}
                   0 & -1 & 0 & 0 \\
                   1 & -1 & 0 & 0 \\
                   0 & 0 & -1 & -1 \\
                   0 & 0 & 0 & 1
                   \end{pmatrix}\;.
\end{align}
\end{subequations}
Just as in the $\Group{SL}{2,\mathds{Z}}_\varrho$ case, one must rotate the coordinates $X_{\mathrm{L}/\mathrm{R}}^\pm$ to compensate for any rotation of the $p_{\mathrm{L}/\mathrm{R}}^\pm$ such that the exponential components of the vertex operator $\exp(X^+\cdot p^- + X^- \cdot p^+)$ remains invariant. 
This once again leads to rotations of the oscillators in compact dimensions. 
Under the two unbroken $\Group{SL}{2,\mathds{Z}}_\tau$ transformation, one finds that
\begin{subequations}\label{eq:app_momentum_rotations}
\begin{align}
     p_{\mathrm{L}/\mathrm{R}}^\pm &\xmapsto{~\eqmathbox[STStau]{\mathsf{S}^2_\tau}~} -p_{\mathrm{L}/\mathrm{R}}^\pm\;,& 
    X_{\mathrm{L}/\mathrm{R}}^\pm &\xmapsto{~\eqmathbox[STStau]{\mathsf{S}^2_\tau}~} -X_{\mathrm{L}/\mathrm{R}}^\pm\;, \\
     p_{\mathrm{L}/\mathrm{R}}^\pm &\xmapsto{~\eqmathbox[STStau]{(\mathsf{S\,T})_\tau}~} \omega^{\pm 1}\, p_{\mathrm{L}/\mathrm{R}}^\pm\;, &
    X_{\mathrm{L}/\mathrm{R}}^\pm &\xmapsto{~\eqmathbox[STStau]{(\mathsf{S\,T})_\tau}~} \omega^{\mp 1}\, X_{\mathrm{L}/\mathrm{R}}^\pm\;.
\end{align}
\end{subequations}
From this, we see $\mathsf{S}^2_\tau$ acts as an order 2 transformation on the untwisted states, and $(\mathsf{S\,T})_\tau$ acts as an order 3 transformation on the untwisted states. 
Together, $(\mathsf{S^3\,T})_\tau$ is an order 6 transformation on the untwisted states,
\begin{subequations}\label{app:Z6_S3T}
\begin{align}  
   p_{\mathrm{L}/\mathrm{R}}^\pm &\xmapsto{~(\mathsf{S^3\,T})_\tau~} -\omega^{\pm 1}\, p_{\mathrm{L}/\mathrm{R}}^\pm\;,\\ 
  X_{\mathrm{L}/\mathrm{R}}^\pm &\xmapsto{~(\mathsf{S^3\,T})_\tau~} -\omega^{\mp 1}\, X_{\mathrm{L}/\mathrm{R}}^\pm\;,
\end{align}  
\end{subequations}
with matrix representation acting on the \acp{KK} and winding numbers as
\begin{equation}
   (\mathsf{S^3\,T})_\tau = \begin{pmatrix}
     0 & 1 & 0 & 0 \\
     -1 & 1 & 0 & 0 \\
     0 & 0 & 1 & 1 \\
     0 & 0 & -1 & 0
   \end{pmatrix}\;.
 \end{equation}

\subsection{\texorpdfstring{$(\mathsf{S\,T})_\tau$}{(ST)_tau} transformation and the orbifold twist $\theta$}
\label{sec:ST_transformation_and_orbifold_twist}

Although the matrix representation of $(\mathsf{ST})_\tau$ and the orbifold twist $\theta$ acting on the momentum and winding numbers are identical (cf.\ \Cref{app:S2_and_ST_tau,eq:orbifold_action_def}, respectively), we stress that the action of these operations on string states are different. 
In particular, the modular transformations are both a symmetry of the mass spectrum \emph{and} the S-matrix. 
This means that when we perform a modular transformation, we shift the value of the appropriate modulus, transform the $n_i$ and $w^i$ to preserve the mass spectrum, and rotate the string coordinates to compensate for any rotation of the left and right moving momenta in the compact dimensions. 
This ensures all \acp{OPE} in the $\mathds{T}^2$ theory are preserved after the modular transformation (if the left or right moving momenta were rotated by the string coordinates were not, then the resulting vertex operator would have a different \ac{OPE} with the gauge bosons at enhanced symmetry points for example, and this transformation would not leave general correlation functions of string states invariant). 
The orbifold twist instead is just a rotation of the string coordinates $X^+ \mapsto \omega^2\, X^+$ and $X^- \mapsto \omega\, X^-$. 
One does not compensate for the rotation of any left or right moving momenta by an additional rotation of the string coordiantes, so in general it will map states to new states which have different left- and right-moving momenta, and consequently different \acp{OPE} with states such as the gauge bosons at symmetry enhanced points. 
Clearly, this transformation is not a symmetry of the S-matrix and consequently the theory. 
If the orbifold twist $\theta$ were a symmetry of the theory on $\mathds{T}^2$, we would not be able to use it to project out states and define a new theory. 
We instead build string states which are invariant under the orbifold transformation, as it is not a symmetry of the theory on $\mathds{T}^2$.

As an explicit example, let us check how the twist $\theta$ and a generic modular transformation of $\Group{SL}{2,\mathds{Z}}_\tau \times \Group{SL}{2,\mathds{Z}}_\varrho$, which we denote as $\gamma$, act on an untwisted string vertex operator.  
Let $\alpha(\gamma)$ denote the phase rotation that the $p^+_{\mathrm{L}/\mathrm{R}}$ recieve under $\gamma$, and $\alpha(\gamma)^{-1}$ is the phase rotation the $p^-_{\mathrm{L}/\mathrm{R}}$ receive.  
For example, from \Cref{eq:app_momentum_rotations} one can see $\alpha\big((\mathsf{S\,T})_\tau\big) = \omega$. 
For concreteness we consider a state with $N_\mathrm{R} = 1/2$, $N_\mathrm{L} = 0$, and left and right moving momenta in the compact dimensions $p^{\pm}_{\mathrm{L}/\mathrm{R}}$. 
These states appear as the \Group{SU}{3} gauge bosons on $\mathds{T}^2$ or the $\Group{U}{1} \times \Group{U}{1}$ gauge bosons on $\mathds{T}^2/\mathds{Z}_3$ when the right moving oscillator is in the non-compact dimensions. 
When the oscillator is in the compact dimension, these are complex scalars which are eaten by the gauge bosons when the gauge symmetry is broken. 
If the oscillator is in the non-compact dimensions, the state transforms as
\begin{subequations}
\begin{align}
     \exp\bigl(X^+ \cdot p^- + X^- \cdot p^+\bigr)\otimes \ii \partial X_\mathrm{R}^\mu &\xmapsto{~\theta~} \exp\bigl(\omega^2\, X^+ \cdot p^- + \omega\, X^- \cdot p^+\bigr)\otimes \ii \partial X_\mathrm{R}^\mu \;, \\
     \exp\bigl(X^+ \cdot p^- + X^- \cdot p^+\bigr)\otimes \ii \partial X_\mathrm{R}^\mu &\xmapsto{~\gamma~} \exp\bigl(X^+ \cdot p^- + X^- \cdot p^+\bigr)\otimes \ii \partial X_\mathrm{R}^\mu\;.
\end{align}
\end{subequations}
This tells us the \Group{SU}{3} gauge bosons on $\mathds{T}^2$ are invariant under modular transformations, while they transform under the orbifold action.
The orbifold invariant states which define the $\Group{U}{1} \times \Group{U}{1}$ gauge bosons on the orbifold would be invariant under both transformations as expected. 
Now consider the case where the oscillator is in the non-compact dimension. 
In this case, the oscillator will be one of the $\ii \partial X_\mathrm{R}^\pm$. 
The transformation of the state is
\begin{subequations}
\begin{align}
     \exp\bigl(X^+ \cdot p^- + X^- \cdot p^+\bigr)\otimes \ii \partial X_\mathrm{R}^\pm &\xmapsto{~\theta~} \exp\bigl(\omega^2\, X^+ \cdot p^- + \omega\, X^- \cdot p^+\bigr)\otimes \bigl(\omega^{\mp 1} \ii \partial X_\mathrm{R}^\pm\bigr) \;, \\
     \exp\bigl(X^+ \cdot p^- + X^- \cdot p^+\bigr)\otimes \ii \partial X_\mathrm{R}^\pm &\xmapsto{~\gamma~} \exp\bigl(X^+ \cdot p^- + X^- \cdot p^+\bigr)\otimes \bigl( \alpha(\gamma)^{\pm 1} \ii \partial X_\mathrm{R}^\pm\bigr)\;,
\end{align}
\end{subequations}
where the $\alpha(\gamma)^{\pm 1}$ reflects the fact that $\ii \partial X^+$ transforms the same way as $p^+$ and vice versa. 
Again, we see these transformations are not identical. 
The orbifold action generically associates the original state to a new state with different left- and right-moving momenta. 
The modular transformation instead multiplies the state by an overall phase $\alpha(\gamma)^{\pm1}$.
This phase is related to the automorphy factor under $\Group{SL}{2,\mathds{Z}}_\tau$ \cite{Ibanez:1990ju}. 

\section{String selection rules}
\label{sec:String_selection_rules}

\begin{wraptable}[10]{r}{4.6cm}
\vspace*{-1em}\centering$\begin{array}{ccc}
  \toprule 
  \text{state} & \mathds{Z}_3^\mathrm{PG} & \mathds{Z}_3^\mathrm{SG}\\
  \midrule 
  X & 1 & 0\\
  Y & 1 & 1\\
  Z & 1 & 2\\
  \midrule 
  \overline{X} & 2 & 0\\
  \overline{Y} & 2 & 2\\
  \overline{Z} & 2 & 1\\
  \bottomrule
\end{array}$
\caption{String selection rules.}
\label{tab:String_selection_rules}
\end{wraptable}

The purpose of this appendix is to collect some basic facts on string selection rules on orbifolds \cite{Hamidi:1986vh,Dixon:1986qv}.  %
Strings on orbifolds can be labeled by elements of the space group $(\theta^{k^{(i)}},\ell^{(i)})$. 
Here, $\theta$ denotes the rotational (or twist) part whereas the second entry, $\ell^{(i)}$, is a lattice translation. 
The string selection rules state that, in order for a coupling to be allowed, the space group elements have to such that their product can yield identity,
\begin{equation}
  \prod\limits_i(\theta^{k^{(i)}},\ell^{(i)})\ni(\mathds{1},0)\;.
\end{equation}
In the $\mathds{Z}_3$ orbifold plane these rules demand that (cf.\ e.g.\ \cite{Kobayashi:2006wq})
\begin{subequations}\label{eq:String_selection_Z_3_x_Z_3}
\begin{align}
 \sum_i k^{(i)}&=0\bmod 3\;,\label{eq:Z_3_PG}\\
 \sum_i \ell^{(i)}&=0\bmod 3\;.\label{eq:Z_3_SG}
\end{align}  
\end{subequations}
These rules can be regarded as the product of $\mathds{Z}_3^\mathrm{PG}\times \mathds{Z}_3^\mathrm{SG}$, where ``PG'' and ``SG'' stand for ``point group'' and ``space group'', respectively.
The corresponding charges of the localization eigenstates are shown in \Cref{tab:String_selection_rules}.

\bibliography{ModularGauge}
\bibliographystyle{utphys}

\begin{acronym}
  \acro{2HDM}{two-Higgs doublet model}  
  \acro{BSM}{beyond the standard model}
  \acro{BU}{bottom-up}
  \acro{EFT}{effective field theory}
  \acro{FCNC}{flavor changing neutral current}
  \acro{FN}{Froggatt--Nielsen}
  \acro{CG}{Clebsch--Gordan}
  \acro{IO}{inverted ordering}
  \acro{KK}{Kaluza--Klein}
  \acro{CFT}{conformal field theory}
  \acro{CSA}{Cartan subalgebra}
  \acro{MIHO}{modular invariant holomorphic observables}
  \acro{MSSM}{minimal supersymmetric standard model}
  \acro{NO}{normal ordering}
  \acro{NS}{Neveu--Schwarz}
  \acro{OPE}{operator product expansion}
  \acro{QFT}{quantum field theory}
  \acro{R}{Ramond}
  \acro{RG}{renormalization group}
  \acro{RGE}{renormalization group equation}
  \acro{SM}{standard model}
  \acro{SUSY}{supersymmetry}
  \acro{TD}{top-down}
  \acro{UV}{ultraviolet}
  \acro{VEV}{vacuum expectation value}
  \acro{VVMF}{vector-valued modular forms}
\end{acronym}
\end{document}